\documentclass[reprint,superscriptaddress,aps,prc]{revtex4-2}
\usepackage{amsmath,amssymb}
\usepackage{bm,graphicx}
\usepackage[colorlinks,citecolor=blue]{hyperref}

\DeclareMathOperator{\sgn}{sgn}
\newcommand{\VF}[2]{\ensuremath{{}^V\!F^{(#1)}_{#2}}}
\newcommand{\AF}[2]{\ensuremath{{}^AF^{(#1)}_{#2}}}
\newcommand{\VAF}[2]{\ensuremath{{}^{V\!/A}F^{(#1)}_{#2}}}

\newcommand{\VM}[2]{\ensuremath{{}^V\!\mathfrak{M}^{(#1)}_{#2}}}
\newcommand{\AM}[2]{\ensuremath{{}^A\mathfrak{M}^{(#1)}_{#2}}}
\newcommand{\VAM}[2]{\ensuremath{{}^{V\!/A}\mathfrak{M}^{(#1)}_{#2}}}

\newcommand{\Vm}[2]{\ensuremath{{}^V\!m^{(#1)}_{#2}}}
\newcommand{\Am}[2]{\ensuremath{{}^Am^{(#1)}_{#2}}}
\newcommand{\VAm}[2]{\ensuremath{{}^{V\!/A}m^{(#1)}_{#2}}}

\begin{document}

\title{Forbidden electron capture on $^{24}$Na and $^{27}$Al in degenerate oxygen-neon cores}

\author{D.~F.~Str\"omberg}
\email{d.fahlinstroemberg@gsi.de}
\affiliation{Institut f{\"u}r Kernphysik (Theoriezentrum),
    Fachbereich Physik, Technische Universit{\"a}t
    Darmstadt, Schlossgartenstra{\ss}e 2, 64298 Darmstadt, Germany}
\affiliation{GSI Helmholtzzentrum f\"ur Schwerionenforschung,
  Planckstra{\ss}e~1, 64291 Darmstadt, Germany}
\affiliation{Helmholtz Forschungsakademie Hessen f\"ur FAIR, GSI
    Helmholtzzentrum f\"ur Schwerionenforschung,
  Planckstra{\ss}e~1,
    64291 Darmstadt, Germany} 
  
\author{G.~Mart\'inez-Pinedo}
\email{g.martinez@gsi.de}
\affiliation{GSI Helmholtzzentrum f\"ur Schwerionenforschung,
  Planckstra{\ss}e~1, 64291 Darmstadt, Germany}
\affiliation{Institut f{\"u}r Kernphysik (Theoriezentrum),
    Fachbereich Physik, Technische Universit{\"a}t
    Darmstadt, Schlossgartenstra{\ss}e 2, 64298 Darmstadt, Germany}
\affiliation{Helmholtz Forschungsakademie Hessen f\"ur FAIR, GSI
    Helmholtzzentrum f\"ur Schwerionenforschung,
  Planckstra{\ss}e~1,
    64291 Darmstadt, Germany} 

\author{F.~Nowacki}
\affiliation{Universit\'e de Strasbourg, CNRS, IPHC UMR 7178,
F-67000 Strasbourg, France} 

\begin{abstract}
  \begin{description}
  \item[Background]
	  Stars with an initial mass of $\sim7$--11 solar masses form
	  degenerate oxygen-neon cores following carbon burning.
	  Electron captures in such cores can trigger runaway oxygen
	  burning, resulting in either a collapse or a thermonuclear
	  explosion. Previous work constrained the contribution of the
	  forbidden $0^+_\text{g.s.}\rightarrow{}2^+_\text{g.s.}$
	  transition to the $^{20}\text{Ne}(e^-,\nu_e)^{20}\text{F}$
	  rate and discussed its significance for the evolution of the
	  core.
  \item[Purpose]
	  We provide a detailed description of the formalism used in previous
	  work and apply it to two further forbidden transitions that
	  are relevant to degenerate oxygen-neon cores: the
	  $4^+_\text{g.s.}\rightarrow{}2^+_1$ transition in
	  $^{24}\text{Na}(e^-,\nu_e)^{24}\text{Ne}$ and the
	  ${5/2^+_{\text{g.s.}} \rightarrow{} 1/2^+_{\text{g.s.}}}$
	  transition in $^{27}\text{Al}(e^-,\nu_e)^{27}\text{Mg}$.
  \item[Method]
	  The relevant nuclear matrix elements are determined through
	  shell model calculations and constraints from CVC theory. We then
	  investigate the astrophysical impact using the stellar evolution code
	  MESA (Modules for Experiment in Stellar Astrophysics) and through
	  timescale arguments.
  \item[Results]
	  In the relevant temperature range, the forbidden transitions substantially
	  reduce the threshold densities for
	  $^{24}\text{Na}(e^-,\nu_e)^{24}\text{Ne}$ and
	  $^{27}\text{Al}(e^-,\nu_e)^{27}\text{Mg}$. In the MESA
	  models, $^{24}\text{Na}(e^-,\nu_e)^{24}\text{Ne}$ now occurs
	  immediately following the onset of
	  $^{24}\text{Mg}(e^-,\nu_e)^{24}\text{Na}$. The impact on the
	  overall evolution is uncertain: this is due to known
	  difficulties in accounting for convective instabilities
	  triggered by the $A=24$ electron captures. The transition
	  between $^{27}\text{Al}$ and $^{27}\text{Mg}$ may have a
	  minor effect on the early evolution but is unlikely to affect
	  the outcome.
  \item[Conclusions]
	  The studied transitions should be included when calculating weak
	  interaction rates between $^{24}$Na and $^{24}$Ne for
	  temperatures $\log_{10}(T[\text{K}])\lesssim8.5$ and between
	  $^{27}$Al and $^{27}$Mg for
	  $\log_{10}(T[\text{K}])\lesssim8.8$.
  \end{description}
\end{abstract}

\maketitle

\section{Introduction}
Stars with an initial mass of roughly 7 to 11 solar masses (M$_\odot$) form
super asymptotic giant branch (Super-AGB) stars following carbon
burning~\cite{doherty2017super,jones2013advanced,takahashi2013evolution}. Such
stars have a degenerate oxygen-neon (ONe) core composed mostly of $^{16}$O and
$^{20}$Ne, with minor amounts of other nuclei such as $^{23}$Na, $^{24}$Mg,
$^{25}$Mg, and $^{27}$Al. Thermal pulses in the surrounding shells add mass to
the core and causes it to contract. Eventually it may reach densities where
exothermic electron capture processes ignite oxygen burning, triggering a
thermal runaway. This event is traditionally known as an electron capture
supernova (ECSN) and is believed to result in either a collapse (cECSN) to a
neutron star~\cite{miyaji1980supernova,leung2020electron} or a thermonuclear
explosion (tECSN) with an ONeFe white dwarf
remnant~\cite{isern1991outcome,jones2016electron}. If the growth of the core
is too slow the Super-AGB star will expel its envelope before an ECSN occurs.
The resulting ONe white dwarf may still reach the threshold density for
electron capture by accreting matter from a binary companion. This will lead to
either a collapse or a thermonuclear explosion, analogous to the cECSN and
tECSN scenarios described above.

The final outcome of an ECSN depends on both the ignition conditions
and the physics of the subsequent runaway. In particular, collapse is
expected if the central density at ignition $\rho_c^{\text{ign}}$ is
larger than a critical density $\rho_c^{\text{crit}}$. Recent
three-dimensional hydrodynamical simulations~\cite{jones2016electron}
indicate that
$\rho_c^\text{crit}\approx(1-2)\times10^{10}$~g~cm$^{-3}$
while similar two-dimensional studies~\cite{leung2020electron} suggest
that
$\rho_c^\text{crit}\approx(7.9-8.9)\times10^{9}$~g~cm$^{-3}$.
Simulations of the evolution of ONe cores up to the point of ignition
have shown that $\rho_c^\text{ign}$ is roughly of the order
$\sim10^{10}$~g~cm$^{-3}$, but there is no consensus on
whether it exceeds the above given values of
$\rho_c^\text{crit}$~\cite{miyaji1980supernova,nomoto1987evolution,
  miyaji1987collapse,canal1992quasi, hashimoto1993type,
  gutierrez1996final, gutierrez2005gravitational,jones2013advanced,
  takahashi2013evolution,schwab2015thermal,*schwab2015erratum,schwab2017importance}.
Substantial uncertainties are related to the role of convection, the
precise composition of the core, and the details of the different
electron capture processes.

The effect of a given electron capture depends on the mass number $A$ of the
parent nucleus. Capture on odd-$A$ nuclei generally results in an Urca process
where cycles of electron capture and $\beta^-$ decay (e.g.
${^{25}\text{Mg}+e^-\rightarrow{}^{25}\text{Na}+\nu_e}$ and
${^{25}\text{Na}\rightarrow{}^{25}\text{Mg}+e^-+\bar{\nu}_e}$) cool through
neutrino-antineutrino emission. On the other hand, for an even-$A$ parent nucleus the energy
threshold for a subsequent capture on the odd-$A$ daughter is typically lower
than for the initial capture. This leads to an exothermic double electron
capture such as
$^{20}\text{Ne}(e^-,\nu_e)^{20}\text{F}(e^-,\nu_e)^{20}\text{O}$.

Prior to ignition the electron capture and $\beta^-$ decay rates are in general
determined by a limited number of transitions. This is because the relatively
low temperatures $({kT\lesssim100~\text{keV}}$) ensure that only low-lying
excited states are thermally populated. The rate
tabulations~\cite{oda1994rate,takahara1989microscopic} used historically only
took allowed transitions into account. However, in
Ref.~\cite{martinez2014astrophysical} it was shown that the second-forbidden
non-unique transition (\mbox{$\Delta J^\pi=2^+$}) between the ground states of
$^{20}$Ne and $^{20}$F could have a substantial effect on the capture rate.
Ref.~\cite{kirsebom2019measurement} constrained the strength of this transition
based on a combination of theory and experiment. The astrophysical impact was
studied in Ref.~\cite{kirsebom2019discovery}, which found that the forbidden
transition caused a moderate reduction of $\rho_c^\text{ign}$ while
simultaneously shifting the point of ignition away from the center of the core.

This paper expands on the work of
Ref.~\cite{kirsebom2019measurement,kirsebom2019discovery} by providing a more
complete account of the formalism, and by applying it to two further
second-forbidden transitions of interest. The first of these is the transition
between the $4^+_\mathrm{g.s.}$ ground state in $^{24}$Na and the $2^+_1$
excited state in $^{24}$Ne. Its possible astrophysical implications has
previously been explored in Ref.~\cite{schwab2017importance} for various
estimates of the transition strength. We also investigate the
$5/2^+_\mathrm{g.s.}\rightarrow{}1/2^+_\mathrm{g.s.}$ transition between the
ground states of $^{27}$Al and $^{27}$Mg: it was ignored in previous
studies~\cite{jones2013advanced,toki2013detailed} of electron capture on
$^{27}$Al in degenerate ONe cores. In addition to the above two transitions, we
also test our approach by computing the previously measured
$4^+_\mathrm{g.s.}\rightarrow{}2^+_1$ transition in the $\beta^-$ decay of
$^{24}$Na.

The paper is organized in the following manner: In Section~\ref{sec:formalism}
we introduce the general expressions needed to calculate the weak interaction
rates, while we discuss the corresponding nuclear matrix elements in
Section~\ref{sec:matrix_elements}. We then present the calculated rates for the
forbidden transitions in Section~\ref{sec:rate} and discuss the astrophysical
ramifications in Section~\ref{sec:impact}. Finally, we state our conclusions in
Section~\ref{sec:conclusions}.

\section{Weak interaction rates}
\label{sec:formalism}
During the evolution leading up to oxygen ignition the core reaches densities
and temperatures of the order
$\rho=10^8$--$10^{10}$~g~cm$^{-3}$ and
$T=10^7$--$10^{9}$~K. This means that the nuclei are
completely ionized and the electrons can be described as a relativistic and
degenerate Fermi gas. The weak interaction rates in such a plasma can be
calculated as detailed by Fuller, Fowler and Newman in
Ref.~\cite{fuller1980stellar}. To also include the contribution of forbidden
transitions we incorporate the formalism of Behrens and
B{\"u}hring~\cite{behrens1971nuclear,behrens1982electron,Bambynek.Behrens.ea:1977,*Bambynek.Behrens.ea:1977err} in our
treatment. We will now present a summary of this approach. The reader can find
a more detailed account in Ref.~\cite{stroemberg2020weak}.

\subsection{Key quantities}
The temperatures that we encounter are high enough to thermally populate
excited nuclear states. Assuming that the nuclei are in thermal equilibrium
with the environment the total electron capture or $\beta^-$ decay rate is
\begin{equation}
	\label{eq:lambda_total}
	\lambda^{\text{EC}/\beta^-}=\frac{1}{G(Z,A,T)}\sum_{if}(2J_i+1)\lambda^{\text{EC}/\beta^-}_{if}e^{-E_i/(kT)}
\end{equation}
where the sum runs over all possible initial $i$ and final $f$ states. $Z$ and
$A$ refer to the atomic and mass numbers of the parent nucleus and the quantity
${G(Z,A,T)=\sum_{i}(2J_i+1)e^{-E_i/(kT)}}$ is the corresponding partition
function.

The partial rates in \eqref{eq:lambda_total} are given by the expressions
\begin{subequations}
\begin{align}
	\label{eq:ecrate}
	\begin{split}
		\lambda^\text{EC}_{if}=\frac{\ln{2}}{K}\int_{w_l}^{\infty}&C(w)wp_e(q_{if}+w)^2F(Z,w)\\
		\times&S_e(w)dw
	\end{split}\\
	\label{eq:brate}
	\begin{split}
		\lambda^{\beta^-}_{if}=\frac{\ln{2}}{K}\int_1^{q_{if}}&C(w)wp_e(q_{if}-w)^2F(Z+1,w)\\
		\times&[1-S_e(w)]dw.
	\end{split}
\end{align}
\end{subequations}
From measurements of superallowed decays the value of the constant $K$ can be
constrained to $K=6144\pm2~\text{s}$~\cite{hardy2009superallowed}. The
integration variable $w$ is the electron energy (including the rest mass) in
units of $m_ec^2$, where $m_e$ is the electron mass. Similarly,
$p_e=\sqrt{w^2-1}$ is the electron momentum in units of $m_ec$. We also define
$q_{if}$ as the difference in energy between the final and initial states
\begin{equation}
	\label{eq:qif}
	q_{if}=\frac{Q_{if}}{m_ec^2}=\frac{M_pc^2-M_dc^2+E_i-E_f}{m_ec^2},
\end{equation}
again in units of $m_ec^2$. Here $M_{p}$ and $M_{d}$ refer to the nuclear
masses of the parent and daughter nuclei, whereas $E_i$ and $E_f$ are the
initial and final excitation energies. For electron capture the lower limit of
integration is
\begin{equation*}
	w_l=
	\begin{cases}
		|q_{if}| & \text{if }q_{if}<-1\\
		1 &  \text{if }q_{if}>-1.
	\end{cases}
\end{equation*}
In this work we only encounter electron capture reactions where $q_{if}<-1$ and
thus ${w_l=-q_{if}}$.

$F(Z,w)$ is the Fermi function that arises from the Coulomb distortion
of the electron wave function around the nucleus. Furthermore, the function
$S_e(w)$ describes the energy distribution of the electrons in the degenerate
Fermi gas. It is given by the Fermi-Dirac distribution
\begin{equation}
	\label{eq:Se}
	S_e(w)=\frac{1}{\exp\left(\frac{wm_ec^2-\mu_e}{kT}\right)+1}
\end{equation}
with $\mu_e$ being the electron chemical potential, again including the rest
mass. We can determine $\mu_e$ from the density $\rho$, electron fraction $Y_e$
and temperature $T$ by solving the equation
\begin{equation}
	\rho Y_e=\frac{m_u}{\pi^2}\left(\frac{m_ec}{\hbar}\right)^3\int_0^\infty(S_e-S_p)p^2dp
\end{equation}
where $m_u$ is the atomic mass unit and $S_p$ is the positron energy
distribution. The latter is like $S_e$ given by a Fermi-Dirac distribution
\eqref{eq:Se}, but with the opposite sign of the chemical potential (i.e.
$\mu_p=-\mu_e$).

All details on the nuclear structure of the involved nuclei are contained in
the shape factor $C(w)$. For allowed transitions it is constant with respect to
$w$. Typically we are only interested in Gamow-Teller transitions in which case
we have
\begin{equation}
	C(w)=g_A^2\frac{|\langle{}f\|\sum_k\mathbf{\sigma}^kt^k_\mp\|i\rangle{}|^2}{2J_i+1}.
\end{equation}
Here $g_A=-1.2762(5)$~\cite{Zyla:2020zbs} is the weak axial coupling constant, $\mathbf{\sigma}^k$ is
the spin operator and $t_\mp^k$ refers to the isospin lowering and raising
operators (corresponding to $\beta^-$ decay and electron capture,
respectively). The summation index $k$ runs over all $A$ nucleons.

For second-forbidden non-unique transitions the shape factor depends on $w$ and
is given by an expression involving eight different nuclear matrix elements. We
discuss the determination of the matrix elements in
Section~\ref{sec:matrix_elements} and present the full expressions for $C(w)$
in Appendix~\ref{sec:app_shape}. The dependence on $w$ of the shape factor is
of the form
\begin{equation}
	\label{eq:Cw}
	C(w)=a_0+\frac{a_{-1}}{w}+a_1w+a_2w^2+a_3w^3+a_4w^4.
\end{equation}

\subsection{Coulomb corrections}
Coulomb interactions between the electrons and ions have a significant impact
on the properties of the high-density plasma. As a consequence we must
introduce corrections to some of the quantities presented in the preceding
section. Here we follow the treatment of Ref.~\cite{bravo1999coulomb,
juodagalvis2010improved}.

The most important effect is an increase in the energy difference between
nuclei with different charges. Consequently we must replace $q_{if}$ in
\eqref{eq:qif} with in-medium values according to
\begin{subequations}
\begin{align}
	\label{eq:qmed_EC}
	q_{if}^{\text{EC,med}} &=q_{if}^{\text{EC}}-\Delta q_{if}(Z)\\
	\label{eq:qmed_beta}
	q_{if}^{\beta^-,\text{med}}&=q_{if}^{\beta^-}+\Delta q_{if}(Z+1).
\end{align}
\end{subequations}
This translates to a larger energy barrier for electron capture, essentially
delaying the onset of capture to higher densities.

A further correction arises due to the fact that the energy of the captured or
emitted electron is modified in the presence of the background electron gas. To
account for this we replace the electron chemical potential in \eqref{eq:Se}
with
\begin{equation}
	\label{eq:mue_screen}
	\mu_e^\text{med}=\mu_e-V_s.
\end{equation}
We direct the reader to Ref.~\cite{bravo1999coulomb, juodagalvis2010improved,
itoh2002screening} for details on the calculation of $\Delta q_{if}(Z)$ and
$V_s$.

\subsection{Specific heating rates}
Under the assumption that the weak interactions are slow compared to the
timescale needed to maintain thermal equilibrium the specific heating rates for
electron capture and $\beta^-$ decay can be written
\begin{subequations}
\begin{align}
	\label{eq:specific_energy_EC}\dot{\epsilon}^{\text{EC}}&=\frac{Y}{m_u}\Bigl[\mu_e+Q_\text{g.s.}^\text{EC}-\langle{}E_\nu\rangle{}^{\text{EC}}\Bigr]\lambda^\text{EC}\\
  \label{eq:specific_energy_betam}\dot{\epsilon}^{\beta^-}&=\frac{Y}{m_u}
\Bigl[Q_\text{g.s.}^{\beta^-}-\mu_e-\langle{}E_\nu\rangle{}^{\beta^-}\Bigr]\lambda^{\beta^-}
                                                            .
\end{align}
\end{subequations}
Here $Y$ is the abundance of the parent nucleus and $m_u$ is the atomic mass
unit. Furthermore, $Q_\text{g.s.}$ denotes the energy difference between the
ground states of the parent and daughter nuclei (including Coulomb
corrections). Note that for the nuclei considered here we have
$Q_\text{g.s.}^{\beta^-}=-Q_\text{g.s.}^{\text{EC}} > 0$.
$\langle{}E_\nu\rangle{}^{\text{EC}/\beta^-}$ is the average energy of the
emitted neutrinos. Finally, $\mu_e$ is the electron chemical potential without
the correction term in \eqref{eq:mue_screen}.

The average neutrino energy is given by the ratio
\begin{equation}
	\langle{}E_\nu\rangle{}^{\text{EC}/\beta^-}=\frac{\xi^{\text{EC}/\beta^-}}{\lambda^{\text{EC}/\beta^-}}
\end{equation}
where $\lambda^{\text{EC}/\beta^-}$ is the reaction rate
\eqref{eq:lambda_total} and $\xi^{\text{EC}/\beta^-}$ is the rate of energy
loss through neutrino emission. The latter is calculated as
\begin{equation}
	\label{eq:xi_total}
	\xi^{\text{EC}/\beta^-}=\frac{1}{G(Z,A,T)}\sum_{if}(2J_i+1)\xi^{\text{EC}/\beta^-}_{if}e^{-E_i/(kT)}
\end{equation}
with the partial neutrino loss rates being
\begin{subequations}
\begin{align}
	\label{eq:ecxi}
	\begin{split}
		\xi^\text{EC}_{if}=m_ec^2\frac{\ln{2}}{K}\int_{w_l}^{\infty}&C(w)wp_e(q_{if}+w)^3\\
		\times&F(Z,w)S_e(w)dw
	\end{split}\\
	\label{eq:bxi}
	\begin{split}
		\xi^{\beta^-}_{if}=m_ec^2\frac{\ln{2}}{K}\int_1^{q_{if}}&C(w)wp_e(q_{if}-w)^3\\
		\times&F(Z+1,w)[1-S_e(w)]dw.
	\end{split}
\end{align}
\end{subequations}
Note that we now have one extra power of the neutrino energy compared to
\eqref{eq:ecrate} and \eqref{eq:brate}.

\section{Nuclear matrix elements}
\label{sec:matrix_elements}
\subsection{Form factor coefficients}
The expressions for the shape factor $C(w)$ contain quantities $\VAF{N}{KLs}$
known as form factor coefficients. In the impulse approximation they correspond
to nuclear matrix elements $\VAM{N}{KLs}$ according to
\begin{subequations}
\begin{align}
	\VF{N}{KLs}&=(-1)^{K-L}\,\VM{N}{KLs}\\
	\AF{N}{KLs}&=-g_A(-1)^{K-L}\,\AM{N}{KLs}.
\end{align}
\end{subequations}
Here $V$ and $A$ denote vector and axial matrix elements, respectively.
Furthermore, $K$ is the rank of the corresponding spherical tensor operator,
which arises by coupling the orbital angular momentum $L$ and spin $s$ of the
leptons. The label $N$ refers to the order in which the coefficients appear
when expanding the corresponding form factors in powers of $qR$, where $q$ is
the momentum transfer and $R$ is the nuclear radius.

In the leading order second-forbidden non-unique transitions depend on a total
of eight form factor coefficients. These include $\VF{0}{211}$, $\VF{0}{220}$,
$\AF{0}{221}$ and $\AF{0}{321}$, as well as four further coefficients denoted
by $\VF{0}{220}(k_e,1,1,1)$ and $\AF{0}{221}(k_e,1,1,1)$, where $k_e=1,2$. The
latter differ from $\VF{0}{220}$ and $\AF{0}{221}$ in that the corresponding
operators contain an additional factor $I(k_e,1,1,1;r)$ that modifies their
radial dependence. Note that only seven form factor coefficients appeared in
Ref.~\cite{kirsebom2019measurement}: The reason is that $\AF{0}{321}$
(corresponding to a rank 3 operator) does not satisfy the triangular selection
rule for the $0^+\rightarrow{}2^+$ transition that occurs in
$^{20}\text{Ne}(e^-,\nu_e)^{20}\text{F}$. This is the opposite of
second-forbidden \emph{unique} transitions ($\Delta J^\pi=3^+$), which only
depend on \AF{0}{321}.

We discuss the $I(k_e,1,1,1;r)$ factors further in Appendix~\ref{sec:app_I} and
list the single-particle matrix elements corresponding to the different form
factor coefficients in Appendix~\ref{sec:app_spme}.

\subsection{Shell model calculations}
To determine the nuclear matrix elements we perform shell model calculations in
the $sd$-shell valence space using the ANTOINE
code~\cite{caurier1999present,caurier2005shell} and the USDB
interaction~\cite{brown2006new}. In our calculations we assume harmonic
oscillator single-particle wave functions. Following the procedure described in
Ref.~\cite{towner1977analogue} we can relate the nuclear radius $R$ and
oscillator length $b$ to measurements of the charge mean-square radius
$\langle{}r^2\rangle{}_\text{exp}$. We take the values of the latter from x-ray
spectroscopy of muonic atoms~\cite{fricke1995nuclear}. For the mass number
$A=24$ we then get $R=3.947$~fm and $b=1.829$~fm, and for $A=27$ we have
$R=3.954$~fm and $b=1.808$~fm.

Our choice of valence space means that we ignore all $2\hbar\omega$
excitations. To account for this we should renormalize the operators in the
nuclear matrix elements. The axial matrix elements may also be affected by the
fact that the axial current is not conserved in the nuclear medium. Analogous
to the approach used for Gamow-Teller and unique forbidden transitions (see
e.g. Ref.~\cite{Warburton:1992,Martinez-Pinedo.Vogel:1998,suhonen2017}) we
attempt to account for both effects by using two different values of the axial
coupling constant: $g_A=-1.276$ (bare) and $g_A=-1.0$ (quenched). We do not
modify the operators contained in the coefficients of the type \VF{0}{220}:
these are essentially of the same type as isovector E2 transitions, which are
usually not renormalized in shell model calculations. Finally, \VF{0}{211} is a
special case which we discuss in the next section.

\subsection{Constraints from CVC theory}
A limitation of our approach is that an $sd$-shell calculation with harmonic
oscillator wave functions yields $\VF{0}{211}=0$ identically. To arrive at a
non-zero value of $\VF{0}{211}$ we can instead relate it to $\VF{0}{220}$ via
conserved vector current (CVC) theory as described in
Ref.~\cite{behrens1982electron}. For $\beta^-$ decay we then have the following
relation:
\begin{equation}
	\label{eq:cvc}
	\VF{0}{211} = -\frac{1}{\sqrt{10}}\left(\frac{E_\gamma}{m_ec^2}\right)R\,\VF{0}{220}.
\end{equation}
Here $E_\gamma$ is the energy of the analog E2 gamma decay in the
$\beta^-$ daughter nucleus and $R$ is the nuclear radius in units of
the reduced electron Compton wavelength
${\lambdabar_e=\hbar/(m_ec)}$. For electron capture the same relation
applies but with a positive sign.

In Ref.~\cite{kirsebom2019measurement} we further related $\VF{0}{220}$ in the
forbidden transition between $^{20}$F and $^{20}$Ne to the experimentally
measured strength $B(E2)$ of the analog gamma decay in $^{20}$Ne. When combined
with \eqref{eq:cvc} and the assumption that the ratios
$\VF{0}{220}(k_e,1,1,1)/\VF{0}{220}$ are well-described by the shell model
calculation this allowed us to determine the entire vector contribution. We do
not use this method for the forbidden transitions between $^{24}\text{Na}$ and
$^{24}\text{Ne}$ or $^{27}\text{Al}$ and $^{27}\text{Mg}$ as the corresponding
$B(E2)$ strengths are poorly constrained. In addition, in the
$^{24}\text{Na}(e^-,\nu_e)^{24}\text{Ne}$ case the axial contribution appears
to play a more important role than for the forbidden transition in
$^{20}\text{Ne}(e^-,\nu_e)^{20}\text{F}$. This means that determining the exact
magnitude of the vector form factor coefficients may not be as helpful in
constraining the electron capture rate.

\section{Calculated rates}
\label{sec:rate}
\subsection{$^{24}\text{Mg}\leftrightarrow{}^{24}\text{Na}$}
\label{sec:exp24}
Apart from the forbidden decay of $^{20}$F studied in
Ref.~\cite{kirsebom2019measurement}, there is only one~\cite{singh1998}
experimentally measured second-forbidden non-unique transition among nuclei
with $A<36$: the $4^+_\mathrm{g.s.}\rightarrow{}2^+_1$ transition in the
$\beta^-$ decay of $^{24}$Na. We have calculated the corresponding form factor
coefficients as presented in Table~\ref{tbl:ffc24exp}. There are two sets of
values: one (SM) taken directly from our shell model calculations and one
(SM+CVC) where we have determined \VF{0}{211} via the CVC relation
\eqref{eq:cvc}. When applying \eqref{eq:cvc} we used the value
$E_\gamma=8.148$~MeV, which we deduced from the energy level data in
Ref.~\cite{firestone2007nuclear}. For this particular transition \AF{0}{321} is
several times larger than the other form factor coefficients. However, this
does not necessarily mean that it is several times more important in
determining the decay rate: the impact of a form factor coefficient also
depends on the factors multiplied by it in the expression for the shape factor
$C(w)$.

\begin{table}[htb]
  \centering
  \caption{\label{tbl:ffc24exp}
	Form factor coefficients for the forbidden transition from the
	$4^+_\mathrm{g.s.}$ state in $^{24}$Na to the $2^+_1$ state in
	$^{24}$Mg. These values have been computed assuming $g_A=-1.276$.}
  \begin{ruledtabular}
    \begin{tabular}{lcc}
      Form factor coefficient	& SM		& SM+CVC\\ \hline
      \VF{0}{211}		& 0 		&  0.00144\footnote{Calculated from \VF{0}{220} via the CVC relation.}\\
      \VF{0}{220}            	& $-$0.0279     &  $-$0.0279\\
      \VF{0}{220}$(1,1,1,1)$	& $-$0.0352     &  $-$0.0352\\
      \VF{0}{220}$(2,1,1,1)$	& $-$0.0339	&  $-$0.0339\\
      \AF{0}{221}       	& $-$0.0361	&  $-$0.0361\\
      \AF{0}{221}$(1,1,1,1)$	& $-$0.0452	&  $-$0.0452\\
      \AF{0}{221}$(2,1,1,1)$	& $-$0.0435	&  $-$0.0435\\
      \AF{0}{321}       	& $-$0.121	&  $-$0.121
    \end{tabular}
  \end{ruledtabular}
\end{table}

\begin{table}[htb]
  \centering
  \caption{\label{tbl:logft24exp}
	$\log ft$ values calculated from Table~\ref{tbl:ffc24exp}, with and
	without quenching of the axial form factor coefficients, compared with
	the experimental value given in Ref.~\cite{turner1951lxvi}.
	}
  \begin{ruledtabular}
    \begin{tabular}{lll}
      Form factor coefficients		&  $g_A$    & $\log ft$ \\ \hline
      SM				&  $-1.276$ & $12.24$ \\
      SM				&  $-1.0$   & $12.39$ \\
      SM+CVC				&  $-1.276$ & $12.48$ \\
      SM+CVC				&  $-1.0$   & $12.68$ \\ \hline
      Exp.				&  --      & $12.7$
    \end{tabular}
  \end{ruledtabular}
\end{table}

From the form factor coefficients and the resulting $\beta^-$ decay shape factors
$C(w)$ we have calculated the $\log(ft)$ values listed in
Table~\ref{tbl:logft24exp}. Our theoretical results can be compared with the
measurement $\log(ft)=12.7$ given in Ref.~\cite{turner1951lxvi}. (Note that
Ref.~\cite{firestone2007nuclear} also lists the value $\log(ft)=11.34$: this is
based on intensity balance arguments and not on any direct measurements.) To
arrive at the total forbidden decay rate in Ref.~\cite{turner1951lxvi} the
authors had to extrapolate from the observed region of the electron energy
spectrum. In doing so, they effectively assumed that the spectrum was shaped like
that of an allowed transition. By their own estimate the transition may in
reality be a factor two weaker, giving  $\log(ft)=13$.

The results in Table~\ref{tbl:logft24exp} indicate that we
overestimate the transition strength by a factor $\sim1$--3. Comparing
with $\log(ft)=13$, we instead get a factor $\sim2$--6. The agreement
is better using the non-zero value of $\VF{0}{211}$ given by the CVC
relation, and the quenched axial contribution. Compared with previous
calculations for second-forbidden \emph{unique} transitions, where
shell-model calculations in the $sd$-shell generally predict
half-lives that are within a factor two
experiment~\cite{Warburton:1992,Martinez-Pinedo.Vogel:1998}, we seem
to do somewhat worse. This is related to the complicated interferences
between different matrix elements. However, the astrophysical
scenarios studied in this work do not appear to be very sensitive to
the exact strength of the forbidden transitions: as we shall see in
Section~\ref{sec:impact} rates that differ by a factor 2--3 produce
very similar results.

\subsection{$^{24}\text{Na}\leftrightarrow{}^{24}\text{Ne}$}
\label{sec:rate24}

\begin{figure}[hbt]
  \centering
  \includegraphics[width=\linewidth]{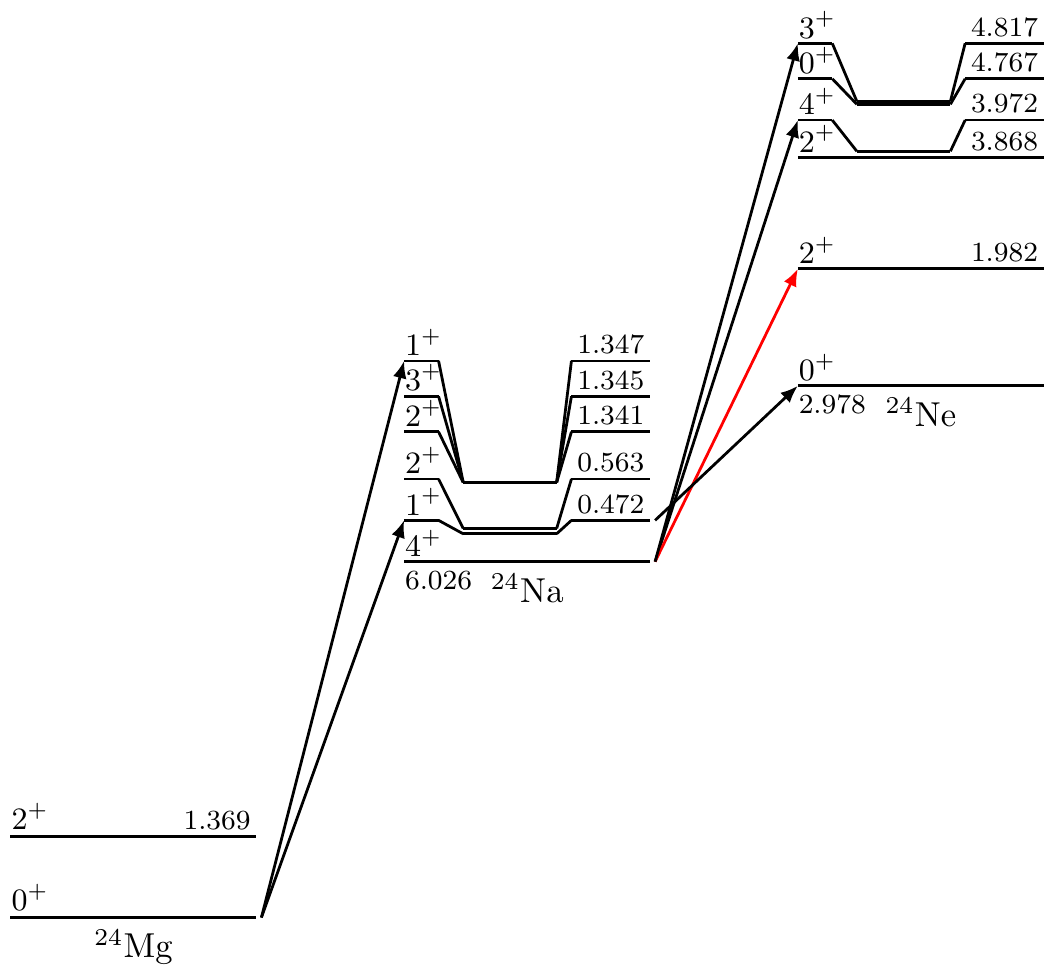}
  \caption{\label{fig:energylevels24}
	Energy level diagram illustrating the
	$^{24}\text{Mg}(e^-,\nu_e)^{24}\text{Na}(e^-,\nu_e)^{24}\text{Ne}$
	double electron capture. We list excitation energies and, for $^{24}$Na
	and $^{24}$Ne, the ground state energy relative to the ground state of
	the parent nucleus. All energies are in units of MeV. The nuclear data
	has been taken from Ref.~\cite{firestone2007nuclear,
	martinez2014astrophysical}. Relevant transitions are shown with arrows,
	with the forbidden transition in red.}
\end{figure}

We illustrate the transitions relevant to the
$^{24}\text{Mg}(e^-,\nu_e)^{24}\text{Na}(e^-,\nu_e)^{24}\text{Ne}$ double
electron capture in Fig.~\ref{fig:energylevels24}. As the transitions between
the ground states are fourth forbidden ($\Delta J^\pi=4^+$) they are too weak
to produce any significant electron capture rate. Instead, the
$^{24}\text{Mg}(e^-,\nu_e)^{24}\text{Na}$ reaction proceeds via the allowed
$0^+_\mathrm{g.s.}\rightarrow{}1^+_1$ transition to the first excited state in
$^{24}$Na at $0.472~\text{MeV}$. A subsequent
$^{24}\text{Na}(e^-,\nu_e)^{24}\text{Ne}$ capture can then occur through
${1^+_1\rightarrow{}0^+_\mathrm{g.s.}}$. Note that the reverse $\beta^-$ decay
rate is Pauli blocked: since $\mu_e$ is at this point substantially larger than
the maximal electron energy, the factor $[1-S_e(w)]$ in \eqref{eq:brate}
vanishes in the interval of integration.

The situation is quite different if the temperature is so low that the
population of the $1^+_1$ state becomes negligible. In this case the rate of
the $^{24}\text{Na}(e^-,\nu_e)^{24}\text{Ne}$ capture is determined by the two
transitions $4^+_\mathrm{g.s.}\rightarrow{}2^+_1$ (second forbidden) and
$4^+_\mathrm{g.s.}\rightarrow{}4^+_1$ (allowed). Note that the energy threshold
of the latter ($\sim6.95~\text{MeV}$) is slightly higher than for the initial
capture on $^{24}$Mg ($\sim6.59~\text{MeV}$). This means that ignoring the
forbidden transition we at low temperatures expect to see two separate capture
reactions: first $^{24}\text{Mg}(e^-,\nu_e)^{24}\text{Na}$ and then
$^{24}\text{Na}(e^-,\nu_e)^{24}\text{Ne}$ at slightly higher densities.

Both of the above reactions are exothermic, i.e. the specific heating rate
$\dot{\epsilon}^{\text{EC}}$ is positive. For
$^{24}\text{Mg}(e^-,\nu_e)^{24}\text{Na}$ this is due to the capture being
delayed until the threshold for the allowed
$0^+_\mathrm{g.s.}\rightarrow{}1^+_1$ transition is reached. At this point we
have (see Ref.~\cite{martinez2014astrophysical})
${\mu_e\approx-Q^{\text{EC}}_\text{g.s.}+E_{1^+}}$ and
$\langle{}E_\nu\rangle{}^{\text{EC}}\approx0$, where
${E_{1^+}=0.472~\text{MeV}}$ is the excitation energy of the $1^+_1$ state.
When inserted into \eqref{eq:specific_energy_EC} the above yields
\begin{equation}
	\dot{\epsilon}^{\text{EC}}\approx\frac{Y}{m_u}\lambda^\text{EC}_{0^+\rightarrow{}1^+}\,E_{1^+}
\end{equation}
which we can interpret as the heating arising from the gamma decay of the
$1^+_1$ state. Similarly, if the $^{24}\text{Na}(e^-,\nu_e)^{24}\text{Ne}$
capture proceeds via the $4^+_\mathrm{g.s.}\rightarrow{}4^+_1$ transition there
is an associated heating effect due to the subsequent gamma decay of the
excited $4^+_1$ state. The situation is somewhat different if electron capture
instead occurs via the ${1^+_1\rightarrow{}0^+_\mathrm{g.s.}}$ or
${4^+_\mathrm{g.s.}\rightarrow{}2^+_1}$ transitions, whose energy thresholds
are lower than for the preceding $^{24}\text{Mg}(e^-,\nu_e)^{24}\text{Na}$
reaction. For such superthreshold captures we naturally have
$\mu_e+Q_\text{g.s.}>0$ but also $\langle{}E_\nu\rangle{}^{\text{EC}}>0$. The
net effect is, however, still exothermic.

A subtlety in the evaluation of the $^{24}\text{Na}(e^-,\nu_e)^{24}\text{Ne}$
rate is the fact that the $1^+_1$ state in $^{24}$Na is an isomer. It decays to
the $4^+_\mathrm{g.s.}$ ground state via an M3 transition with a half-life of
$20.18~\text{ms}$~\cite{firestone2007nuclear}. As this is many orders of
magnitude slower than typical gamma decays one might ask whether the population
of the $1^+_1$ state still follows a Boltzmann distribution as assumed in
\eqref{eq:lambda_total}. In Appendix~\ref{sec:app_isomer} we argue that this
assumption is still appropriate.

To determine the effect of the forbidden transition on the electron
capture rate we have calculated the corresponding form factor
coefficients as listed in Table~\ref{tbl:ffc24}. In contrast to the
preceding section, here we list the form factor coefficients in the
electron capture direction. As before we either let $\VF{0}{211}=0$
(SM) or use \eqref{eq:cvc} to relate it to $\VF{0}{220}$
(SM+CVC). Since there is no experimental value for $E_\gamma$ in
\eqref{eq:cvc}, i.e. the excitation energy of the isobaric analog of
the $^{24}$Ne $2^+_1$ state in $^{24}$Na, we estimate it based on the
Coulomb displacement energies between isobaric analog
states~\cite{Antony.Pape.Britz:1997}:
\begin{subequations}
\begin{align}
  \label{eq:1}
  E_\gamma&= Q + \Delta E_C - (m_n
                  c^2 - m_p c^2), \\
        \Delta E_C &= 1.4136(1)\bar{Z}/A^{1/3} - 0.91338(11)\ \text{MeV}
\end{align}
\end{subequations}
Here $Q=4.959$~MeV is the energy difference between the initial and final
states in the forbidden transition, and $m_n$ and $m_p$ are the neutron and
proton masses. We have also introduced the quantity $\bar{Z}=(Z_i+Z_f)/2$.
The above equations yield $E_\gamma = 7.90$~MeV.

\begin{table}[htb]
  \centering
  \caption{\label{tbl:ffc24}
	Form factor coefficients for the forbidden transition from the
	$4^+_\mathrm{g.s.}$ state in $^{24}$Na to the $2^+_1$ state in
	$^{24}$Ne. We use $g_A=-1.276$ when computing the axial
	coefficients.}
  \begin{ruledtabular}
    \begin{tabular}{lcc}
      Form factor coefficient	& SM			& SM+CVC\\ \hline
      \VF{0}{211}		&  0 			&    0.00225\footnote{Calculated from \VF{0}{220} via the CVC relation.}	\\
      \VF{0}{220}            	&  0.0451               &    0.0451			\\
      \VF{0}{220}$(1,1,1,1)$	&  0.0630		&    0.0630			\\
      \VF{0}{220}$(2,1,1,1)$	&  0.0620		&    0.0620			\\
      \AF{0}{221}       	& $-$0.109		& $-$0.109			\\
      \AF{0}{221}$(1,1,1,1)$	& $-$0.128		& $-$0.128			\\
      \AF{0}{221}$(2,1,1,1)$	& $-$0.122		& $-$0.122			\\
      \AF{0}{321}       	& $-$0.00020		& $-$0.00020
    \end{tabular}
  \end{ruledtabular}
\end{table}

\begin{table}[htb]
  \centering
  \caption{\label{tbl:logft24}
	$\log ft$ values for laboratory $\beta^-$ decay corresponding to the
	form factor coefficients in Table~\ref{tbl:ffc24}, with and without
	quenching applied to the axial coefficients.}
  \begin{ruledtabular}
    \begin{tabular}{lll}
      Form factor coefficients		&  $g_A$    & $\log ft$ \\ \hline
      SM				&  $-1.276$ & $11.57$ \\
      SM				&  $-1.0$   & $11.69$ \\
      SM+CVC				&  $-1.276$ & $12.14$ \\
      SM+CVC				&  $-1.0$   & $12.38$
    \end{tabular}
  \end{ruledtabular}
\end{table}

From Table~\ref{tbl:ffc24} we can derive the corresponding electron capture
shape factors, which are trivially related to the ones of the reverse
$\beta^-$ decay (see Appendix~\ref{sec:app_shape}). To give an indication of
the transition strength we list the predicted $\log ft$ values for the
$\beta^-$ decay in Table~\ref{tbl:logft24}. In addition to the SM and SM+CVC
results we also include the values obtained using quenched axial matrix
elements. We should note that a lower $\log(ft)$ value does not always imply a
higher electron capture rate: the $\beta^-$ decay rate is determined by the
behaviour of the shape factor for electron energies below $Q^{\beta^-}$, while
electron capture occurs for energies above this value.

\begin{figure}[htb]
  \centering
  \includegraphics[width=\linewidth]{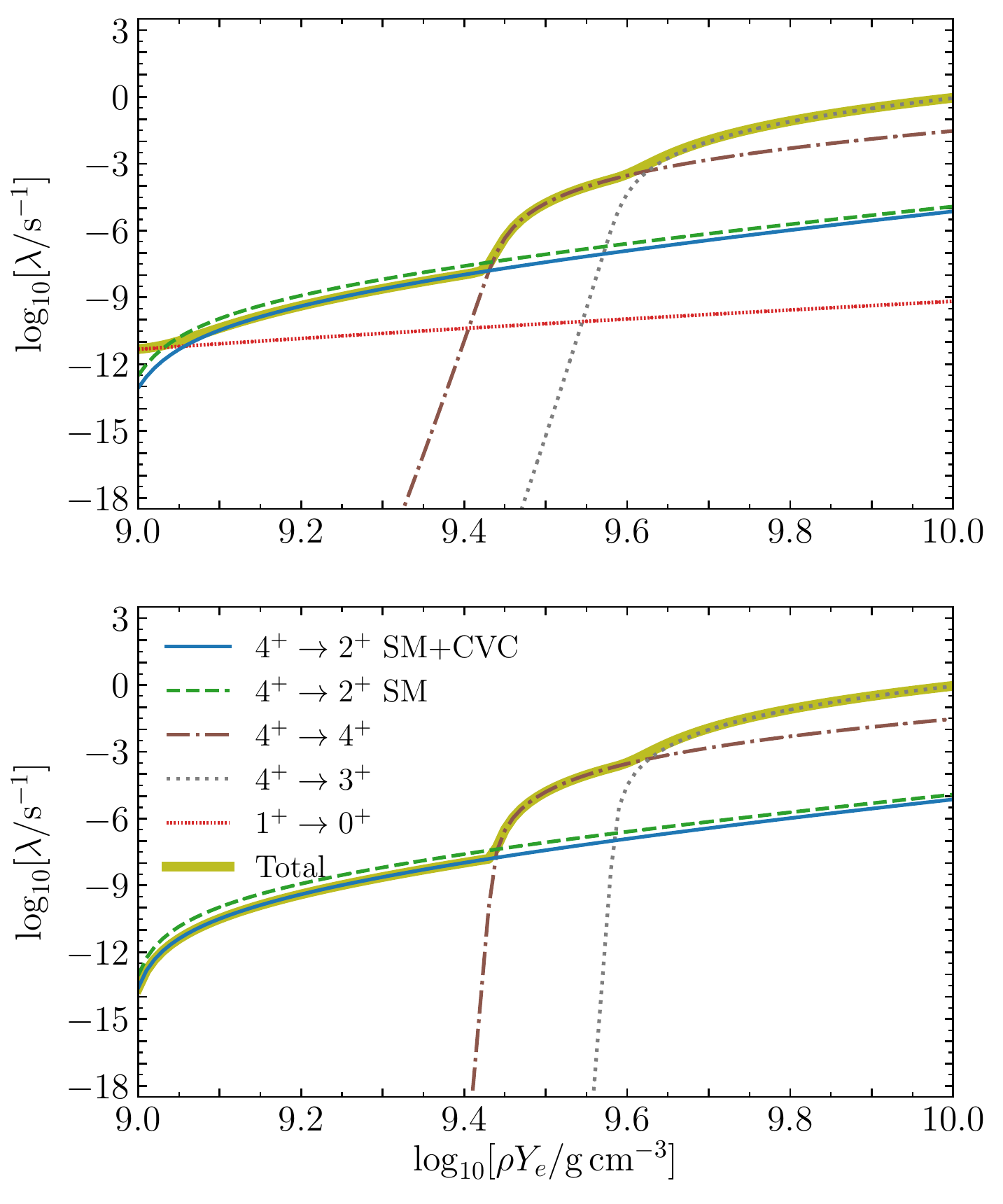}
  \caption{\label{fig:rates24}
	Electron capture rate on $^{24}$Na as a function of density with
	Coulomb corrections included. We plot the total rate as well as the
	individual contributions from the four transitions that dominate the
	capture rate. The upper panel corresponds to
	$\log_{10}(T[\text{K}])=8.4$ and the lower panel to
	$\log_{10}(T[\text{K}])=7.8$.}
\end{figure}

We have evaluated the forbidden electron capture rate for both the SM and
SM+CVC cases as shown in Fig.~\ref{fig:rates24}. To calculate the allowed rates
in the plot we used the transition strengths listed in
Ref.~\cite{martinez2014astrophysical}. In the relevant density range (roughly
$\log_{10}(\rho{}Y_e [\text{g~cm$^{-3}$}])=9.3$--9.4) the SM rate is 2.5--3
times higher than the SM+CVC rate. Quenching of the axial matrix elements
results in rates (not shown in the figure) that are $30\%$ to $50\%$ lower. In
the upper panel of Fig.~\ref{fig:rates24} we have assumed a temperature of
$\log_{10}(T[\text{K}])=8.4$, which is roughly what is seen in simulations
without Urca cooling (see Ref.~\cite{schwab2015thermal}). In this case the
forbidden transition increases the rate by up to 3 orders of magnitude for
$\log_{10}(\rho{}Y_e[\text{g~cm$^{-3}$}])\lesssim9.4$, but even without this
contribution there would still be a substantial capture rate at low densities
via the $1^+\rightarrow{}0^+$ transition. This is not the case in the lower
panel, where we have plotted the rates for the temperature
$\log_{10}(T[\text{K}])=7.8$. Such low temperatures are seen in models with
substantial Urca cooling (see Ref.~\cite{schwab2017importance}). The population
of the $1^+$ state in $^{24}$Na is now so low that the contribution from the
$1^+\rightarrow{}0^+$ transition is negligible and we would see no electron
capture for $\log_{10}(\rho{}Y_e[\text{g~cm$^{-3}$}])\lesssim9.4$ if we did not
include the forbidden transition. In general, the forbidden transition will
have a significant contribution to the total capture rate for temperatures
$\log_{10}(T[\text{K}])\lesssim8.5$.

\subsection{$^{27}\text{Al}\leftrightarrow{}^{27}\text{Mg}$}
\label{sec:rate27}

\begin{figure}[htb]
	\centering
	\includegraphics[width=0.619\linewidth]{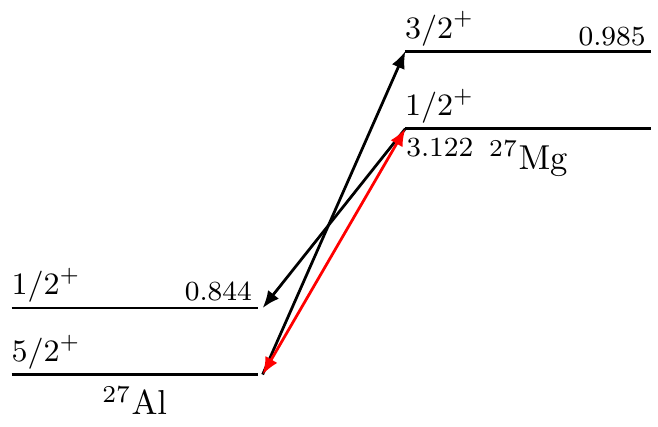}
	\caption{\label{fig:energylevels27}
		Energy level diagram illustrating the transitions that
		determine the rate of the
		$^{27}\text{Al}(e^-,\nu_e)^{27}\text{Mg}$ electron capture as
		well as the reverse beta decay. The forbidden transition is in
		red. We list excitation energies as well as the ground state
		energy of $^{27}$Mg relative to the ground state of $^{27}$Al.
		Both are in units of MeV. The nuclear data has been taken from
		Ref.~\cite{basunia2011nuclear}.
		}
\end{figure}

In Fig.~\ref{fig:energylevels27} we show the transitions that determine the
$^{27}\text{Al}(e^-,\nu_e)^{27}\text{Mg}$ electron capture rate as well as the
rate of the reverse $\beta^-$ decay. As $^{27}$Al and $^{27}$Mg are odd-even
nuclei we would expect electron capture to trigger an Urca process. However, at
low temperatures this can only occur via the second-forbidden transition
between the two ground states. If we instead only consider allowed transitions
electron capture is delayed until the threshold for the
$5/2^+_\mathrm{g.s.}\rightarrow{}3/2^+_1$ transition is reached. At this point
the reverse $\beta^-$ decay rate of $^{27}$Mg is negligible: there is no
significant thermal population of the excited $3/2^+$ state and the
$1/2^+_\mathrm{g.s.}\rightarrow{}1/2^+_1$ transition from the ground state to
the first excited state in $^{27}$Al has become Pauli blocked. Without
$\beta^-$ decay there cannot be any Urca cycles and instead of a cooling effect
we would see heating from the gamma decay of the excited $3/2^+$ state (see
our discussion in the preceding section). Note that at higher temperatures
($\log_{10}(T[\text{K}])\gtrsim8.8$) Urca processes can occur via allowed
transitions between excited states. This is demonstrated in
Ref.~\cite{toki2013detailed}, which did not include any forbidden transitions.

To establish whether any significant Urca cooling occurs via the forbidden
transition between the ground states we must first constrain its strength.
Following the same procedure as in the preceding sections we have calculated
the form factor coefficients using shell model (SM) calculations and by also
applying the CVC relation \eqref{eq:cvc} (SM+CVC), where we used
$E_\gamma=6.814~\text{MeV}$~\cite{basunia2011nuclear} for the latter. We list
the results in Table~\ref{tbl:ffc27}. From these coefficients (with and without
quenching of the axial part) we derive the corresponding shape factors. We show
the resulting $\log ft$ values of the reverse $\beta^-$ decay in
Table~\ref{tbl:logft27}.

\begin{table}[htb]
  \centering
  \caption{\label{tbl:ffc27}
	Form factor coefficients for the forbidden transition from the
	$5/2^+_\mathrm{g.s.}$ state in $^{27}$Al to the $1/2^+_\mathrm{g.s.}$
	state in $^{27}$Mg. For the axial coefficients we assumed $g_A=-1.276$.}
  \begin{ruledtabular}
    \begin{tabular}{lcc}
      Form factor coefficient	& SM			& SM+CVC\\ \hline
      \VF{0}{211}		&    0     		&    0.0116\footnote{Calculated from \VF{0}{220} via the CVC relation.}	\\
	\VF{0}{220}            	&    0.269		&    0.269 \\
	\VF{0}{220}$(1,1,1,1)$	&    0.316		&    0.316 \\
	\VF{0}{220}$(2,1,1,1)$	&    0.300		&    0.300 \\
	\AF{0}{221}       	&    0.125		&    0.125 \\
	\AF{0}{221}$(1,1,1,1)$	&    0.139		&    0.139 \\
	\AF{0}{221}$(2,1,1,1)$	&    0.130		&    0.130 \\
	\AF{0}{321}       	& $-$0.368      	& $-$0.368
    \end{tabular}
  \end{ruledtabular}
\end{table}

\begin{table}[htb]
	\centering
	\caption{\label{tbl:logft27}
		$\log ft$ values for laboratory $\beta^-$ decay corresponding
		to the form factor coefficients in Table~\ref{tbl:ffc27}, with
		and without quenching applied to the axial coefficients.
		}
	\begin{ruledtabular}
		\begin{tabular}{lll}
			Form factor coefficients	& $g_A$    & $\log ft$ \\ \hline
			SM				& $-1.276$ & $11.33$ \\
			SM				& $-1.0$   & $11.32$ \\
			SM+CVC				& $-1.276$ & $11.13$ \\
			SM+CVC				& $-1.0$   & $11.23$
		\end{tabular}
	\end{ruledtabular}
\end{table}

\begin{figure}[htb]
  \centering
  \includegraphics[width=\linewidth]{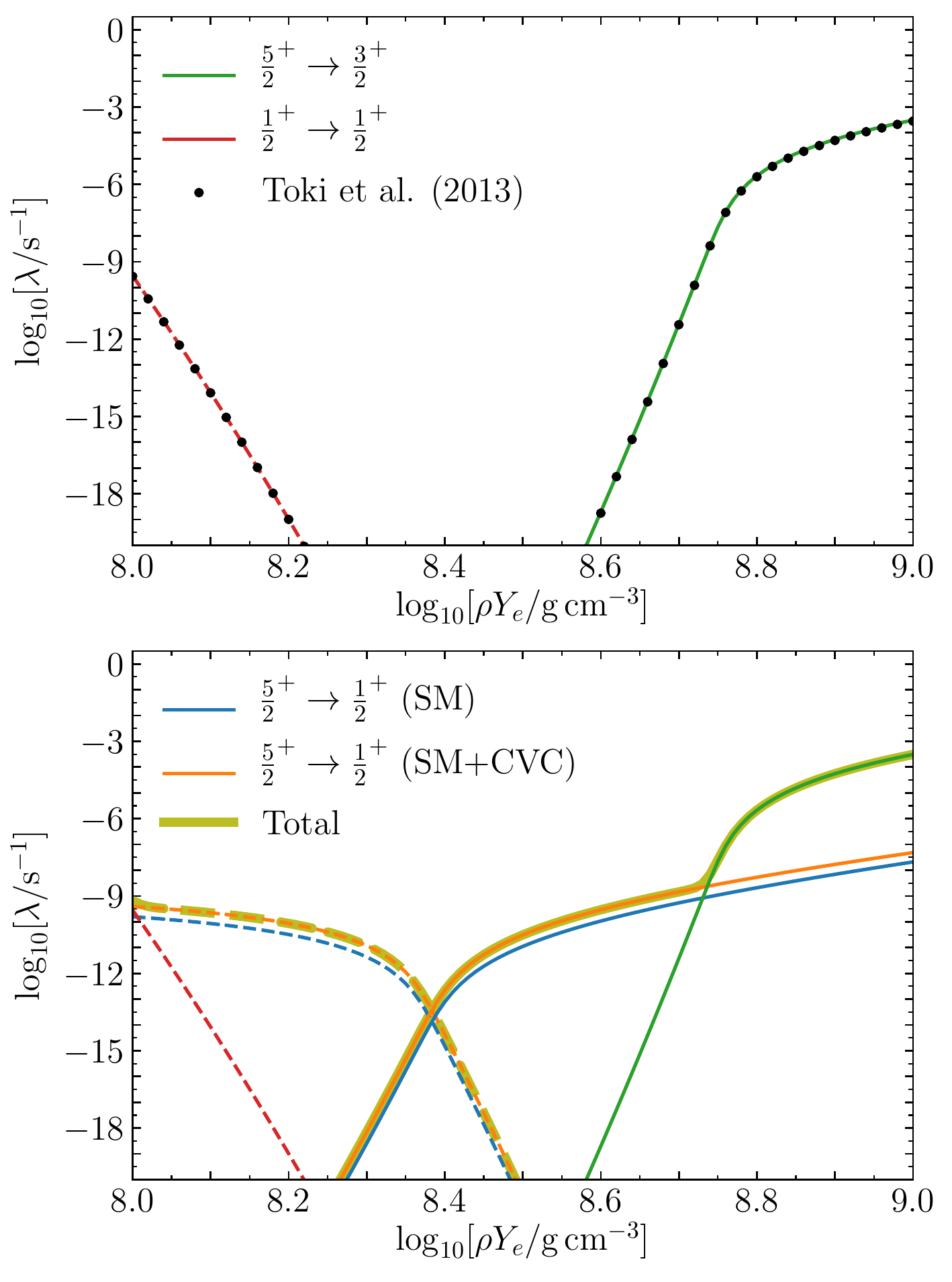}
  \caption{\label{fig:rates27}
	Electron capture rates on $^{27}$Al (solid) and the reverse $\beta^-$
	decay rates of $^{27}$Mg (dashed) as a function of density with Coulomb
	corrections included. We have assumed a temperature of
	$\log_{10}(T[\text{K}])=8.3$. In the upper panel we only include
	allowed transitions: this reproduces the rate from
	Ref.~\cite{toki2013detailed}. In the lower panel we show the impact of
	the forbidden transition. Note that the total electron capture and
	$\beta^-$ decay rates are plotted assuming the SM+CVC rate.}
\end{figure}

In Fig.~\ref{fig:rates27} we plot the electron capture and $\beta^-$ decay
rates as a function of the density at a temperature of
$\log_{10}(T[\text{K}])=8.3$. For the allowed transition between the
$5/2^+_\mathrm{g.s.}$ and $3/2^+_1$ states we used a transition strength
corresponding to $\log ft=5.51$ in the $\beta^-$ decay direction. We determined
this value through a shell model calculation. As seen in the upper panel we are
able to reproduce the rate from Ref.~\cite{toki2013detailed} by solely
including allowed transitions. Per our earlier discussion there is no Urca
cooling in this case: excited states are not thermally populated at this
temperature and the $\beta^-$ decay of the $^{27}$Mg ground state becomes Pauli
blocked before electron capture on $^{27}$Al can occur. In contrast, in the
lower panel we also include the contribution from the forbidden transition.
This significantly alters both the electron capture and $\beta^-$ decay rates
and the density ranges in which the two processes occur now overlap. The decay
and capture rates become roughly equal at $\log_{10}(\rho
Y_e[\text{g~cm$^{-3}$}])\approx8.4$. This is when we would expect the Urca
process to be the most efficient. However, at this point the rates are only of
the order $10^{-14}-10^{-13}$~s$^{-1}$. As a consequence, the Urca cooling is
marginal in our models (see Section~\ref{sec:27text}). Nevertheless, the
forbidden transition has a large impact on both $\beta^-$ decay and electron
capture rates at temperatures $\log_{10}(T[\text{K}])\lesssim8.8$, and we
recommend its inclusion in stellar evolution modeling.

\section{Astrophysical impact}
\label{sec:impact}
\subsection{$^{24}\text{Na}\leftrightarrow{}^{24}\text{Ne}$}
To illustrate the impact of the forbidden transition between $^{24}$Na and
$^{24}$Ne we use the one-dimensional stellar evolution code
MESA~\cite{paxton2010modules} (Modules for Experiments in Stellar Astrophysics,
version 10398). We use MESA's ability to calculate weak interaction rates on
the fly, which we have extended to support second-forbidden transitions as
described in Appendix~\ref{sec:app_calcrate}.

In our simulations we use the set up of
Refs.~\cite{schwab2015thermal,schwab2017importance} to which we refer
the reader for the full details. In short, a degenerate ONe core is
prepared at a density of $\rho=0.40\times10^9$~g~cm$^{-3}$ and mass is
then added at a constant rate $\dot{M}$ to simulate the core
growth. For the initial composition we assume the default set of mass
fractions used in Ref.~\cite{schwab2017importance}:
${X(^{16}\text{O})=0.50}$, ${X(^{20}\text{Ne})=0.39}$,
${X(^{23}\text{Na})=0.05}$, ${X(^{24}\text{Mg})=0.05}$ and
${X(^{25}\text{Mg})=0.01}$. Furthermore, for the growth rate we use
the values $\dot{M}=10^{-7}$~M$_\odot$~yr$^{-1}$,
$10^{-6}$~M$_\odot$~yr$^{-1}$ or $10^{-5}$~M$_\odot$~yr$^{-1}$. This
is in line with simulations of thermally stable hydrogen and helium
burning which predict core growth rates 
$(4\text{--}7)\times10^{-7}$~M$_\odot$~yr$^{-1}$~\cite{wolf2013hydrogen}
and
$(1.5\text{--}4.5)\times10^{-6}$~M$_\odot$~yr$^{-1}$~\cite{brooks2016carbon},
respectively.

\begin{figure}[htb]
	\centering
	\includegraphics[width=\linewidth]{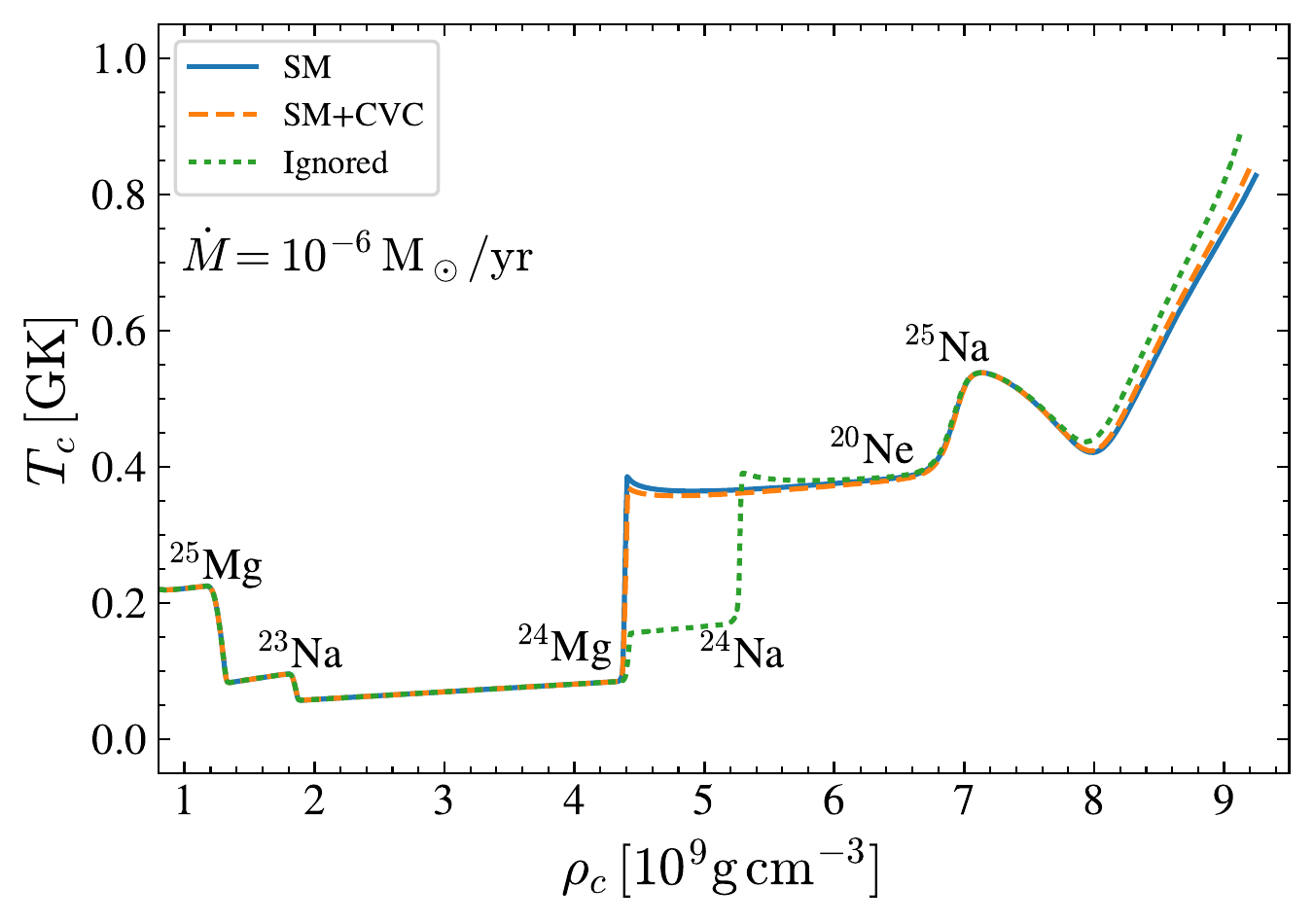}
	\caption{\label{fig:rhoT24}
		Central temperature as a function of central density in MESA
		simulations with three different treatments of the forbidden
		transition between $^{24}$Na and $^{24}$Ne: included using the
		SM rate, using the SM+CVC rate, or ignored completely. The
		onset of each electron capture is labeled with the parent
		nucleus. Note that for simulations including the forbidden
		transition capture on $^{24}$Na does not occur separately from
		capture on $^{24}$Mg.
		}
\end{figure}

In Fig.~\ref{fig:rhoT24} we show the evolution of the central temperature and
density for the MESA simulations with the growth rate
$\dot{M}=10^{-6}$~M$_\odot$~yr$^{-1}$. The initial evolution is identical, with
substantial cooling from the $^{25}\text{Mg}\leftrightarrow{}^{25}\text{Na}$
and $^{23}\text{Na}\leftrightarrow{}^{23}\text{Ne}$ Urca cycles. Differences
between the simulations appear once the threshold density for capture on
$^{24}$Mg is reached. If the forbidden transition is ignored capture occurs
first on $^{24}$Mg and later on $^{24}$Na, as explained in
Section~\ref{sec:rate24}. If it is included we get the double electron capture
$^{24}\text{Mg}(e^-,\nu_e)^{24}\text{Na}(e^-,\nu_e)^{24}\text{Ne}$ when the
electron chemical potential reaches the threshold for capture on $^{24}$Mg. The
difference between using the SM and SM+CVC rates is minimal. Furthermore, the
simulations converge for densities $\rho_c \approx 6\times 10^9$ independently
of the inclusion of the forbidden transition. As explained in
Ref.~\cite{schwab2017importance} this is due to the core returning to the
trajectory defined by the balance between compressional heating and losses from
thermal neutrinos.

The subsequent evolution follows the results presented in
Ref.~\cite{kirsebom2019discovery}: Electron capture on $^{20}$Ne sets in at
$\rho_c\approx6.8\times10^9$~g~cm$^{-3}$ via the second-forbidden transition
between the ground states of $^{20}$Ne and $^{20}$F. As the resulting rate is
comparatively slow the heating from the
$^{20}\text{Ne}(e^-,\nu_e)^{20}\text{F}(e^-,\nu_e)^{20}\text{O}$ double
electron capture is gradual and it is temporary canceled by the cooling from
the $^{25}\text{Na}\leftrightarrow{}^{25}\text{Ne}$ Urca process. After the
exhaustion of $^{25}$Na in the center the heating resumes and eventually
ignites runaway oxygen burning: Our simulations stop when the energy from the
oxygen burning exceeds the thermal neutrino losses. Prior to this a substantial
fraction of the central $^{20}$Ne has been converted into $^{20}$O. If the
depletion is large enough ignition occurs mildly off-center where there is more
$^{20}$Ne left.

\begin{table}[htbp]
  \centering
  \caption{\label{tbl:rhoRign24}
    Ignition densities and radii with and without the forbidden
    transition between $^{24}$Na and $^{24}$Ne (assuming the SM+CVC
    rate).}
  \begin{ruledtabular}
    \begin{tabular}{c|cc|cc}
      $\dot{M}$~(M$_\odot$~yr$^{-1}$)  &
      \multicolumn{2}{c}{$\rho^{\text{ign}}_c$~(g~cm$^{-3}$} &
      \multicolumn{2}{c}{$R_{\text{ign}}$~(km)}\\ 
					& Without	& With		&  Without	& With\\
			\hline
			$10^{-7}$	& $9.43$	& $9.52$	& $58$		& $62$ \\
			$10^{-6}$	& $9.13$	& $9.21$	& $35$  	& $42$ \\
			$10^{-5}$	& $8.64$	& $8.64$	& $<10$ 	& $<10$
		\end{tabular}
	\end{ruledtabular}
\end{table}

In the final phase the simulations with and without the forbidden
transition between $^{24}$Na and $^{24}$Ne again diverge, although
only slightly. We summarize the effects on the ignition density and
radii for the three different growth rates in
Table~\ref{tbl:rhoRign24}. The only discernible change occurs in the
simulations with off-center ignition: in these cases the ignition
density is slightly reduced while the ignition radius grows by
4--7~km.

\begin{figure}[htb]
	\centering
	\includegraphics[width=\linewidth]{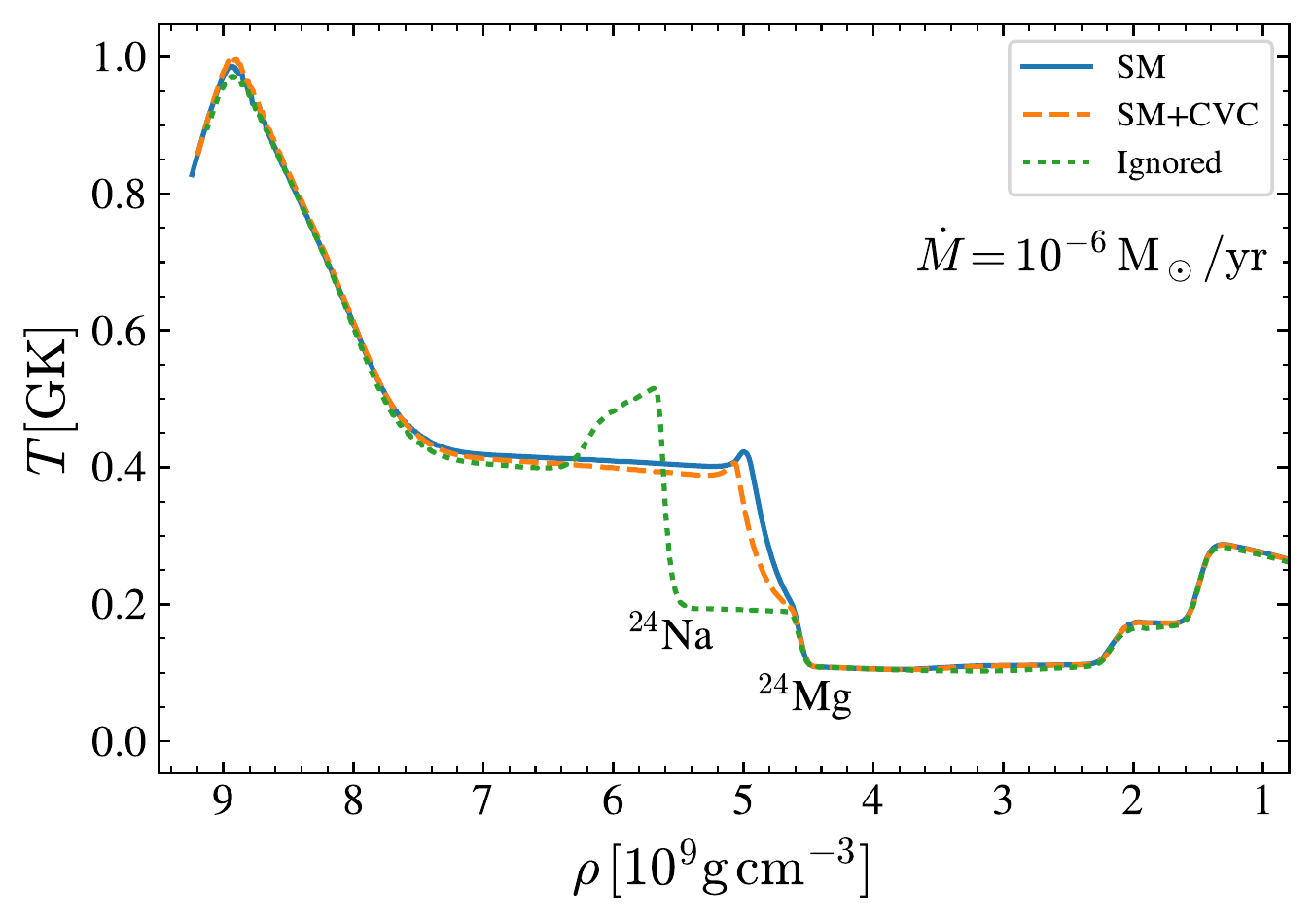}
	\caption{\label{fig:rhoT24profile}
		Temperature profiles at ignition for the models in
		Fig.~\ref{fig:rhoT24}. Note that we plot the temperature as a
		function of density as opposed to radius. We label the
		approximate locations where electron capture on $^{24}$Mg and
		(when occurring separately) $^{24}$Na is taking place.}
\end{figure}

To also illustrate the off-center differences between the models
Fig.~\ref{fig:rhoT24profile} presents the temperature profiles at the end of
the simulations. The spatial variation in temperature partially mirrors the
temporal evolution in Fig.~\ref{fig:rhoT24}. Note that electron capture
processes that have ceased in the center are still taking place further out
where the density is lower. We have marked the approximate locations of
electron capture on $^{24}$Mg and $^{24}$Na. As before the inclusion of the
forbidden transition means that the reactions occur in close succession,
whereas without this transition the increase in temperature occurs in two
separate steps. Further towards the center the differences between the models
are minimal.

The above results would suggest that the forbidden transition only has a minor
impact on the ignition conditions. However, as described in
Ref.~\cite{schwab2017importance} the steep temperature gradient arising from
electron capture on $^{24}$Mg and $^{24}$Na induces convective instabilities
that cannot be modeled in the present version of MESA. Further work is needed
to determine their implications. As the forbidden transition affects the
appearance of the temperature gradients it should be included in future work.

\begin{figure}[htb]
	\centering
	\includegraphics[width=\linewidth]{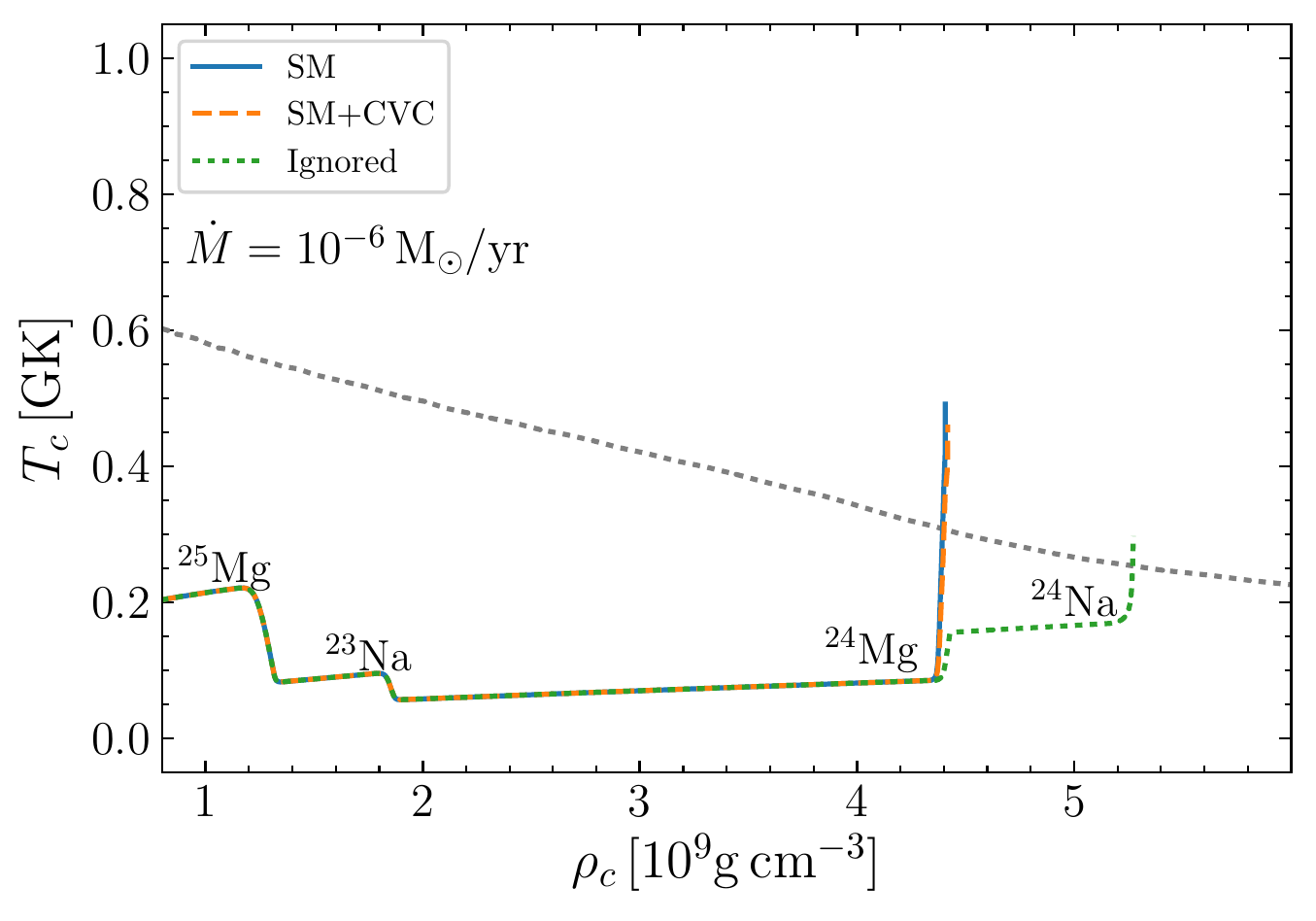}
	\caption{\label{fig:carbonrhoT24}
		Evolution of central temperature and density as in
		Fig.~\ref{fig:rhoT24}, but with the core having a residual
		carbon mass fraction of $X(^{12}\text{C})=0.01$. The dotted gray
		line is the carbon ignition curve taken from Fig.~6 in
		Ref.~\cite{schwab2019residual}. Along this curve energy
		generation from carbon burning equals thermal neutrino losses.
		}
\end{figure}

The evolution of the core is of course sensitive to the details of the
composition, which is determined by the preceding carbon burning. A
special case arises when there is a substantial amount of residual
$^{12}$C in the core. This has recently been investigated in
Ref.~\cite{schwab2019residual}. Following the same procedure, we have
run MESA simulations presented in Fig.~\ref{fig:carbonrhoT24}. The
composition is taken as $X(^{12}\text{C})=0.01$ and
$X(^{20}\text{Ne})=0.38$ with other mass fractions remaining the same
as before. In all three cases, the electron captures ignite runaway
carbon burning, but when the forbidden transition is included this
occurs at significantly lower densities than before. The difference
between using the SM and SM+CVC rates is as before negligible. Due to
numerical difficulties resulting from convection triggered by the carbon burning
(see Ref.~\cite{schwab2019residual} for details) we are unable to
follow the subsequent evolution of the core towards oxygen ignition.
Nevertheless, the above results indicate that the forbidden transition
is of significance for cores with residual carbon.

\subsection{$^{27}\text{Al}\leftrightarrow{}^{27}\text{Mg}$}
\label{sec:27text}

An Urca process affects the temperature if it has enough time to
release a significant amount of neutrinos before it ceases to
operate. In quantitative terms, the cooling timescale $t_\text{cool}$
must be shorter than the timescale $t_\text{cross}$ it takes to cross
the density range where the Urca process is
active. Ref.~\cite{schwab2017importance} derives the following
approximations
\begin{align}
	\begin{split}
		t_\text{cool}\approx&4\times10^2\left(\frac{T}{10^8~\text{K}}\right)^{-3}\left(\frac{X}{0.01}\right)^{-1}\\
				&\times\left(\frac{Q}{5~\text{MeV}}\right)^{-2}\left(\frac{ft}{10^5~\text{s}}\right)~\text{yrs}
	\end{split}\\
	t_\text{cross}\approx&2\times10^{-2}\left(\frac{T}{10^8~\text{K}}\right)\left(\frac{\rho}{10^9~\text{g~cm$^{-3}$}}\right)^{-1/3}t_\rho.
\end{align}
Here $X$ is the initial mass fraction of the parent nucleus, $Q$ is the energy
difference between the initial and final states, and $ft$ describes the strength
of the transition. Furthermore, ${t_\rho=(d\ln\rho_c/dt)^{-1}}$ is the
compression timescale that sets the pace of the density growth. From
Ref.~\cite{schwab2015thermal} we have the expression
\begin{equation}
	\label{eq:t_rho}
	\begin{split}
		t_\rho\approx5\times10^4&\left(\frac{\rho}{10^9~\text{g~cm$^{-3}$}}\right)^{-0.55}\\
                \times&\left(\frac{\dot{M}}{10^{-6}~\text{M}_\odot\text{yr}^{-1}
                  }\right)^{-1}
                \ \text{yr}
	\end{split}
\end{equation}
which was derived by modeling the core as a zero-temperature white
dwarf in hydrostatic equilibrium with $Y_e\approx0.5$ and
$\rho\sim10^9$--$10^{10}$~g~cm$^{-3}$. While the threshold density
for capture on $^{27}$Al is somewhat lower than this
($\rho\approx5\times10^{8}$~g~cm$^{-3}$) the approximation in
\eqref{eq:t_rho} is still acceptable for our purposes.

For the forbidden transition, we have $Q\approx{3~\text{MeV}}$ and
$\log{}ft\approx11$. Furthermore,
simulations~\cite{jones2013advanced,takahashi2013evolution} of the carbon
burning stage indicate that $X<0.01$. Given a slow growth rate of
$\dot{M}=10^{-7}$~M$_\odot$~yr$^{-1}$ we then have that
$t_\text{cool}<t_\text{cross}$ at the threshold density only if
$T\gtrsim1.6~\text{GK}$. For larger values of $\dot{M}$ the required
temperature is even higher. As we expect that $T\lesssim1~\text{GK}$ prior to
oxygen ignition we can conclude that the Urca cooling due to the forbidden
transition is negligible. At such high temperatures the electron capture and
$\beta^-$ decay rates would in any case be dominated by allowed transitions.

Despite the weakness of the forbidden transition the timescale
$(\lambda^\text{EC}_{\text{forbidden}})^{-1}$ of the corresponding electron
capture rate will still fall below the compression timescale $t_\rho$. This
means that the onset of $^{27}\text{Al}(e^-,\nu_e)^{27}\text{Mg}$ is not
delayed until the threshold density for the allowed
$5/2^+_\mathrm{g.s.}\rightarrow{}3/2^+_1$ transition is reached. As a
consequence the mild heating due to the gamma decay of the $3/2^+$ state, as
seen in Fig.~8 of Ref.~\cite{schwab2019residual}, would likely not occur. This
effect is at any rate small and occurs comparatively early in the evolution of
the core. It is thus reasonable to conclude that the forbidden transition does
not have any significant impact on the ignition conditions.

\section{Conclusions}
\label{sec:conclusions}

We have provided a more detailed account of the formalism used in
Ref.~\cite{kirsebom2019measurement,kirsebom2019discovery}. Furthermore, we have
investigated two additional second-forbidden non-unique transitions ($\Delta J^\pi=2^+$) of
relevance to degenerate ONe cores: the ${4^+_\text{g.s.}\rightarrow{}2^+_1}$
transition from $^{24}$Na to $^{24}$Ne, and the
${5/2^+_\text{g.s.}\rightarrow{}1/2^+_\text{g.s.}}$ transition from $^{27}$Al
to $^{27}$Mg. The necessary nuclear matrix elements have been determined
through shell-model calculations (SM) in combination with constraints from CVC
theory (SM+CVC). We have calculated the axial matrix elements using both bare
($g_A=-1.276$) and quenched ($g_A=-1.0$) values of the axial coupling constant.
To test our approach we have applied it to the previously measured
$4^+_\mathrm{g.s.}\rightarrow{}2^+_1$ transition in the $\beta^-$ decay of
$^{24}$Na.

For the decay of $^{24}$Na our calculated rate is within a factor 1 to 3 of the
value given by Ref.~\cite{turner1951lxvi}. However, when extrapolating from the
measured region of the electron energy spectrum the authors of
Ref.~\cite{turner1951lxvi} had assumed an allowed spectrum shape: in reality
the difference between the theoretical and measured rates may be as high as a
factor 3 to 6. We arrived at the best agreement by using the SM+CVC approach,
with quenching applied to the axial nuclear matrix elements.

The forbidden transition between $^{24}$Na and $^{24}$Ne has a substantial
impact on the electron capture rate for temperatures
$\log_{10}(T[\text{K}])\lesssim8.5$. It lowers the threshold density for
$^{24}\text{Na}(e^-,\nu_e)^{24}\text{Ne}$ so that it can occur immediately
following the onset of $^{24}\text{Mg}(e^-,\nu_e)^{24}\text{Na}$. To illustrate
the astrophysical impact we performed simulations using the stellar evolution
code MESA according to the procedures detailed in
Ref.~\cite{schwab2015thermal,schwab2017importance,schwab2019residual}. The
effects do not appear to be very sensitive to uncertainties in the rate: the SM
and SM+CVC rates produce very similar results despite being up to a factor 3
different. In general the impact on the ignition conditions appears negligible,
but there is a change in the radial temperature gradients arising from the
$A=24$ captures. These gradients are known to cause convectional instabilities
that are not accounted for in our simulations and future work
should include the forbidden transition. Finally, we note that in models where
burning of residual $^{12}$C occurs the forbidden transition reduces the carbon
ignition density.

The forbidden transition between the ground states of $^{27}$Al to $^{27}$Mg
permits the onset of electron capture at lower densities, precluding the minor
heating associated with capture into excited states. However, the transition
strength is too small to produce any significant cooling during the time that
the stellar core spends in the density range where Urca cycles occur.
Considering the early onset and small impact of this electron capture reaction,
we expect its significance for the ignition conditions to be minimal. However,
we still recommend that this transition is included when calculating the
electron capture and $\beta^-$ decay rates between $^{27}$Al and $^{27}$Mg for
conditions $\log_{10}(T [\text{K}])\lesssim8.8$.

\begin{acknowledgments}
	The authors thank Oliver S. Kirsebom for his comments on the
	manuscript. DFS and GMP acknowledge the support of the Deutsche
	Forschungsgemeinschaft (DFG, German Research Foundation) --
	Project-ID 279384907 -- SFB 1245 ``Nuclei: From Fundamental
	Interactions to Structure and Stars'', the Helmholtz
	Forschungsakademie Hessen f\"ur FAIR, and the ChETEC COST action
	(CA16117), funded by COST (European Cooperation in Science and
	Technology).
\end{acknowledgments}

\appendix
\section{Shape factors}
\label{sec:app_shape}
For second-forbidden non-unique transitions the shape factor can be written
(see Ref.~\cite{behrens1982electron}) as the sum
\begin{align}
	\label{eq:shapefactor_secforb}
		C(w)=\sum_{k_e+k_\nu=3}\lambda_{k_e}\Big\{&M_2^2(k_e,k_\nu)+m_2^2(k_e,k_\nu)\notag\\
		-\frac{2\mu_{k_e}\gamma_{k_e}}{k_ew}&M_2(k_e,k_\nu)m_2(k_e,k_\nu)\Big\}\\
		+\sum_{k_e+k_\nu=4}\lambda_{k_e}\{&\tilde{M}_2^2(k_e,k_\nu)+\tilde{M}_3^2(k_e,k_\nu)\}\notag
\end{align}
where the indices $k_e$ and $k_\nu$ run over the positive integers.
Furthermore, we have $\gamma_k=\sqrt{k^2-(\alpha Z)^2}$ and assume
$\lambda_{k_e}\approx1$ and $\mu_{k_e}\approx1$. The two latter quantities
arise due to the Coulomb interaction between the nucleus and the electron and
the approximations are valid for the small $Z$ we encounter in this work.
$M_K(k_e,k_\nu)$ and $m_K(k_e,k_\nu)$ contain the form factor coefficients and
are given by
\begin{widetext}
	\begin{align}
		\begin{split}
			\label{eq:secforb_M2_1}
			&M_2(k_e,k_\nu)=\frac{2(p_eR)^{k_e-1}(p_\nu R)^{k_\nu-1}}{\sqrt{15(2k_e-1)!(2k_\nu-1)!}}\bigg\{-\sqrt{\frac{5}{2}}\VF{0}{211}\mp\frac{\alpha Z}{2k_e+1}\VF{0}{220}(k_e,1,1,1)\\
			&\mp\bigg[\frac{wR}{2k_e+1}\pm\frac{p_\nu R}{2k_\nu+1}\bigg]\VF{0}{220}
			-\frac{\alpha Z}{2k_e+1}\sqrt{\frac{3}{2}}\AF{0}{221}(k_e,1,1,1)
			-\bigg[\frac{wR}{2k_e+1}\mp\frac{p_\nu R}{2k_\nu+1}\bigg]\sqrt{\frac{3}{2}}\AF{0}{221}\Bigg\}
		\end{split}\\
		\label{eq:secforb_m2}
		&m_2(k_e,k_\nu)=\mp\frac{2(p_e R)^{k_e-1}(p_\nu R)^{k_\nu-1}}{\sqrt{15(2k_e-1)!(2k_\nu-1)!}}\frac{R}{2k_e+1}\Bigg\{\VF{0}{220}\pm\sqrt{\frac{3}{2}}\AF{0}{221}\Bigg\}\\
		\label{eq:secforb_M2_2}
		&\tilde{M}_2(k_e,k_\nu)=\pm2\frac{\sqrt{2}(p_e R)^{k_e-1}(p_\nu R)^{k_\nu-1}}{\sqrt{5(2k_e-1)(2k_\nu-1)(2k_e-1)!(2k_\nu-1)!}}\Bigg\{\VF{0}{220}\mp\frac{k_e-k_\nu}{3}\sqrt{\frac{3}{2}}\AF{0}{221}\Bigg\}\\
		\label{eq:secforb_M3}
		&\tilde{M}_3(k_e,k_\nu)=-2\frac{2(p_e R)^{k_e-1}(p_\nu R)^{k_\nu-1}}{\sqrt{15(2k_e-1)!(2k_\nu-1)!}}\AF{0}{321}.
	\end{align}
\end{widetext}
In these expressions $p_e$ and $p_\nu$ are the electron and neutrino momenta in
units of $m_ec$ and $R$ refers to the nuclear radius in units of the reduced
electron Compton wavelength $\lambdabar_e=\hbar/(m_ec)$. Upper signs apply to
$\beta^-$ decay while lower signs correspond to electron capture. Note that the
latter case was derived in Ref.~\cite{Bambynek.Behrens.ea:1977,*Bambynek.Behrens.ea:1977err,behrens1982electron}
assuming orbital electron capture. We can with some modifications apply this
derivation to continuum electron capture as shown in
Ref.~\cite{stroemberg2020weak}.

One can compare the shape factor for $\beta^-$ decay from an initial state $i$
to a final state $f$ with the shape factor of the reverse electron capture from
$f$ to $i$. For a given electron energy $w$ the neutrino momentum has opposite
signs in the two cases:
\begin{equation}
	\label{eq:q_betam_ec}
	p_\nu^{\beta^-}=q_{if}^{\beta^-}-w=-(q_{fi}^\text{EC}+w)=-p_\nu^\text{EC}.
\end{equation}
Furthermore, the form factor coefficients follow the relation
\begin{equation}
	\label{eq:ffc_betam_ec}
	\begin{split}
		\left[\VAF{N}{KLs}\right]^\text{EC}_{f\rightarrow{}i}=&(-1)^{K-s+J_i-J_f}\\
		\times&\frac{\sqrt{2J_i+1}}{\sqrt{2J_f+1}}\left[\VAF{N}{KLs}\right]^{\beta^-}_{i\rightarrow{}f}
	\end{split}
\end{equation}
which of course also applies to coefficients of the type
$\VAF{N}{KLs}(k_e,m,n,\rho)$. Note that
Ref.~\cite{behrens1982electron,Bambynek.Behrens.ea:1977,*Bambynek.Behrens.ea:1977err}
arrived at a similar equation when exchanging initial and final states in the
reduced lepton matrix elements. As shown in Ref.~\cite{weidenmuller1961} the
sign change in \eqref{eq:ffc_betam_ec} ultimately arises from the conjugation
properties of the corresponding spherical tensor operators.

If we insert \eqref{eq:ffc_betam_ec} and \eqref{eq:q_betam_ec} into
\eqref{eq:shapefactor_secforb}--\eqref{eq:secforb_M3} we arrive at
\begin{equation}
	\label{eq:CEC_Cbeta}
	C^\text{EC}_{f\rightarrow{}i}(w)=\frac{2J_i+1}{2J_f+1}C^{\beta^-}_{i\rightarrow{}f}(w).
\end{equation}
In other words the shape factors for $\beta^-$ decay and for the reverse
electron capture are given by the same expression, up to a trivial factor
correcting for the exchange of the initial and final angular momenta. The key
difference instead lies in the domain of the two shape factors: for $\beta^-$
decay the electron energy cannot exceed the decay $Q$ value (${w<q_{if}}$)
whereas for electron capture the energy must instead be larger than this value
($w>q_{if}$, assuming $q_{if} > 1$). One can view $C^{\text{EC}}(w)$ as an
extension of $C^{\beta^-}(w)$ to higher energies.

One should note that \eqref{eq:CEC_Cbeta} does not apply when comparing stellar
electron capture to $\beta^-$ decay measured in the laboratory. This is due to
the Coulomb correction~\eqref{eq:qmed_EC} which means that
$q_{if}^\text{EC,med}\neq-q_{fi}^{\beta^-}$, thus
invalidating~\eqref{eq:q_betam_ec}. The relationship used in
Ref.~\cite{kirsebom2019measurement} between the forbidden
$^{20}\text{Ne}\rightarrow{}^{20}\text{F}$ electron capture to the measured
$\beta^-$ decay of $^{20}$F corrected for this fact by first computing the
electron capture shape factor for arbitrary $q_{if}^\text{EC,med}$ and then
rescaling it by the same amount as the fitted $\beta^-$ decay shape factor so
that \eqref{eq:CEC_Cbeta} was recovered in the limit of no Coulomb correction.

\section{$I(k_e,m,n,\rho)$ factors}
\label{sec:app_I}
In addition to the regular matrix elements $\VAM{N}{KLs}$, we also need to
calculate $\VAM{N}{KLs}(k_e,m,n,\rho)$ where the operator is multiplied by a
factor $I(k_e,m,n,\rho;r)$. Although this factor depends on how the nuclear
charge distribution is parameterized, it has been
shown~\cite{behrens1971nuclear} that it is not very sensitive to the precise
form. For simplicity, we can then assume a spherical uniform distribution for
which the factors we need in this work can be written
\begin{equation}
	\begin{split}
		&I(k_e,1,1,1;r)=\\
		&\begin{cases}
			\displaystyle{\frac{3}{2}-\frac{2k_e+1}{2(2k_e+3)}\biggl(\frac{r}{R}\biggr)^2,\;0\leq{}r\leq{}R}\\[4mm]
			\displaystyle{\frac{2k_e+1}{2k_e}\frac{R}{r}-\frac{3}{2k_e(2k_e+3)}\biggl(\frac{R}{r}\biggr)^{2k_e+1},r\geq R.}
		\end{cases}
	\end{split}
\end{equation}
Furthermore, one can show that we have the general property $I(k_e,m,n,0)=1$.
This means that we can put matrix elements with and without these factors on an
equal footing by writing
\begin{equation*}
	\VAM{N}{KLs}=\VAM{N}{KLs}(k_e,m,n,0).
\end{equation*}

\section{Single-particle matrix elements}
\label{sec:app_spme}
The nuclear matrix elements can be decomposed as
\begin{equation}
	\begin{split}
		&\VAM{N}{KLs}(k_e,n,n,\rho)=\\
		&\frac{1}{\sqrt{2J_i+1}}\sum_{\alpha\beta}\frac{\langle{}\psi_f\|[a_\alpha^\dagger\otimes\tilde{a}_\beta]^K\|\psi_i\rangle{}}{\sqrt{2K+1}}\\
		& \times\VAm{N}{KLs}(k_e,m,n,\rho)(\alpha\beta)
	\end{split}
\end{equation}
where $\psi_i$ and $\psi_f$ are the initial and final nuclear states and the
summation indices $\alpha$ and $\beta$ run over all single-nucleon states in
the valence space. Here
${\langle{}\psi_f\|[a_\alpha^\dagger\otimes\tilde{a}_\beta]^K\|\psi_i\rangle{}}$
is the reduced one-body transition density (which we get from our shell model
calculations) and $\VAm{0}{KLs}(k_e,m,n,\rho)(\alpha\beta)$ is the associated
single-particle matrix element. Note that for $\beta^-$ decay $\alpha$ denotes
a proton and $\beta$ denotes a neutron, with the opposite being the case for
electron capture.

A derivation of the single-particle matrix elements can be found in
Ref.~\cite{behrens1982electron}. It should be noted that the authors follow the
Biedenharn-Rose (BR) phase convention~\cite{biedenharn1953theory} while our
shell model calculations use the Condon-Shortley (CS)
convention~\cite{condon1951theory}. This means that expressions for the
single-particle matrix elements must be modified so that they can be used with
one-body transition densities in the CS convention.

The formalism in Ref.~\cite{behrens1982electron} is fully relativistic and the
single-nucleon states are thus given by spinors. If we follow the CS convention
but otherwise adhere to the definitions given in
Ref.~\cite{behrens1982electron} we can write these states as
\begin{equation}
	\label{eq:phi_CS}
	\phi_{\kappa\mu}(\mathbf{r})=
	\begin{pmatrix}
		-i f_\kappa(r)\chi_{-\kappa\mu}\\
		g_\kappa(r)\chi_{\kappa\mu}
	\end{pmatrix}
\end{equation}
where the quantum number $\kappa$ determines both $l$ and $j$ and is
defined by
\begin{equation*}
	\kappa=
	\begin{cases}
		\quad\, j+\frac{1}{2}\,&\text{for}\ l=j+\frac{1}{2}\\
		-(j+\frac{1}{2})\,&\text{for}\ l=j-\frac{1}{2}
	\end{cases}
\end{equation*}
and where $\chi_{\kappa\mu}$ is the spin-angular part
\begin{equation}
	\label{eq:chi_CS}
		\chi_{\kappa\mu}=\left[Y_l(\hat{r})\otimes\chi_{1/2}\right]_{\kappa\mu}.
\end{equation}
The functions $g_\kappa(r)$ and $f_\kappa(r)$ are solutions to the
radial Dirac equation. As described in
Ref.~\cite{behrens1982electron} we can relate these functions to
non-relativistic nuclear models (such as the shell model) by taking
the non-relativistic limit of the radial equations. $g_\kappa(r)$ then
becomes the solution of the corresponding radial Schr{\"o}dinger
equation whereas $f_\kappa(r)$ is given by
\begin{equation}
	\label{eq:nonrel_f_g}
	f_\kappa(r)=\frac{1}{2M_N} \left(\frac{d}{dr} +
          \frac{\kappa+1}{r} \right) g_\kappa(r) 
\end{equation}
with $M_N$ being the nucleon mass in units of $m_e$. For the harmonic
oscillator basis we use in this work we have
\begin{align}
	\begin{split}
		g_{n\kappa}(r)=&\sqrt{\frac{2n!}{b^3(n+l+1/2)!}}\bigg(\frac{r}{b}\bigg)^l\\
				&\times\exp(-r^2/2b^2)L_{n}^{l+\frac{1}{2}}(r^2/b^2)
	\end{split}\\
	\begin{split}
		f_{n\kappa}(r)=&\frac{1}{2M_N}\Bigg[\bigg(\frac{1+\kappa+l}{r}+\frac{r}{b^2}\bigg)g_{nl}(r)\\
				&-\frac{2}{b}\sqrt{n+l+\frac{3}{2}}g_{nl+1}(r)\Bigg]
	\end{split}
\end{align}
where $L_{n}^{l+\frac{1}{2}}(x)$ is the associated Laguerre polynomial and $b$
is the oscillator length. Note that $r$ and $b$ are like the nuclear radius $R$
given in units of the reduced electron Compton wavelength
$\lambdabar_e=\hbar/(m_ec)$.

To arrive at the single-particle matrix elements in the CS convention we
replace the BR single-particle states in Ref.~\cite{behrens1982electron} with
our new CS states given by \eqref{eq:phi_CS}. We then have to factor out a
global complex phase factor to keep our single-particle matrix elements real.
We describe this in more detail in Ref.~\cite{stroemberg2020weak}. The final
expressions are as follows:
\begin{widetext}
  \begin{subequations}
	\begin{align}
		\label{eq:sp_vMKK0_CS}
		\Vm{N}{KK0}(k_e,m,n,\rho)(\alpha\beta)=&\sqrt{2}\bigg\{G_{KK0}(\kappa_\alpha,\kappa_\beta)\int_0^\infty g_{n_\alpha,\kappa_\alpha}(r)\bigg(\frac{r}{R}\bigg)^{K+2N}I(k_e,m,n,\rho;r)g_{n_\beta,\kappa_\beta}(r)r^2dr  \notag \\
			&+G_{KK0}(-\kappa_\alpha,-\kappa_\beta)\int_0^\infty f_{n_\alpha,\kappa_\alpha}(r)\bigg(\frac{r}{R}\bigg)^{K+2N}I(k_e,m,n,\rho;r)f_{n_\beta,\kappa_\beta}(r)r^2dr\bigg\}\\
		\label{eq:sp_aMKL1_CS}
		\Am{N}{KL1}(k_e,m,n,\rho)(\alpha\beta)=&\sgn\left(K-L+1/2\right)\notag \\
		&\times\sqrt{2}\bigg\{G_{KL1}(\kappa_\alpha,\kappa_\beta)\int_0^\infty g_{n_\alpha,\kappa_\alpha}(r)\bigg(\frac{r}{R}\bigg)^{L+2N}I(k_e,m,n,\rho;r)g_{n_\beta,\kappa_\beta}(r)r^2dr \notag \\
		&+G_{KL1}(-\kappa_\alpha,-\kappa_\beta)\int_0^\infty f_{n_\alpha,\kappa_\alpha}(r)\bigg(\frac{r}{R}\bigg)^{L+2N}I(k_e,m,n,\rho;r)f_{n_\beta,\kappa_\beta}(r)r^2dr\bigg\}\\
		\label{eq:sp_vMKL1_CS}
		\Am{N}{KK0}(k_e,m,n,\rho)(\alpha\beta)=&\sqrt{2}\bigg\{G_{KK0}(\kappa_\alpha,-\kappa_\beta)\int_0^\infty g_{n_\alpha,\kappa_\alpha}(r)\bigg(\frac{r}{R}\bigg)^{K+2N}I(k_e,m,n,\rho;r)f_{n_\beta,\kappa_\beta}(r)r^2dr \notag \\
			&-G_{KK0}(-\kappa_\alpha,\kappa_\beta)\int_0^\infty f_{n_\alpha,\kappa_\alpha}(r)\bigg(\frac{r}{R}\bigg)^{K+2N}I(k_e,m,n,\rho;r)g_{n_\beta,\kappa_\beta}(r)r^2dr\bigg\}\\
		\label{eq:sp_aMKK0_CS}
		\Vm{N}{KL1}(k_e,m,n,\rho)(\alpha\beta)=&\sgn\left(L-K+1/2\right)\notag\\
		&\times\sqrt{2}\bigg\{G_{KL1}(\kappa_\alpha,-\kappa_\beta)\int_0^\infty g_{n_\alpha,\kappa_\alpha}(r)\bigg(\frac{r}{R}\bigg)^{L+2N}I(k_e,m,n,\rho;r)f_{n_\beta,\kappa_\beta}(r)r^2dr \notag \\
		&-G_{KL1}(-\kappa_\alpha,\kappa_\beta)\int_0^\infty f_{n_\alpha,\kappa_\alpha}(r)\bigg(\frac{r}{R}\bigg)^{L+2N}I(k_e,m,n,\rho;r)g_{n_\beta,\kappa_\beta}(r)r^2dr\bigg\}.
	\end{align}
      \end{subequations}
\end{widetext}
Our expressions contain the quantity $G_{KLs}(\kappa_1,\kappa_2)$ which holds
the spin-angular matrix element. In the CS convention it is given by the
expression
\begin{equation*}
	\begin{split}
		&G_{KLs}(\kappa_1,\kappa_2)=(-1)^{j_2-j_1+l_1}\sqrt{(2s+1)(2K+1)}\\
		&\times\sqrt{(2j_1+1)(2j_2+1)}\sqrt{(2l_1+1)(2l_2+1)}\\
		&\times
		{\Big\langle}{l_1 l_2 0 0\Big|L 0\Big\rangle}
		\begin{Bmatrix}
			K& s& L\\
			j_1& \frac{1}{2}& l_1\\
			j_2& \frac{1}{2}& l_2
		\end{Bmatrix}.
	\end{split}
\end{equation*}

\section{Population of isomeric state in $^{24}$Na}
\label{sec:app_isomer}
Ref.~\cite{schwab2015thermal} argues that the population of the
isomeric state in $^{24}$Na would follow a thermal distribution since
the gamma decay rate, while slow, would still be substantially faster
the electron capture rate. Here, we reexamine this issue by describing
the population of the different states as the steady-state solution to
a set of differential equations following the approach of
Ref.~\cite{ward1980thermalization}.

Note that we only need to consider the first three states
($4^+_\text{g.s.}$, $1^+_1$ and $2^+_1$) since all others have
excitation energies well above $1~\text{MeV}$. We denote the number
density of $^{24}$Na nuclei in the $4^+_\text{g.s.}$ state by $n_1$
and let $n_2$ and $n_3$ be the corresponding quantities for the
$1^+_1$ and $2^+_1$ states. Given this we can write
\begin{align}
	\label{eq:ward_equation_dn1dt}\frac{dn_1}{dt}&=-(\lambda^\gamma_{12}+\lambda^\gamma_{13}+\lambda^\text{EC}_{1d})n_1+\lambda^\gamma_{21}n_2+\lambda^\gamma_{31}n_3\\
	\label{eq:ward_equation_dn2dt}
	\begin{split}
		\frac{dn_2}{dt}&=\lambda^\gamma_{12}n_1-(\lambda^\gamma_{21}+\lambda^\gamma_{23}+\lambda^\text{EC}_{2d})n_2+\lambda^\gamma_{32}n_3\\
				&+\lambda^\text{EC}_{p2}n_p
	\end{split}\\
	\label{eq:ward_equation_dn3dt}\frac{dn_3}{dt}&=\lambda^\gamma_{13}n_1+\lambda^\gamma_{23}n_2-(\lambda^\gamma_{31}+\lambda^\gamma_{32})n_3.
\end{align}
Here $\lambda^\text{EC}_{id}$ is the sum of all partial electron capture
rates from a state $i$ to any state in the daughter nucleus $^{24}$Ne.
Furthermore, $n_p$ is the number density of the parent nucleus $^{24}$Mg and
$\lambda^\text{EC}_{pi}$ is the total electron capture rate from $^{24}$Mg to
a state $i$ in $^{24}$Na. Note that in
\eqref{eq:ward_equation_dn1dt}--\eqref{eq:ward_equation_dn3dt} we only include
the electron capture transitions shown in Fig.~\ref{fig:energylevels24}.
Finally, $\lambda^\gamma_{ij}$ refers to the gamma decay or excitation rate
from a state $i$ to a state $j$. We get the decay rates
$\lambda^\gamma_{21}=3.43\times10^1$~s$^{-1}$,
$\lambda^\gamma_{31}=6.78\times10^{8}$~s$^{-1}$ and
$\lambda^\gamma_{32}=1.86\times10^{10}$~s$^{-1}$ from
Ref.~\cite{firestone2007nuclear} and the inverse excitation rates via detailed
balance as
\begin{equation}
	\label{eq:detailed_balance}
	\lambda^\gamma_{ji}=\frac{2J_i+1}{2J_j+1}\exp\left(-\frac{E_i-E_j}{kT}\right)\lambda^\gamma_{ij}.
\end{equation}

The solution to \eqref{eq:ward_equation_dn1dt}--\eqref{eq:ward_equation_dn3dt}
is in general time-dependent, but will approach a steady-state solution as time
goes by. In this limit the ratio between the probabilities of occupying the
$1^+_1$ and $4^+_\text{g.s.}$ states is given by
\begin{equation}
	\label{eq:p2p1_ratio_general}
	\frac{P_2}{P_1}=\frac{n_2}{n_1}=\frac{(\lambda^\gamma_{12}+\lambda^\gamma_{13}+\lambda^\text{EC}_{1d})(\lambda^\gamma_{31}+\lambda^\gamma_{32})-\lambda^\gamma_{13}\lambda^\gamma_{31}}
			{\lambda^\gamma_{21}(\lambda^\gamma_{31}+\lambda^\gamma_{32})+\lambda^\gamma_{23}\lambda^\gamma_{31}}.
\end{equation}
Note that if the electron captures are much slower than the internal
transitions we have
\begin{equation}
	\label{eq:ec_slow_limit}
	\lambda^\text{EC}_{1d}\ll\lambda^\gamma_{12}+\lambda^\gamma_{13}
\end{equation}
in which case \eqref{eq:p2p1_ratio_general} can be shown to equal the
thermal-equilibrium ratio
\begin{equation}
	\label{eq:p2p1_ratio_thermal}
	\frac{P_2}{P_1}=\frac{2J_2+1}{2J_1+1}\exp\left(-\frac{E_2}{kT}\right).
\end{equation}
This corresponds to the Boltzmann distribution assumed in
\eqref{eq:lambda_total} and \eqref{eq:xi_total}.

\begin{figure}[htb]
	\centering
	\includegraphics[width=\linewidth]{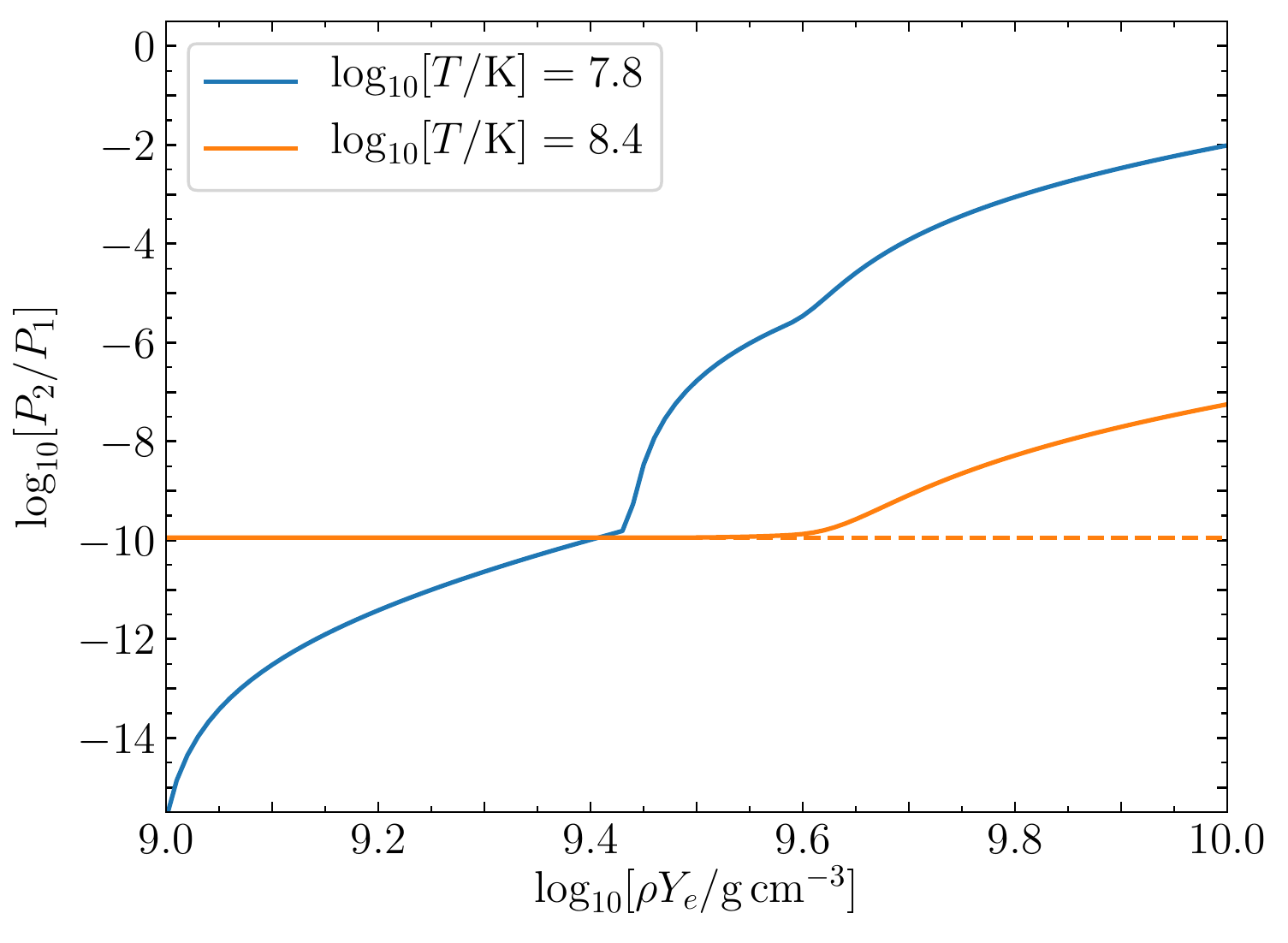}
	\caption{\label{fig:p2p1_ratio}
		Ratio between the probabilities of occupying the $1^+_1$  and
		$4^+_\text{g.s.}$ states as a function of the density. Solid
		lines corresponds to the steady-state expression
		\eqref{eq:p2p1_ratio_general} and dashed lines to the
		thermal-equilibrium limit \eqref{eq:p2p1_ratio_thermal}. Note
		that there is no dashed blue line as
		\eqref{eq:p2p1_ratio_thermal} is negligible
		(${\log_{10}(P_2/P_1)<-38}$) for $\log_{10}(T[\text{K}])=7.8$.
		}
\end{figure}

We plot \eqref{eq:p2p1_ratio_general} and \eqref{eq:p2p1_ratio_thermal} as a
function of the density in Fig.~\ref{fig:p2p1_ratio}. For
${\log_{10}(T[\text{K}])=8.4}$ the two coincide as long as
$\log_{10}(\rho{}Y_e[\text{g~cm$^{-3}$}])\lesssim9.5$. At higher densities
$\lambda^\text{EC}_{1d}$ is so large that \eqref{eq:ec_slow_limit} no longer
applies. As a consequence the thermal-equilibrium limit now underestimates the
population of the $1^+_1$ state. However, in our MESA simulations $^{24}$Na has
already been depleted when this density is reached. This means that it is still
appropriate to describe the population of the excited states using a Boltzmann
distribution at this temperature. The situation is quite different for
${\log_{10}(T[\text{K}])=7.8}$. At such low temperatures $\lambda^\gamma_{12}$
and $\lambda^\gamma_{13}$ are so small that \eqref{eq:ec_slow_limit} is
violated for all relevant densities, meaning that we cannot expect the
thermal-equilibrium limit to be a good approximation to the steady-state
solution at all.  Indeed, \eqref{eq:p2p1_ratio_thermal} underestimates the
$P_2/P_1$ ratio by tens of orders of magnitude compared to
\eqref{eq:p2p1_ratio_general}. Note that for
$\log_{10}(\rho{}Y_e[\text{g~cm$^{-3}$}])\gtrsim9.4$ the population of the
$1^+_1$ state is lower at ${\log_{10}(T[\text{K}])=8.4}$ than at
${\log_{10}(T[\text{K}])=7.8}$. This is because higher temperatures favors
excitation from the $1^+_1$ state to the $2^+_1$ state. The latter is not an
isomer and quickly decays to the ground state.

\begin{figure}[htb]
	\centering
	\includegraphics[width=\linewidth]{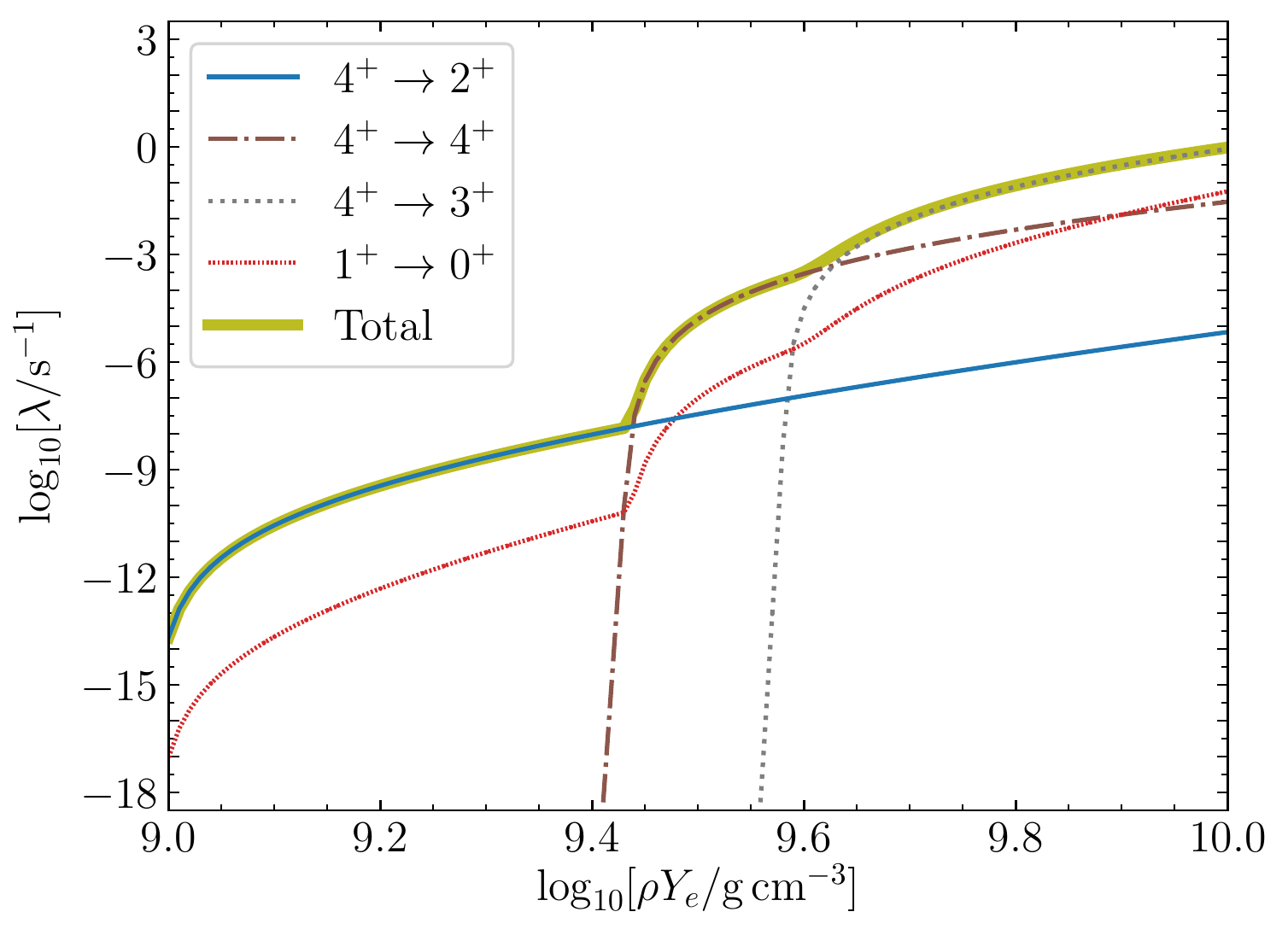}
	\caption{\label{fig:p2p1_rate_impact}
		Electron capture rates on $^{24}$Na at
		${\log_{10}(T[\text{K}])=7.8}$ assuming that the relative
		population of the $1^+_1$ state is given by
		\eqref{eq:p2p1_ratio_general}.
		}
\end{figure}

In Fig.~\ref{fig:p2p1_rate_impact} we illustrate the impact of the non-thermal
population at ${\log_{10}(T[\text{K}])=7.8}$ on the
$^{24}\text{Na}(e^-,\nu_e)^{24}\text{Ne}$ rate. Note that while the population
of the $1^+_1$ state is dramatically higher relative to the thermal equilibrium
the probability of a nucleus occupying the $4^+_\text{g.s.}$ ground state is
still very close to $1$. Consequently the electron capture rate via the
$4^+_\mathrm{g.s.}\rightarrow{}2^+_1$, $4^+_\mathrm{g.s.}\rightarrow{}3^+_1$
and $4^+_\mathrm{g.s.}\rightarrow{}4^+_1$ transitions are practically the same
as when assuming thermal equilibrium (i.e. as in the lower panel of
Fig.~\ref{fig:rates24}). In contrast, the contribution from the
$1^+_1\rightarrow{}0^+_\mathrm{g.s.}$ transition is now much larger than
before. However, it is still dwarfed by the rate of electron capture on the
ground state and the total rate is virtually unaffected. This would also be
true even if we ignored the forbidden transition: $\lambda^\text{EC}_{1d}$
would then be essentially zero for
$\log_{10}(\rho{}Y_e[\text{g~cm$^{-3}$}])\lesssim9.4$ and
\eqref{eq:ec_slow_limit} would still apply in this range. Given the above we
conclude that assuming thermal equilibrium will still give us the correct total
electron capture rate.

\begin{figure}[htb]
	\centering
	\includegraphics[width=\linewidth]{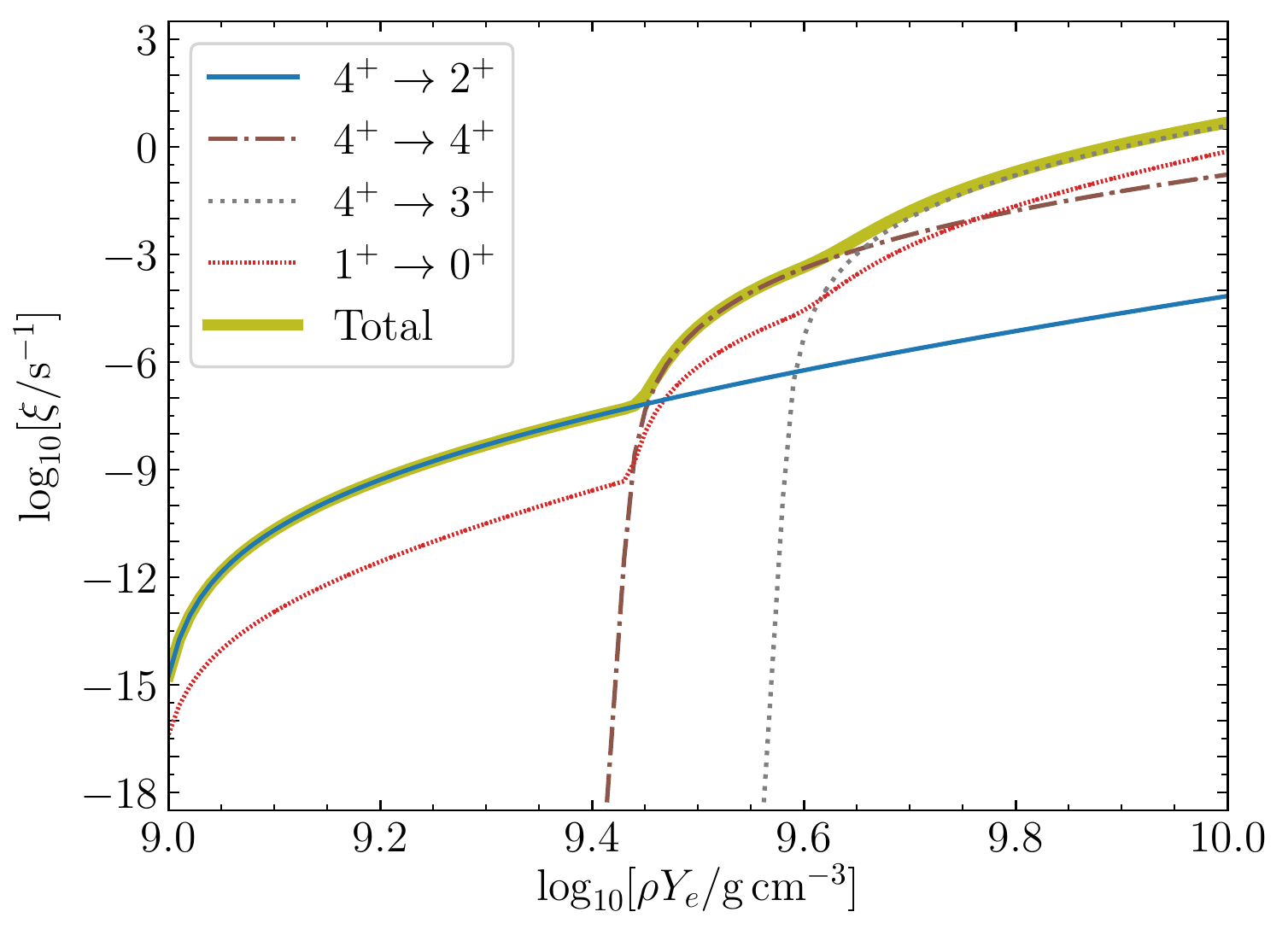}
	\caption{\label{fig:p2p1_neurate_impact}
		Neutrino loss rates for the electron capture on $^{24}$Na at
		${\log_{10}(T[\text{K}])=7.8}$, assuming that the relative
		population of the $1^+_1$ state is given by
		\eqref{eq:p2p1_ratio_general}.
		}
\end{figure}

The reader may ask whether the non-thermal population could affect the neutrino
loss rate \eqref{eq:xi_total}. We have computed this quantity at
${\log_{10}(T[\text{K}])=7.8}$ as shown in Fig.~\ref{fig:p2p1_neurate_impact}.
In this case the enhancement of the contribution from the
$1^+_1\rightarrow{}0^+_\mathrm{g.s.}$ transition appears to be larger than in
Fig.~\ref{fig:p2p1_rate_impact}. However, the neutrino loss rate is still
dominated by transitions from the ground state, which are not affected by the
higher relative population of the $1^+_1$ state. In conclusion, a thermal
distribution also appears to yield the correct neutrino loss rate.

\section{Calculation of the forbidden electron capture rate in MESA}
\label{sec:app_calcrate}
For allowed transitions MESA is able to calculate $\beta^-$ decay and electron
capture rates directly. This capability was introduced in
Ref.~\cite{schwab2015thermal,paxton2015modules,*paxton2016erratum}
and allows to avoid the interpolating errors associated with using tabulated
rates. In practice, the implementation is based on rewriting the rate
expressions in terms of Fermi integrals
\begin{equation}
	\label{eq:fermi}
	F_k(\eta)=\int_0^\infty\frac{x^k}{\exp(x-\eta)+1}dx
\end{equation}
as MESA already includes efficient routines to evaluate such quantities. We
wish to do the same for the electron capture via second-forbidden transitions
and we will now provide the corresponding expressions.

As in Ref.~\cite{schwab2015thermal} we start by writing
\begin{equation}
	\frac{p_e}{w}F(Z,w)\approx\exp(\pi\alpha Z)
\end{equation}
which is a valid approximation for light nuclei (${\alpha{}Z\ll1}$) and
ultra-relativistic electrons (${p_e\approx{}w}$). After inserting the above
into \eqref{eq:ecrate} and \eqref{eq:ecxi} we arrive at
\begin{align}
	\label{eq:lambda_approx}
	\begin{split}
		\lambda^\text{EC}_{if}=&\exp(\pi\alpha Z)\frac{\ln{2}}{K}\\
		&\times\int_{-q_{if}}^{\infty}\frac{C(w)w^2(q_{if}+w)^2}{1+\exp[\beta(w-\mu_e)]}dw
	\end{split}\\
	\label{eq:neuloss_approx}
	\begin{split}
		\xi^\text{EC}_{if}=&m_ec^2\exp(\pi\alpha Z)\frac{\ln{2}}{K}\\
		&\times\int_{-q_{if}}^{\infty}\frac{C(w)w^2(q_{if}+w)^3}{1+\exp[\beta(w-\mu_e)]}dw
	\end{split}
\end{align}
where we have used  $\beta=1/(kT)$. If we then replace $C(w)$
with the general form of a second-forbidden shape factor \eqref{eq:Cw} we get
\begin{align}
	\begin{split}
		\lambda^\text{EC}_{if}=&\exp(\pi\alpha Z)\frac{\ln{2}}{K}\\
						\times\bigg(&a_{-1}I_{-1}+a_0I_0+a_1I_1\\
						+&a_2I_2+a_3I_3+a_4I_4\bigg)
	\end{split}\\
	\begin{split}
		\xi^\text{EC}_{if}=&m_ec^2\exp(\pi\alpha Z)\frac{\ln{2}}{K}\\
						\times\bigg(&a_{-1}J_{-1}+a_0J_0+a_1J_1\\
						+&a_2J_2+a_3J_3+a_4J_4\bigg)
	\end{split}
\end{align}
where we have the integrals
\begin{align}
	I_n&=\frac{w^{2+n}(q_{if}+w)^2}{1+\exp[\beta(w-\mu_e)]}\\
	J_n&=\frac{w^{2+n}(q_{if}+w)^3}{1+\exp[\beta(w-\mu_e)]}.
\end{align}

We can express $I_n$ in terms of Fermi integrals by first introducing the
quantities $\eta=\beta\mu_e$ and $\zeta=\beta q_{if}$ and then making the
variable substitution $x=\beta(w+q_{if})$. The result is
\begingroup
\allowdisplaybreaks
\begin{align*}
	I_{-1}=\frac{1}{\beta^4}\big[&F_3(\eta+\zeta)-\zeta F_{2}(\eta+\zeta)\big]\\
	I_0=\frac{1}{\beta^5}\big[&F_4(\eta+\zeta)-2\zeta F_{3}(\eta+\zeta)+\zeta^2F_{2}(\eta+\zeta)\big]\\
	\begin{split}
		I_1=\frac{1}{\beta^6}\big[&F_5(\eta+\zeta)-3\zeta F_{4}(\eta+\zeta) +3\zeta^2 F_{3}(\eta+\zeta)\\
		-&\zeta^3F_{2}(\eta+\zeta)\big]
	\end{split}\\
	\begin{split}
		I_2=\frac{1}{\beta^7}\big[&F_6(\eta+\zeta)-4\zeta F_{5}(\eta+\zeta) +6\zeta^2F_{4}(\eta+\zeta)\\
		-&4\zeta^3F_{3}(\eta+\zeta) +\zeta^4F_{2}(\eta+\zeta)\big]
	\end{split}\\
	\begin{split}
		I_3=\frac{1}{\beta^8}\big[&F_7(\eta+\zeta)-5\zeta F_{6}(\eta+\zeta)+10\zeta^2 F_{5}(\eta+\zeta)\\
		-&10\zeta^3F_{4}(\eta+\zeta)+5\zeta^4F_{3}(\eta+\zeta) -\zeta^5F_{2}(\eta+\zeta)\big]
	\end{split}\\
	\begin{split}
		I_4=\frac{1}{\beta^9}\big[&F_8(\eta+\zeta)-6\zeta F_{7}(\eta+\zeta)+15\zeta^2 F_{6}(\eta+\zeta)\\
		-&20\zeta^3F_{5}(\eta+\zeta)+15\zeta^4F_{4}(\eta+\zeta)-6\zeta^5F_{3}(\eta+\zeta)\\
		+&\zeta^6F_{2}(\eta+\zeta)\big].
	\end{split}
\end{align*}
\endgroup
Applying the same approach to $J_n$ yields
\begingroup
\allowdisplaybreaks
\begin{align*}
	\begin{split}
		J_{-1}=\frac{1}{\beta^5}\big[&F_4(\eta+\zeta)-\zeta F_{3}(\eta+\zeta)\big]
	\end{split}\\
	\begin{split}
		J_0=\frac{1}{\beta^6}\big[&F_5(\eta+\zeta)-2\zeta F_{4}(\eta+\zeta)+\zeta^2F_{3}(\eta+\zeta)\big]
	\end{split}\\
	\begin{split}
		J_1=\frac{1}{\beta^7}\big[&F_6(\eta+\zeta)-3\zeta F_{5}(\eta+\zeta)+3\zeta^2 F_{4}(\eta+\zeta)\\
		-&\zeta^3F_{3}(\eta+\zeta)\big]
	\end{split}\\
	\begin{split}
		J_2=\frac{1}{\beta^8}\big[&F_7(\eta+\zeta)-4\zeta F_{6}(\eta+\zeta)+6\zeta^2 F_{5}(\eta+\zeta)\\
		-&4\zeta^3F_{4}(\eta+\zeta)+\zeta^4F_{3}(\eta+\zeta)\big]
	\end{split}\\
	\begin{split}
		J_3=\frac{1}{\beta^9}\big[&F_8(\eta+\zeta)-5\zeta F_{7}(\eta+\zeta)+10\zeta^2 F_{6}(\eta+\zeta)\\
		-&10\zeta^3F_{5}(\eta+\zeta)+5\zeta^4F_{4}(\eta+\zeta)-\zeta^5F_{3}(\eta+\zeta)\big]
	\end{split}\\
	\begin{split}
		J_4=\frac{1}{\beta^{10}}\big[&F_9(\eta+\zeta)-6\zeta F_{8}(\eta+\zeta)+15\zeta^2 F_{7}(\eta+\zeta)\\
		-&20\zeta^3F_{6}(\eta+\zeta)+15\zeta^4F_{5}(\eta+\zeta) -6\zeta^5F_{4}(\eta+\zeta)\\
		+&\zeta^6F_{3}(\eta+\zeta)\big].
	\end{split}
\end{align*}
\endgroup

We have implemented the above expressions in the rate calculation routines in
MESA. Note that the shape factor $C(w)$ must be reevaluated throughout the
simulation as the Coulomb correction \eqref{eq:qmed_EC} varies with density
and temperature. (This quantity enters
\eqref{eq:secforb_M2_1}--\eqref{eq:secforb_M3} via the neutrino momentum
${p_\nu^\text{EC}=q_{if}^\text{EC,med}+w}$.) As a consequence we must write
the coefficients $a_{-1},a_0\dots{}a_4$ in \eqref{eq:Cw} as functions of
$q_{if}^\text{EC,med}$ so that MESA can compute their values at each
simulation step. For the $0^+_\text{g.s.}\rightarrow2^+_\text{g.s.}$
transition in $^{20}\text{Ne}(e^-,\nu_e)^{20}\text{F}$ we use the SM+CVC+E2
fit from Ref.~\cite{kirsebom2019measurement}, which gives us
\begin{align*}
	a_{-1}=&(-5.191 - 0.3228q + 8.411q^2 + 0.3222q^3)\times10^{-11}\\
	\begin{split}
	a_0=&(-659.9 - 65.33q + 624.3q^2 + 48.01q^3\\
	    &+ 1.009q^4)\times10^{-11}
	\end{split}\\
	a_1=&(-63.35 + 1248q + 135.6q^2 + 3.712q^3)\times10^{-11}\\
	a_2=&(1283 + 209.3q + 8.446q^2)\times10^{-11}\\
	a_3=&(116.5 + 9.469q)\times10^{-11}\\
	a_4=&4.734\times10^{-11}.
\end{align*}
Here we have written $q=q_{if}^\text{EC,med}$ for brevity (note that $q<0$).
Similarly, for the ${4^+_\text{g.s.}\rightarrow{}2^+_1}$ transition in
$^{24}\text{Na}(e^-,\nu_e)^{24}\text{Ne}$ the coefficients in the SM+CVC
shape factor are
\begin{align*}
	a_{-1} =& (-5.690 + 10.10q - 130.7q^2 - 10.08q^3)\times10^{-13}\\
	\begin{split}
	a_0    =& (15.48 - 256.8q + 199.2q^2 + 38.81q^3\\
		& + 6.702q^4)\times10^{-13}
	\end{split}\\
	a_1    =& (-11.47 + 40.85q + 24.76q^2 + 3.692q^3)\times10^{-12}\\
	a_2    =& (18.71 + 37.41q + 9.289q^2)\times10^{-12}\\
	a_3    =& (1.596 + 1.119q)\times10^{-11}\\
	a_4    =& 5.597\times10^{-12}.
\end{align*}

\bibliography{refs}

\begin{thebibliography}{56}%
\makeatletter
\providecommand \@ifxundefined [1]{%
 \@ifx{#1\undefined}
}%
\providecommand \@ifnum [1]{%
 \ifnum #1\expandafter \@firstoftwo
 \else \expandafter \@secondoftwo
 \fi
}%
\providecommand \@ifx [1]{%
 \ifx #1\expandafter \@firstoftwo
 \else \expandafter \@secondoftwo
 \fi
}%
\providecommand \natexlab [1]{#1}%
\providecommand \enquote  [1]{``#1''}%
\providecommand \bibnamefont  [1]{#1}%
\providecommand \bibfnamefont [1]{#1}%
\providecommand \citenamefont [1]{#1}%
\providecommand \href@noop [0]{\@secondoftwo}%
\providecommand \href [0]{\begingroup \@sanitize@url \@href}%
\providecommand \@href[1]{\@@startlink{#1}\@@href}%
\providecommand \@@href[1]{\endgroup#1\@@endlink}%
\providecommand \@sanitize@url [0]{\catcode `\\12\catcode `\$12\catcode
  `\&12\catcode `\#12\catcode `\^12\catcode `\_12\catcode `\%12\relax}%
\providecommand \@@startlink[1]{}%
\providecommand \@@endlink[0]{}%
\providecommand \url  [0]{\begingroup\@sanitize@url \@url }%
\providecommand \@url [1]{\endgroup\@href {#1}{\urlprefix }}%
\providecommand \urlprefix  [0]{URL }%
\providecommand \Eprint [0]{\href }%
\providecommand \doibase [0]{https://doi.org/}%
\providecommand \selectlanguage [0]{\@gobble}%
\providecommand \bibinfo  [0]{\@secondoftwo}%
\providecommand \bibfield  [0]{\@secondoftwo}%
\providecommand \translation [1]{[#1]}%
\providecommand \BibitemOpen [0]{}%
\providecommand \bibitemStop [0]{}%
\providecommand \bibitemNoStop [0]{.\EOS\space}%
\providecommand \EOS [0]{\spacefactor3000\relax}%
\providecommand \BibitemShut  [1]{\csname bibitem#1\endcsname}%
\let\auto@bib@innerbib\@empty
\bibitem [{\citenamefont {Doherty}\ \emph {et~al.}(2017)\citenamefont
  {Doherty}, \citenamefont {Gil-Pons}, \citenamefont {Siess},\ and\
  \citenamefont {Lattanzio}}]{doherty2017super}%
  \BibitemOpen
  \bibfield  {author} {\bibinfo {author} {\bibfnamefont {C.~L.}\ \bibnamefont
  {Doherty}}, \bibinfo {author} {\bibfnamefont {P.}~\bibnamefont {Gil-Pons}},
  \bibinfo {author} {\bibfnamefont {L.}~\bibnamefont {Siess}},\ and\ \bibinfo
  {author} {\bibfnamefont {J.~C.}\ \bibnamefont {Lattanzio}},\ }\bibfield
  {title} {\bibinfo {title} {{Super-AGB Stars and their Role as Electron
  Capture Supernova Progenitors}},\ }\href
  {https://doi.org/10.1017/pasa.2017.52} {\bibfield  {journal} {\bibinfo
  {journal} {Publ. Astron. Soc. Austr.}\ }\textbf {\bibinfo {volume} {34}},\
  \bibinfo {pages} {e056} (\bibinfo {year} {2017})}\BibitemShut {NoStop}%
\bibitem [{\citenamefont {{Jones}}\ \emph {et~al.}(2013)\citenamefont
  {{Jones}}, \citenamefont {{Hirschi}}, \citenamefont {{Nomoto}}, \citenamefont
  {{Fischer}}, \citenamefont {{Timmes}}, \citenamefont {{Herwig}},
  \citenamefont {{Paxton}}, \citenamefont {{Toki}}, \citenamefont {{Suzuki}},
  \citenamefont {{Mart{\'{\i}}nez-Pinedo}}, \citenamefont {{Lam}},\ and\
  \citenamefont {{Bertolli}}}]{jones2013advanced}%
  \BibitemOpen
  \bibfield  {author} {\bibinfo {author} {\bibfnamefont {S.}~\bibnamefont
  {{Jones}}}, \bibinfo {author} {\bibfnamefont {R.}~\bibnamefont {{Hirschi}}},
  \bibinfo {author} {\bibfnamefont {K.}~\bibnamefont {{Nomoto}}}, \bibinfo
  {author} {\bibfnamefont {T.}~\bibnamefont {{Fischer}}}, \bibinfo {author}
  {\bibfnamefont {F.~X.}\ \bibnamefont {{Timmes}}}, \bibinfo {author}
  {\bibfnamefont {F.}~\bibnamefont {{Herwig}}}, \bibinfo {author}
  {\bibfnamefont {B.}~\bibnamefont {{Paxton}}}, \bibinfo {author}
  {\bibfnamefont {H.}~\bibnamefont {{Toki}}}, \bibinfo {author} {\bibfnamefont
  {T.}~\bibnamefont {{Suzuki}}}, \bibinfo {author} {\bibfnamefont
  {G.}~\bibnamefont {{Mart{\'{\i}}nez-Pinedo}}}, \bibinfo {author}
  {\bibfnamefont {Y.~H.}\ \bibnamefont {{Lam}}},\ and\ \bibinfo {author}
  {\bibfnamefont {M.~G.}\ \bibnamefont {{Bertolli}}},\ }\bibfield  {title}
  {\bibinfo {title} {{Advanced Burning Stages and Fate of 8-10 M$_\odot$
  Stars}},\ }\href {https://doi.org/10.1088/0004-637X/772/2/150} {\bibfield
  {journal} {\bibinfo  {journal} {Astrophys. J.}\ }\textbf {\bibinfo {volume}
  {772}},\ \bibinfo {eid} {150} (\bibinfo {year} {2013})}\BibitemShut {NoStop}%
\bibitem [{\citenamefont {Takahashi}\ \emph {et~al.}(2013)\citenamefont
  {Takahashi}, \citenamefont {Yoshida},\ and\ \citenamefont
  {Umeda}}]{takahashi2013evolution}%
  \BibitemOpen
  \bibfield  {author} {\bibinfo {author} {\bibfnamefont {K.}~\bibnamefont
  {Takahashi}}, \bibinfo {author} {\bibfnamefont {T.}~\bibnamefont {Yoshida}},\
  and\ \bibinfo {author} {\bibfnamefont {H.}~\bibnamefont {Umeda}},\ }\bibfield
   {title} {\bibinfo {title} {{Evolution of progenitors for electron capture
  supernovae}},\ }\href {https://doi.org/10.1088/0004-637X/771/1/28} {\bibfield
   {journal} {\bibinfo  {journal} {Astrophys. J.}\ }\textbf {\bibinfo {volume}
  {771}},\ \bibinfo {pages} {28} (\bibinfo {year} {2013})}\BibitemShut
  {NoStop}%
\bibitem [{\citenamefont {Miyaji}\ \emph {et~al.}(1980)\citenamefont {Miyaji},
  \citenamefont {Nomoto}, \citenamefont {Yokoi},\ and\ \citenamefont
  {Sugimoto}}]{miyaji1980supernova}%
  \BibitemOpen
  \bibfield  {author} {\bibinfo {author} {\bibfnamefont {S.}~\bibnamefont
  {Miyaji}}, \bibinfo {author} {\bibfnamefont {K.}~\bibnamefont {Nomoto}},
  \bibinfo {author} {\bibfnamefont {K.}~\bibnamefont {Yokoi}},\ and\ \bibinfo
  {author} {\bibfnamefont {D.}~\bibnamefont {Sugimoto}},\ }\bibfield  {title}
  {\bibinfo {title} {{Supernova triggered by electron captures}},\ }\href@noop
  {} {\bibfield  {journal} {\bibinfo  {journal} {Publ. Astron. Soc. Japan}\
  }\textbf {\bibinfo {volume} {32}},\ \bibinfo {pages} {303} (\bibinfo {year}
  {1980})}\BibitemShut {NoStop}%
\bibitem [{\citenamefont {Leung}\ \emph {et~al.}(2020)\citenamefont {Leung},
  \citenamefont {Nomoto},\ and\ \citenamefont {Suzuki}}]{leung2020electron}%
  \BibitemOpen
  \bibfield  {author} {\bibinfo {author} {\bibfnamefont {S.-C.}\ \bibnamefont
  {Leung}}, \bibinfo {author} {\bibfnamefont {K.}~\bibnamefont {Nomoto}},\ and\
  \bibinfo {author} {\bibfnamefont {T.}~\bibnamefont {Suzuki}},\ }\bibfield
  {title} {\bibinfo {title} {{Electron-capture Supernovae of Super-{AGB} Stars:
  Sensitivity on Input Physics}},\ }\href
  {https://doi.org/10.3847/1538-4357/ab5d2f} {\bibfield  {journal} {\bibinfo
  {journal} {Astrophys. J.}\ }\textbf {\bibinfo {volume} {889}},\ \bibinfo
  {pages} {34} (\bibinfo {year} {2020})}\BibitemShut {NoStop}%
\bibitem [{\citenamefont {Isern}\ \emph {et~al.}(1991)\citenamefont {Isern},
  \citenamefont {Canal},\ and\ \citenamefont {Labay}}]{isern1991outcome}%
  \BibitemOpen
  \bibfield  {author} {\bibinfo {author} {\bibfnamefont {J.}~\bibnamefont
  {Isern}}, \bibinfo {author} {\bibfnamefont {R.}~\bibnamefont {Canal}},\ and\
  \bibinfo {author} {\bibfnamefont {J.}~\bibnamefont {Labay}},\ }\bibfield
  {title} {\bibinfo {title} {{The outcome of explosive ignition of ONeMg cores
  - Supernovae, neutron stars, or "iron" white dwarfs?}},\ }\href
  {https://doi.org/10.1086/186029} {\bibfield  {journal} {\bibinfo  {journal}
  {Astrophys. J.}\ }\textbf {\bibinfo {volume} {372}},\ \bibinfo {pages} {L83}
  (\bibinfo {year} {1991})}\BibitemShut {NoStop}%
\bibitem [{\citenamefont {Jones}\ \emph {et~al.}(2016)\citenamefont {Jones},
  \citenamefont {R{\"o}pke}, \citenamefont {Pakmor}, \citenamefont
  {Seitenzahl}, \citenamefont {Ohlmann},\ and\ \citenamefont
  {Edelmann}}]{jones2016electron}%
  \BibitemOpen
  \bibfield  {author} {\bibinfo {author} {\bibfnamefont {S.}~\bibnamefont
  {Jones}}, \bibinfo {author} {\bibfnamefont {F.~K.}\ \bibnamefont
  {R{\"o}pke}}, \bibinfo {author} {\bibfnamefont {R.}~\bibnamefont {Pakmor}},
  \bibinfo {author} {\bibfnamefont {I.~R.}\ \bibnamefont {Seitenzahl}},
  \bibinfo {author} {\bibfnamefont {S.~T.}\ \bibnamefont {Ohlmann}},\ and\
  \bibinfo {author} {\bibfnamefont {P.~V.}\ \bibnamefont {Edelmann}},\
  }\bibfield  {title} {\bibinfo {title} {{Do electron-capture supernovae make
  neutron stars? - First multidimensional hydrodynamic simulations of the
  oxygen deflagration}},\ }\href {https://doi.org/10.1051/0004-6361/201628321}
  {\bibfield  {journal} {\bibinfo  {journal} {Astron. \& Astrophys.}\ }\textbf
  {\bibinfo {volume} {593}},\ \bibinfo {pages} {A72} (\bibinfo {year}
  {2016})}\BibitemShut {NoStop}%
\bibitem [{\citenamefont {Nomoto}(1987)}]{nomoto1987evolution}%
  \BibitemOpen
  \bibfield  {author} {\bibinfo {author} {\bibfnamefont {K.}~\bibnamefont
  {Nomoto}},\ }\bibfield  {title} {\bibinfo {title} {{Evolution of 8-10 solar
  mass stars toward electron capture supernovae. II-Collapse of an O+Ne+Mg
  core}},\ }\href {https://doi.org/10.1086/165716} {\bibfield  {journal}
  {\bibinfo  {journal} {Astrophys. J.}\ }\textbf {\bibinfo {volume} {322}},\
  \bibinfo {pages} {206} (\bibinfo {year} {1987})}\BibitemShut {NoStop}%
\bibitem [{\citenamefont {Miyaji}\ and\ \citenamefont
  {Nomoto}(1987)}]{miyaji1987collapse}%
  \BibitemOpen
  \bibfield  {author} {\bibinfo {author} {\bibfnamefont {S.}~\bibnamefont
  {Miyaji}}\ and\ \bibinfo {author} {\bibfnamefont {K.}~\bibnamefont
  {Nomoto}},\ }\bibfield  {title} {\bibinfo {title} {{On the Collapse of 8--10
  M$_{sun}$ Stars Due to Electron Capture}},\ }\href
  {https://doi.org/10.1086/165368} {\bibfield  {journal} {\bibinfo  {journal}
  {Astrophys. J.}\ }\textbf {\bibinfo {volume} {318}},\ \bibinfo {pages} {307}
  (\bibinfo {year} {1987})}\BibitemShut {NoStop}%
\bibitem [{\citenamefont {Canal}\ \emph {et~al.}(1992)\citenamefont {Canal},
  \citenamefont {Isern},\ and\ \citenamefont {Labay}}]{canal1992quasi}%
  \BibitemOpen
  \bibfield  {author} {\bibinfo {author} {\bibfnamefont {R.}~\bibnamefont
  {Canal}}, \bibinfo {author} {\bibfnamefont {J.}~\bibnamefont {Isern}},\ and\
  \bibinfo {author} {\bibfnamefont {J.}~\bibnamefont {Labay}},\ }\bibfield
  {title} {\bibinfo {title} {{The quasi-static evolution of ONeMg cores -
  Explosive ignition densities and the collapse/explosion alternative}},\
  }\href {https://doi.org/10.1086/186574} {\bibfield  {journal} {\bibinfo
  {journal} {Astrophys. J.}\ }\textbf {\bibinfo {volume} {398}},\ \bibinfo
  {pages} {L49} (\bibinfo {year} {1992})}\BibitemShut {NoStop}%
\bibitem [{\citenamefont {Hashimoto}\ \emph {et~al.}(1993)\citenamefont
  {Hashimoto}, \citenamefont {Iwamoto},\ and\ \citenamefont
  {Nomoto}}]{hashimoto1993type}%
  \BibitemOpen
  \bibfield  {author} {\bibinfo {author} {\bibfnamefont {M.}~\bibnamefont
  {Hashimoto}}, \bibinfo {author} {\bibfnamefont {K.}~\bibnamefont {Iwamoto}},\
  and\ \bibinfo {author} {\bibfnamefont {K.}~\bibnamefont {Nomoto}},\
  }\bibfield  {title} {\bibinfo {title} {{Type II Supernovae from 8--10
  M$_{sun}$ Asymptotic Giant Branch Stars}},\ }\href
  {https://doi.org/10.1086/187007} {\bibfield  {journal} {\bibinfo  {journal}
  {Astrophys. J.}\ }\textbf {\bibinfo {volume} {414}},\ \bibinfo {pages} {L105}
  (\bibinfo {year} {1993})}\BibitemShut {NoStop}%
\bibitem [{\citenamefont {Gutierrez}\ \emph {et~al.}(1996)\citenamefont
  {Gutierrez}, \citenamefont {Garcia-Berro}, \citenamefont {Iben~Jr},
  \citenamefont {Isern}, \citenamefont {Labay},\ and\ \citenamefont
  {Canal}}]{gutierrez1996final}%
  \BibitemOpen
  \bibfield  {author} {\bibinfo {author} {\bibfnamefont {J.}~\bibnamefont
  {Gutierrez}}, \bibinfo {author} {\bibfnamefont {E.}~\bibnamefont
  {Garcia-Berro}}, \bibinfo {author} {\bibfnamefont {I.}~\bibnamefont
  {Iben~Jr}}, \bibinfo {author} {\bibfnamefont {J.}~\bibnamefont {Isern}},
  \bibinfo {author} {\bibfnamefont {J.}~\bibnamefont {Labay}},\ and\ \bibinfo
  {author} {\bibfnamefont {R.}~\bibnamefont {Canal}},\ }\bibfield  {title}
  {\bibinfo {title} {{The final evolution of ONeMg electron-degenerate
  cores}},\ }\href {https://doi.org/10.1086/176934} {\bibfield  {journal}
  {\bibinfo  {journal} {Astrophys. J.}\ }\textbf {\bibinfo {volume} {459}},\
  \bibinfo {pages} {701} (\bibinfo {year} {1996})}\BibitemShut {NoStop}%
\bibitem [{\citenamefont {Guti{\'e}rrez}\ \emph {et~al.}(2005)\citenamefont
  {Guti{\'e}rrez}, \citenamefont {Canal},\ and\ \citenamefont
  {Garcia-Berro}}]{gutierrez2005gravitational}%
  \BibitemOpen
  \bibfield  {author} {\bibinfo {author} {\bibfnamefont {J.}~\bibnamefont
  {Guti{\'e}rrez}}, \bibinfo {author} {\bibfnamefont {R.}~\bibnamefont
  {Canal}},\ and\ \bibinfo {author} {\bibfnamefont {E.}~\bibnamefont
  {Garcia-Berro}},\ }\bibfield  {title} {\bibinfo {title} {{The gravitational
  collapse of ONe electron-degenerate cores and white dwarfs: The role of Mg
  and C revisited}},\ }\href {https://doi.org/10.1051/0004-6361:20042254}
  {\bibfield  {journal} {\bibinfo  {journal} {Astron. \& Astrophys.}\ }\textbf
  {\bibinfo {volume} {435}},\ \bibinfo {pages} {231} (\bibinfo {year}
  {2005})}\BibitemShut {NoStop}%
\bibitem [{\citenamefont {Schwab}\ \emph {et~al.}(2015)\citenamefont {Schwab},
  \citenamefont {Quataert},\ and\ \citenamefont
  {Bildsten}}]{schwab2015thermal}%
  \BibitemOpen
  \bibfield  {author} {\bibinfo {author} {\bibfnamefont {J.}~\bibnamefont
  {Schwab}}, \bibinfo {author} {\bibfnamefont {E.}~\bibnamefont {Quataert}},\
  and\ \bibinfo {author} {\bibfnamefont {L.}~\bibnamefont {Bildsten}},\
  }\bibfield  {title} {\bibinfo {title} {{Thermal runaway during the evolution
  of ONeMg cores towards accretion-induced collapse}},\ }\href
  {https://doi.org/10.1093/mnras/stv1804} {\bibfield  {journal} {\bibinfo
  {journal} {Mon. Not. Roy. Astron. Soc.}\ }\textbf {\bibinfo {volume} {453}},\
  \bibinfo {pages} {1910} (\bibinfo {year} {2015})}\BibitemShut {NoStop}%
\bibitem [{\citenamefont {Schwab}\ \emph {et~al.}(2016)\citenamefont {Schwab},
  \citenamefont {Quataert},\ and\ \citenamefont
  {Bildsten}}]{schwab2015erratum}%
  \BibitemOpen
  \bibfield  {author} {\bibinfo {author} {\bibfnamefont {J.}~\bibnamefont
  {Schwab}}, \bibinfo {author} {\bibfnamefont {E.}~\bibnamefont {Quataert}},\
  and\ \bibinfo {author} {\bibfnamefont {L.}~\bibnamefont {Bildsten}},\
  }\bibfield  {title} {\bibinfo {title} {{Erratum: Thermal runaway during the
  evolution of ONeMg cores towards accretion-induced collapse}},\ }\href
  {https://doi.org/10.1093/mnras/stw560} {\bibfield  {journal} {\bibinfo
  {journal} {Mon. Not. Roy. Astron. Soc.}\ }\textbf {\bibinfo {volume} {458}},\
  \bibinfo {pages} {3613} (\bibinfo {year} {2016})}\BibitemShut {NoStop}%
\bibitem [{\citenamefont {Schwab}\ \emph {et~al.}(2017)\citenamefont {Schwab},
  \citenamefont {Bildsten},\ and\ \citenamefont
  {Quataert}}]{schwab2017importance}%
  \BibitemOpen
  \bibfield  {author} {\bibinfo {author} {\bibfnamefont {J.}~\bibnamefont
  {Schwab}}, \bibinfo {author} {\bibfnamefont {L.}~\bibnamefont {Bildsten}},\
  and\ \bibinfo {author} {\bibfnamefont {E.}~\bibnamefont {Quataert}},\
  }\bibfield  {title} {\bibinfo {title} {{The importance of Urca-process
  cooling in accreting ONe white dwarfs}},\ }\href
  {https://doi.org/10.1093/mnras/stx2169} {\bibfield  {journal} {\bibinfo
  {journal} {Mon. Not. Roy. Astron. Soc.}\ }\textbf {\bibinfo {volume} {472}},\
  \bibinfo {pages} {3390} (\bibinfo {year} {2017})}\BibitemShut {NoStop}%
\bibitem [{\citenamefont {Oda}\ \emph {et~al.}(1994)\citenamefont {Oda},
  \citenamefont {Hino}, \citenamefont {Muto}, \citenamefont {Takahara},\ and\
  \citenamefont {Sato}}]{oda1994rate}%
  \BibitemOpen
  \bibfield  {author} {\bibinfo {author} {\bibfnamefont {T.}~\bibnamefont
  {Oda}}, \bibinfo {author} {\bibfnamefont {M.}~\bibnamefont {Hino}}, \bibinfo
  {author} {\bibfnamefont {K.}~\bibnamefont {Muto}}, \bibinfo {author}
  {\bibfnamefont {M.}~\bibnamefont {Takahara}},\ and\ \bibinfo {author}
  {\bibfnamefont {K.}~\bibnamefont {Sato}},\ }\bibfield  {title} {\bibinfo
  {title} {{Rate tables for the weak processes of sd-shell nuclei in stellar
  matter}},\ }\href {https://doi.org/10.1006/adnd.1994.1007} {\bibfield
  {journal} {\bibinfo  {journal} {At. Data Nucl. Data Tables}\ }\textbf
  {\bibinfo {volume} {56}},\ \bibinfo {pages} {231} (\bibinfo {year}
  {1994})}\BibitemShut {NoStop}%
\bibitem [{\citenamefont {Takahara}\ \emph {et~al.}(1989)\citenamefont
  {Takahara}, \citenamefont {Hino}, \citenamefont {Oda}, \citenamefont {Muto},
  \citenamefont {Wolters}, \citenamefont {Glaudemans},\ and\ \citenamefont
  {Sato}}]{takahara1989microscopic}%
  \BibitemOpen
  \bibfield  {author} {\bibinfo {author} {\bibfnamefont {M.}~\bibnamefont
  {Takahara}}, \bibinfo {author} {\bibfnamefont {M.}~\bibnamefont {Hino}},
  \bibinfo {author} {\bibfnamefont {T.}~\bibnamefont {Oda}}, \bibinfo {author}
  {\bibfnamefont {K.}~\bibnamefont {Muto}}, \bibinfo {author} {\bibfnamefont
  {A.}~\bibnamefont {Wolters}}, \bibinfo {author} {\bibfnamefont
  {P.}~\bibnamefont {Glaudemans}},\ and\ \bibinfo {author} {\bibfnamefont
  {K.}~\bibnamefont {Sato}},\ }\bibfield  {title} {\bibinfo {title}
  {{Microscopic calculation of the rates of electron captures which induce the
  collapse of O+Ne+Mg cores}},\ }\href
  {https://doi.org/10.1016/0375-9474(89)90288-1} {\bibfield  {journal}
  {\bibinfo  {journal} {Nucl. Phys. A}\ }\textbf {\bibinfo {volume} {504}},\
  \bibinfo {pages} {167} (\bibinfo {year} {1989})}\BibitemShut {NoStop}%
\bibitem [{\citenamefont {Mart{\'i}nez-Pinedo}\ \emph
  {et~al.}(2014)\citenamefont {Mart{\'i}nez-Pinedo}, \citenamefont {Lam},
  \citenamefont {Langanke}, \citenamefont {Zegers},\ and\ \citenamefont
  {Sullivan}}]{martinez2014astrophysical}%
  \BibitemOpen
  \bibfield  {author} {\bibinfo {author} {\bibfnamefont {G.}~\bibnamefont
  {Mart{\'i}nez-Pinedo}}, \bibinfo {author} {\bibfnamefont {Y.~H.}\
  \bibnamefont {Lam}}, \bibinfo {author} {\bibfnamefont {K.}~\bibnamefont
  {Langanke}}, \bibinfo {author} {\bibfnamefont {R.~G.~T.}\ \bibnamefont
  {Zegers}},\ and\ \bibinfo {author} {\bibfnamefont {C.}~\bibnamefont
  {Sullivan}},\ }\bibfield  {title} {\bibinfo {title} {{Astrophysical
  weak-interaction rates for selected $A=20$ and $A=24$ nuclei}},\ }\href
  {https://doi.org/10.1103/PhysRevC.89.045806} {\bibfield  {journal} {\bibinfo
  {journal} {Phys. Rev. C}\ }\textbf {\bibinfo {volume} {89}},\ \bibinfo
  {pages} {045806} (\bibinfo {year} {2014})}\BibitemShut {NoStop}%
\bibitem [{\citenamefont {Kirsebom}\ \emph
  {et~al.}(2019{\natexlab{a}})\citenamefont {Kirsebom}, \citenamefont
  {Hukkanen}, \citenamefont {Kankainen}, \citenamefont {Trzaska}, \citenamefont
  {Str{\"o}mberg}, \citenamefont {Mart{\'i}nez-Pinedo}, \citenamefont
  {Andersen}, \citenamefont {Bodewits}, \citenamefont {Brown}, \citenamefont
  {Canete}, \citenamefont {Cederk\"all}, \citenamefont {Enqvist}, \citenamefont
  {Eronen}, \citenamefont {Fynbo}, \citenamefont {Geldhof}, \citenamefont
  {de~Groote}, \citenamefont {Jenkins}, \citenamefont {Jokinen}, \citenamefont
  {Joshi}, \citenamefont {Khanam}, \citenamefont {Kostensalo}, \citenamefont
  {Kuusiniemi}, \citenamefont {Langanke}, \citenamefont {Moore}, \citenamefont
  {Munch}, \citenamefont {Nesterenko}, \citenamefont {Ovejas}, \citenamefont
  {Penttil\"a}, \citenamefont {Pohjalainen}, \citenamefont {Reponen},
  \citenamefont {Rinta-Antila}, \citenamefont {Riisager}, \citenamefont
  {de~Roubin}, \citenamefont {Schotanus}, \citenamefont {Srivastava},
  \citenamefont {Suhonen}, \citenamefont {Swartz}, \citenamefont {Tengblad},
  \citenamefont {Vilen}, \citenamefont {V\'{\i}nals},\ and\ \citenamefont
  {\"Ayst\"o}}]{kirsebom2019measurement}%
  \BibitemOpen
  \bibfield  {author} {\bibinfo {author} {\bibfnamefont {O.~S.}\ \bibnamefont
  {Kirsebom}}, \bibinfo {author} {\bibfnamefont {M.}~\bibnamefont {Hukkanen}},
  \bibinfo {author} {\bibfnamefont {A.}~\bibnamefont {Kankainen}}, \bibinfo
  {author} {\bibfnamefont {W.~H.}\ \bibnamefont {Trzaska}}, \bibinfo {author}
  {\bibfnamefont {D.~F.}\ \bibnamefont {Str{\"o}mberg}}, \bibinfo {author}
  {\bibfnamefont {G.}~\bibnamefont {Mart{\'i}nez-Pinedo}}, \bibinfo {author}
  {\bibfnamefont {K.}~\bibnamefont {Andersen}}, \bibinfo {author}
  {\bibfnamefont {E.}~\bibnamefont {Bodewits}}, \bibinfo {author}
  {\bibfnamefont {B.~A.}\ \bibnamefont {Brown}}, \bibinfo {author}
  {\bibfnamefont {L.}~\bibnamefont {Canete}}, \bibinfo {author} {\bibfnamefont
  {J.}~\bibnamefont {Cederk\"all}}, \bibinfo {author} {\bibfnamefont
  {T.}~\bibnamefont {Enqvist}}, \bibinfo {author} {\bibfnamefont
  {T.}~\bibnamefont {Eronen}}, \bibinfo {author} {\bibfnamefont {H.~O.~U.}\
  \bibnamefont {Fynbo}}, \bibinfo {author} {\bibfnamefont {S.}~\bibnamefont
  {Geldhof}}, \bibinfo {author} {\bibfnamefont {R.}~\bibnamefont {de~Groote}},
  \bibinfo {author} {\bibfnamefont {D.~G.}\ \bibnamefont {Jenkins}}, \bibinfo
  {author} {\bibfnamefont {A.}~\bibnamefont {Jokinen}}, \bibinfo {author}
  {\bibfnamefont {P.}~\bibnamefont {Joshi}}, \bibinfo {author} {\bibfnamefont
  {A.}~\bibnamefont {Khanam}}, \bibinfo {author} {\bibfnamefont
  {J.}~\bibnamefont {Kostensalo}}, \bibinfo {author} {\bibfnamefont
  {P.}~\bibnamefont {Kuusiniemi}}, \bibinfo {author} {\bibfnamefont
  {K.}~\bibnamefont {Langanke}}, \bibinfo {author} {\bibfnamefont
  {I.}~\bibnamefont {Moore}}, \bibinfo {author} {\bibfnamefont
  {M.}~\bibnamefont {Munch}}, \bibinfo {author} {\bibfnamefont {D.~A.}\
  \bibnamefont {Nesterenko}}, \bibinfo {author} {\bibfnamefont {J.~D.}\
  \bibnamefont {Ovejas}}, \bibinfo {author} {\bibfnamefont {H.}~\bibnamefont
  {Penttil\"a}}, \bibinfo {author} {\bibfnamefont {I.}~\bibnamefont
  {Pohjalainen}}, \bibinfo {author} {\bibfnamefont {M.}~\bibnamefont
  {Reponen}}, \bibinfo {author} {\bibfnamefont {S.}~\bibnamefont
  {Rinta-Antila}}, \bibinfo {author} {\bibfnamefont {K.}~\bibnamefont
  {Riisager}}, \bibinfo {author} {\bibfnamefont {A.}~\bibnamefont {de~Roubin}},
  \bibinfo {author} {\bibfnamefont {P.}~\bibnamefont {Schotanus}}, \bibinfo
  {author} {\bibfnamefont {P.~C.}\ \bibnamefont {Srivastava}}, \bibinfo
  {author} {\bibfnamefont {J.}~\bibnamefont {Suhonen}}, \bibinfo {author}
  {\bibfnamefont {J.~A.}\ \bibnamefont {Swartz}}, \bibinfo {author}
  {\bibfnamefont {O.}~\bibnamefont {Tengblad}}, \bibinfo {author}
  {\bibfnamefont {M.}~\bibnamefont {Vilen}}, \bibinfo {author} {\bibfnamefont
  {S.}~\bibnamefont {V\'{\i}nals}},\ and\ \bibinfo {author} {\bibfnamefont
  {J.}~\bibnamefont {\"Ayst\"o}},\ }\bibfield  {title} {\bibinfo {title}
  {{Measurement of the ${2}^{+}\ensuremath{\rightarrow}{0}^{+}$ ground-state
  transition in the $\ensuremath{\beta}$ decay of $^{20}\mathrm{F}$}},\ }\href
  {https://doi.org/10.1103/PhysRevC.100.065805} {\bibfield  {journal} {\bibinfo
   {journal} {Phys. Rev. C}\ }\textbf {\bibinfo {volume} {100}},\ \bibinfo
  {pages} {065805} (\bibinfo {year} {2019}{\natexlab{a}})}\BibitemShut
  {NoStop}%
\bibitem [{\citenamefont {Kirsebom}\ \emph
  {et~al.}(2019{\natexlab{b}})\citenamefont {Kirsebom}, \citenamefont {Jones},
  \citenamefont {Str{\"o}mberg}, \citenamefont {Mart{\'i}nez-Pinedo},
  \citenamefont {Langanke}, \citenamefont {R\"opke}, \citenamefont {Brown},
  \citenamefont {Eronen}, \citenamefont {Fynbo}, \citenamefont {Hukkanen},
  \citenamefont {Idini}, \citenamefont {Jokinen}, \citenamefont {Kankainen},
  \citenamefont {Kostensalo}, \citenamefont {Moore}, \citenamefont {M\"oller},
  \citenamefont {Ohlmann}, \citenamefont {Penttil\"a}, \citenamefont
  {Riisager}, \citenamefont {Rinta-Antila}, \citenamefont {Srivastava},
  \citenamefont {Suhonen}, \citenamefont {Trzaska},\ and\ \citenamefont
  {\"Ayst\"o}}]{kirsebom2019discovery}%
  \BibitemOpen
  \bibfield  {author} {\bibinfo {author} {\bibfnamefont {O.~S.}\ \bibnamefont
  {Kirsebom}}, \bibinfo {author} {\bibfnamefont {S.}~\bibnamefont {Jones}},
  \bibinfo {author} {\bibfnamefont {D.~F.}\ \bibnamefont {Str{\"o}mberg}},
  \bibinfo {author} {\bibfnamefont {G.}~\bibnamefont {Mart{\'i}nez-Pinedo}},
  \bibinfo {author} {\bibfnamefont {K.}~\bibnamefont {Langanke}}, \bibinfo
  {author} {\bibfnamefont {F.~K.}\ \bibnamefont {R\"opke}}, \bibinfo {author}
  {\bibfnamefont {B.~A.}\ \bibnamefont {Brown}}, \bibinfo {author}
  {\bibfnamefont {T.}~\bibnamefont {Eronen}}, \bibinfo {author} {\bibfnamefont
  {H.~O.~U.}\ \bibnamefont {Fynbo}}, \bibinfo {author} {\bibfnamefont
  {M.}~\bibnamefont {Hukkanen}}, \bibinfo {author} {\bibfnamefont
  {A.}~\bibnamefont {Idini}}, \bibinfo {author} {\bibfnamefont
  {A.}~\bibnamefont {Jokinen}}, \bibinfo {author} {\bibfnamefont
  {A.}~\bibnamefont {Kankainen}}, \bibinfo {author} {\bibfnamefont
  {J.}~\bibnamefont {Kostensalo}}, \bibinfo {author} {\bibfnamefont
  {I.}~\bibnamefont {Moore}}, \bibinfo {author} {\bibfnamefont
  {H.}~\bibnamefont {M\"oller}}, \bibinfo {author} {\bibfnamefont {S.~T.}\
  \bibnamefont {Ohlmann}}, \bibinfo {author} {\bibfnamefont {H.}~\bibnamefont
  {Penttil\"a}}, \bibinfo {author} {\bibfnamefont {K.}~\bibnamefont
  {Riisager}}, \bibinfo {author} {\bibfnamefont {S.}~\bibnamefont
  {Rinta-Antila}}, \bibinfo {author} {\bibfnamefont {P.~C.}\ \bibnamefont
  {Srivastava}}, \bibinfo {author} {\bibfnamefont {J.}~\bibnamefont {Suhonen}},
  \bibinfo {author} {\bibfnamefont {W.~H.}\ \bibnamefont {Trzaska}},\ and\
  \bibinfo {author} {\bibfnamefont {J.}~\bibnamefont {\"Ayst\"o}},\ }\bibfield
  {title} {\bibinfo {title} {{Discovery of an Exceptionally Strong
  $\ensuremath{\beta}$-Decay Transition of $^{20}\mathrm{F}$ and Implications
  for the Fate of Intermediate-Mass Stars}},\ }\href
  {https://doi.org/10.1103/PhysRevLett.123.262701} {\bibfield  {journal}
  {\bibinfo  {journal} {Phys. Rev. Lett.}\ }\textbf {\bibinfo {volume} {123}},\
  \bibinfo {pages} {262701} (\bibinfo {year} {2019}{\natexlab{b}})}\BibitemShut
  {NoStop}%
\bibitem [{\citenamefont {Toki}\ \emph {et~al.}(2013)\citenamefont {Toki},
  \citenamefont {Suzuki}, \citenamefont {Nomoto}, \citenamefont {Jones},\ and\
  \citenamefont {Hirschi}}]{toki2013detailed}%
  \BibitemOpen
  \bibfield  {author} {\bibinfo {author} {\bibfnamefont {H.}~\bibnamefont
  {Toki}}, \bibinfo {author} {\bibfnamefont {T.}~\bibnamefont {Suzuki}},
  \bibinfo {author} {\bibfnamefont {K.}~\bibnamefont {Nomoto}}, \bibinfo
  {author} {\bibfnamefont {S.}~\bibnamefont {Jones}},\ and\ \bibinfo {author}
  {\bibfnamefont {R.}~\bibnamefont {Hirschi}},\ }\bibfield  {title} {\bibinfo
  {title} {{Detailed $\beta$-transition rates for URCA nuclear pairs in 8--10
  solar-mass stars}},\ }\href {https://doi.org/10.1103/PhysRevC.88.015806}
  {\bibfield  {journal} {\bibinfo  {journal} {Phys. Rev. C}\ }\textbf {\bibinfo
  {volume} {88}},\ \bibinfo {pages} {015806} (\bibinfo {year}
  {2013})}\BibitemShut {NoStop}%
\bibitem [{\citenamefont {Fuller}\ \emph {et~al.}(1980)\citenamefont {Fuller},
  \citenamefont {Fowler},\ and\ \citenamefont {Newman}}]{fuller1980stellar}%
  \BibitemOpen
  \bibfield  {author} {\bibinfo {author} {\bibfnamefont {G.~M.}\ \bibnamefont
  {Fuller}}, \bibinfo {author} {\bibfnamefont {W.~A.}\ \bibnamefont {Fowler}},\
  and\ \bibinfo {author} {\bibfnamefont {M.~J.}\ \bibnamefont {Newman}},\
  }\bibfield  {title} {\bibinfo {title} {{Stellar weak-interaction rates for
  sd-shell nuclei. I. Nuclear matrix element systematics with application to
  $^{26}$Al and selected nuclei of importance to the supernova problem}},\
  }\href {https://doi.org/10.1086/190657} {\bibfield  {journal} {\bibinfo
  {journal} {Astrophys. J. Suppl.}\ }\textbf {\bibinfo {volume} {42}},\
  \bibinfo {pages} {447} (\bibinfo {year} {1980})}\BibitemShut {NoStop}%
\bibitem [{\citenamefont {Behrens}\ and\ \citenamefont
  {B{\"u}hring}(1971)}]{behrens1971nuclear}%
  \BibitemOpen
  \bibfield  {author} {\bibinfo {author} {\bibfnamefont {H.}~\bibnamefont
  {Behrens}}\ and\ \bibinfo {author} {\bibfnamefont {W.}~\bibnamefont
  {B{\"u}hring}},\ }\bibfield  {title} {\bibinfo {title} {{Nuclear beta
  decay}},\ }\href {https://doi.org/10.1016/0375-9474(71)90489-1} {\bibfield
  {journal} {\bibinfo  {journal} {Nucl. Phys. A}\ }\textbf {\bibinfo {volume}
  {162}},\ \bibinfo {pages} {111} (\bibinfo {year} {1971})}\BibitemShut
  {NoStop}%
\bibitem [{\citenamefont {Behrens}\ and\ \citenamefont
  {B{\"u}hring}(1982)}]{behrens1982electron}%
  \BibitemOpen
  \bibfield  {author} {\bibinfo {author} {\bibfnamefont {H.}~\bibnamefont
  {Behrens}}\ and\ \bibinfo {author} {\bibfnamefont {W.}~\bibnamefont
  {B{\"u}hring}},\ }\href@noop {} {\emph {\bibinfo {title} {{Electron Radial
  Wave Functions and Nuclear Beta-decay}}}}\ (\bibinfo  {publisher} {Clarendon,
  Oxford},\ \bibinfo {year} {1982})\BibitemShut {NoStop}%
\bibitem [{\citenamefont {Bambynek}\ \emph
  {et~al.}(1977{\natexlab{a}})\citenamefont {Bambynek}, \citenamefont
  {Behrens}, \citenamefont {Chen}, \citenamefont {Crasemann}, \citenamefont
  {Fitzpatrick}, \citenamefont {Ledingham}, \citenamefont {Genz}, \citenamefont
  {Mutterer},\ and\ \citenamefont {Intemann}}]{Bambynek.Behrens.ea:1977}%
  \BibitemOpen
  \bibfield  {author} {\bibinfo {author} {\bibfnamefont {W.}~\bibnamefont
  {Bambynek}}, \bibinfo {author} {\bibfnamefont {H.}~\bibnamefont {Behrens}},
  \bibinfo {author} {\bibfnamefont {M.~H.}\ \bibnamefont {Chen}}, \bibinfo
  {author} {\bibfnamefont {B.}~\bibnamefont {Crasemann}}, \bibinfo {author}
  {\bibfnamefont {M.~L.}\ \bibnamefont {Fitzpatrick}}, \bibinfo {author}
  {\bibfnamefont {K.~W.~D.}\ \bibnamefont {Ledingham}}, \bibinfo {author}
  {\bibfnamefont {H.}~\bibnamefont {Genz}}, \bibinfo {author} {\bibfnamefont
  {M.}~\bibnamefont {Mutterer}},\ and\ \bibinfo {author} {\bibfnamefont
  {R.~L.}\ \bibnamefont {Intemann}},\ }\bibfield  {title} {\bibinfo {title}
  {{Orbital electron capture by the nucleus}},\ }\href
  {https://doi.org/10.1103/RevModPhys.49.77} {\bibfield  {journal} {\bibinfo
  {journal} {Rev. Mod. Phys.}\ }\textbf {\bibinfo {volume} {49}},\ \bibinfo
  {pages} {77} (\bibinfo {year} {1977}{\natexlab{a}})}\BibitemShut {NoStop}%
\bibitem [{\citenamefont {Bambynek}\ \emph
  {et~al.}(1977{\natexlab{b}})\citenamefont {Bambynek}, \citenamefont
  {Behrens}, \citenamefont {Chen}, \citenamefont {Crasemann}, \citenamefont
  {Fitzpatrick}, \citenamefont {Ledingham}, \citenamefont {Genz}, \citenamefont
  {Mutterer},\ and\ \citenamefont {Intemann}}]{Bambynek.Behrens.ea:1977err}%
  \BibitemOpen
  \bibfield  {author} {\bibinfo {author} {\bibfnamefont {W.}~\bibnamefont
  {Bambynek}}, \bibinfo {author} {\bibfnamefont {H.}~\bibnamefont {Behrens}},
  \bibinfo {author} {\bibfnamefont {M.~H.}\ \bibnamefont {Chen}}, \bibinfo
  {author} {\bibfnamefont {B.}~\bibnamefont {Crasemann}}, \bibinfo {author}
  {\bibfnamefont {M.~L.}\ \bibnamefont {Fitzpatrick}}, \bibinfo {author}
  {\bibfnamefont {K.~W.~D.}\ \bibnamefont {Ledingham}}, \bibinfo {author}
  {\bibfnamefont {H.}~\bibnamefont {Genz}}, \bibinfo {author} {\bibfnamefont
  {M.}~\bibnamefont {Mutterer}},\ and\ \bibinfo {author} {\bibfnamefont
  {R.~L.}\ \bibnamefont {Intemann}},\ }\bibfield  {title} {\bibinfo {title}
  {{Erratum: Orbital electron capture by the nucleus}},\ }\href
  {https://doi.org/10.1103/RevModPhys.49.961} {\bibfield  {journal} {\bibinfo
  {journal} {Rev. Mod. Phys.}\ }\textbf {\bibinfo {volume} {49}},\ \bibinfo
  {pages} {961} (\bibinfo {year} {1977}{\natexlab{b}})}\BibitemShut {NoStop}%
\bibitem [{\citenamefont {F.~Str{\"o}mberg}(2020)}]{stroemberg2020weak}%
  \BibitemOpen
  \bibfield  {author} {\bibinfo {author} {\bibfnamefont {D.}~\bibnamefont
  {F.~Str{\"o}mberg}},\ }\emph {\bibinfo {title} {{Weak interactions in
  degenerate oxygen-neon cores}}},\ \href
  {https://doi.org/10.25534/tuprints-00013302} {Ph.D. thesis},\ \bibinfo
  {school} {Technische Universit{\"a}t Darmstadt} (\bibinfo {year}
  {2020})\BibitemShut {NoStop}%
\bibitem [{\citenamefont {Hardy}\ and\ \citenamefont
  {Towner}(2009)}]{hardy2009superallowed}%
  \BibitemOpen
  \bibfield  {author} {\bibinfo {author} {\bibfnamefont {J.~C.}\ \bibnamefont
  {Hardy}}\ and\ \bibinfo {author} {\bibfnamefont {I.}~\bibnamefont {Towner}},\
  }\bibfield  {title} {\bibinfo {title} {{Superallowed $0^+\rightarrow{}0^+$
  nuclear $\beta$ decays: A new survey with precision tests of the conserved
  vector current hypothesis and the standard model}},\ }\href
  {https://doi.org/10.1103/PhysRevC.79.055502} {\bibfield  {journal} {\bibinfo
  {journal} {Phys. Rev. C}\ }\textbf {\bibinfo {volume} {79}},\ \bibinfo
  {pages} {055502} (\bibinfo {year} {2009})}\BibitemShut {NoStop}%
\bibitem [{\citenamefont {Zyla}\ \emph {et~al.}(2020)\citenamefont {Zyla} \emph
  {et~al.}}]{Zyla:2020zbs}%
  \BibitemOpen
  \bibfield  {author} {\bibinfo {author} {\bibfnamefont {P.~A.}\ \bibnamefont
  {Zyla}} \emph {et~al.} (\bibinfo {collaboration} {Particle Data Group}),\
  }\bibfield  {title} {\bibinfo {title} {{Review of Particle Physics}},\ }\href
  {https://doi.org/10.1093/ptep/ptaa104} {\bibfield  {journal} {\bibinfo
  {journal} {PTEP}\ }\textbf {\bibinfo {volume} {2020}},\ \bibinfo {pages}
  {083C01} (\bibinfo {year} {2020})}\BibitemShut {NoStop}%
\bibitem [{\citenamefont {Bravo}\ and\ \citenamefont
  {Garc{\'\i}a-Senz}(1999)}]{bravo1999coulomb}%
  \BibitemOpen
  \bibfield  {author} {\bibinfo {author} {\bibfnamefont {E.}~\bibnamefont
  {Bravo}}\ and\ \bibinfo {author} {\bibfnamefont {D.}~\bibnamefont
  {Garc{\'\i}a-Senz}},\ }\bibfield  {title} {\bibinfo {title} {{Coulomb
  corrections to the equation of state of nuclear statistical equilibrium
  matter: implications for SNIa nucleosynthesis and the accretion-induced
  collapse of white dwarfs}},\ }\href
  {https://doi.org/10.1046/j.1365-8711.1999.02694.x} {\bibfield  {journal}
  {\bibinfo  {journal} {Mon. Not. Roy. Astron. Soc.}\ }\textbf {\bibinfo
  {volume} {307}},\ \bibinfo {pages} {984} (\bibinfo {year}
  {1999})}\BibitemShut {NoStop}%
\bibitem [{\citenamefont {Juodagalvis}\ \emph {et~al.}(2010)\citenamefont
  {Juodagalvis}, \citenamefont {Langanke}, \citenamefont {Hix}, \citenamefont
  {Mart{\'\i}nez-Pinedo},\ and\ \citenamefont
  {Sampaio}}]{juodagalvis2010improved}%
  \BibitemOpen
  \bibfield  {author} {\bibinfo {author} {\bibfnamefont {A.}~\bibnamefont
  {Juodagalvis}}, \bibinfo {author} {\bibfnamefont {K.}~\bibnamefont
  {Langanke}}, \bibinfo {author} {\bibfnamefont {W.~R.}\ \bibnamefont {Hix}},
  \bibinfo {author} {\bibfnamefont {G.}~\bibnamefont {Mart{\'\i}nez-Pinedo}},\
  and\ \bibinfo {author} {\bibfnamefont {J.~M.}\ \bibnamefont {Sampaio}},\
  }\bibfield  {title} {\bibinfo {title} {{Improved estimate of electron capture
  rates on nuclei during stellar core collapse}},\ }\href
  {https://doi.org/10.1016/j.nuclphysa.2010.09.012} {\bibfield  {journal}
  {\bibinfo  {journal} {Nucl. Phys. A}\ }\textbf {\bibinfo {volume} {848}},\
  \bibinfo {pages} {454} (\bibinfo {year} {2010})}\BibitemShut {NoStop}%
\bibitem [{\citenamefont {Itoh}\ \emph {et~al.}(2002)\citenamefont {Itoh},
  \citenamefont {Tomizawa}, \citenamefont {Tamamura}, \citenamefont {Wanajo},\
  and\ \citenamefont {Nozawa}}]{itoh2002screening}%
  \BibitemOpen
  \bibfield  {author} {\bibinfo {author} {\bibfnamefont {N.}~\bibnamefont
  {Itoh}}, \bibinfo {author} {\bibfnamefont {N.}~\bibnamefont {Tomizawa}},
  \bibinfo {author} {\bibfnamefont {M.}~\bibnamefont {Tamamura}}, \bibinfo
  {author} {\bibfnamefont {S.}~\bibnamefont {Wanajo}},\ and\ \bibinfo {author}
  {\bibfnamefont {S.}~\bibnamefont {Nozawa}},\ }\bibfield  {title} {\bibinfo
  {title} {{Screening corrections to the electron capture rates in dense stars
  by the relativistically degenerate electron liquid}},\ }\href
  {https://doi.org/10.1086/342726} {\bibfield  {journal} {\bibinfo  {journal}
  {Astrophys. J.}\ }\textbf {\bibinfo {volume} {579}},\ \bibinfo {pages} {380}
  (\bibinfo {year} {2002})}\BibitemShut {NoStop}%
\bibitem [{\citenamefont {Caurier}\ and\ \citenamefont
  {Nowacki}(1999)}]{caurier1999present}%
  \BibitemOpen
  \bibfield  {author} {\bibinfo {author} {\bibfnamefont {E.}~\bibnamefont
  {Caurier}}\ and\ \bibinfo {author} {\bibfnamefont {F.}~\bibnamefont
  {Nowacki}},\ }\bibfield  {title} {\bibinfo {title} {{Present status of shell
  model techniques}},\ }\href@noop {} {\bibfield  {journal} {\bibinfo
  {journal} {Act. Phys. Pol. B}\ }\textbf {\bibinfo {volume} {30}},\ \bibinfo
  {pages} {705} (\bibinfo {year} {1999})}\BibitemShut {NoStop}%
\bibitem [{\citenamefont {Caurier}\ \emph {et~al.}(2005)\citenamefont
  {Caurier}, \citenamefont {Martinez-Pinedo}, \citenamefont {Nowacki},
  \citenamefont {Poves},\ and\ \citenamefont {Zuker}}]{caurier2005shell}%
  \BibitemOpen
  \bibfield  {author} {\bibinfo {author} {\bibfnamefont {E.}~\bibnamefont
  {Caurier}}, \bibinfo {author} {\bibfnamefont {G.}~\bibnamefont
  {Martinez-Pinedo}}, \bibinfo {author} {\bibfnamefont {F.}~\bibnamefont
  {Nowacki}}, \bibinfo {author} {\bibfnamefont {A.}~\bibnamefont {Poves}},\
  and\ \bibinfo {author} {\bibfnamefont {A.}~\bibnamefont {Zuker}},\ }\bibfield
   {title} {\bibinfo {title} {{The shell model as a unified view of nuclear
  structure}},\ }\href {https://doi.org/10.1103/RevModPhys.77.427} {\bibfield
  {journal} {\bibinfo  {journal} {Rev. Mod. Phys.}\ }\textbf {\bibinfo {volume}
  {77}},\ \bibinfo {pages} {427} (\bibinfo {year} {2005})}\BibitemShut
  {NoStop}%
\bibitem [{\citenamefont {Brown}\ and\ \citenamefont
  {Richter}(2006)}]{brown2006new}%
  \BibitemOpen
  \bibfield  {author} {\bibinfo {author} {\bibfnamefont {B.~A.}\ \bibnamefont
  {Brown}}\ and\ \bibinfo {author} {\bibfnamefont {W.}~\bibnamefont
  {Richter}},\ }\bibfield  {title} {\bibinfo {title} {{New ``USD'' Hamiltonians
  for the sd shell}},\ }\href {https://doi.org/10.1103/PhysRevC.74.034315}
  {\bibfield  {journal} {\bibinfo  {journal} {Phys. Rev. C}\ }\textbf {\bibinfo
  {volume} {74}},\ \bibinfo {pages} {034315} (\bibinfo {year}
  {2006})}\BibitemShut {NoStop}%
\bibitem [{\citenamefont {Towner}\ \emph {et~al.}(1977)\citenamefont {Towner},
  \citenamefont {Hardy},\ and\ \citenamefont {Harvey}}]{towner1977analogue}%
  \BibitemOpen
  \bibfield  {author} {\bibinfo {author} {\bibfnamefont {I.}~\bibnamefont
  {Towner}}, \bibinfo {author} {\bibfnamefont {J.}~\bibnamefont {Hardy}},\ and\
  \bibinfo {author} {\bibfnamefont {M.}~\bibnamefont {Harvey}},\ }\bibfield
  {title} {\bibinfo {title} {{Analogue symmetry breaking in superallowed fermi
  $\beta$-decay}},\ }\href {https://doi.org/10.1016/0375-9474(77)90123-3}
  {\bibfield  {journal} {\bibinfo  {journal} {Nucl. Phys. A}\ }\textbf
  {\bibinfo {volume} {284}},\ \bibinfo {pages} {269} (\bibinfo {year}
  {1977})}\BibitemShut {NoStop}%
\bibitem [{\citenamefont {Fricke}\ \emph {et~al.}(1995)\citenamefont {Fricke},
  \citenamefont {Bernhardt}, \citenamefont {Heilig}, \citenamefont {Schaller},
  \citenamefont {Schellenberg}, \citenamefont {Shera},\ and\ \citenamefont
  {Dejager}}]{fricke1995nuclear}%
  \BibitemOpen
  \bibfield  {author} {\bibinfo {author} {\bibfnamefont {G.}~\bibnamefont
  {Fricke}}, \bibinfo {author} {\bibfnamefont {C.}~\bibnamefont {Bernhardt}},
  \bibinfo {author} {\bibfnamefont {K.}~\bibnamefont {Heilig}}, \bibinfo
  {author} {\bibfnamefont {L.}~\bibnamefont {Schaller}}, \bibinfo {author}
  {\bibfnamefont {L.}~\bibnamefont {Schellenberg}}, \bibinfo {author}
  {\bibfnamefont {E.}~\bibnamefont {Shera}},\ and\ \bibinfo {author}
  {\bibfnamefont {C.}~\bibnamefont {Dejager}},\ }\bibfield  {title} {\bibinfo
  {title} {{Nuclear ground state charge radii from electromagnetic
  interactions}},\ }\href {https://doi.org/10.1006/adnd.1995.1007} {\bibfield
  {journal} {\bibinfo  {journal} {At. Data Nucl. Data Tables}\ }\textbf
  {\bibinfo {volume} {60}},\ \bibinfo {pages} {177} (\bibinfo {year}
  {1995})}\BibitemShut {NoStop}%
\bibitem [{\citenamefont {Warburton}(1992)}]{Warburton:1992}%
  \BibitemOpen
  \bibfield  {author} {\bibinfo {author} {\bibfnamefont {E.~K.}\ \bibnamefont
  {Warburton}},\ }\bibfield  {title} {\bibinfo {title} {{Second-forbidden
  unique \ensuremath{\beta} decays of $^{10}\mathrm{Be}$, $^{22}\mathrm{Na}$,
  and $^{26}\mathrm{Al}$}},\ }\href {https://doi.org/10.1103/PhysRevC.45.463}
  {\bibfield  {journal} {\bibinfo  {journal} {Phys. Rev. C}\ }\textbf {\bibinfo
  {volume} {45}},\ \bibinfo {pages} {463} (\bibinfo {year} {1992})}\BibitemShut
  {NoStop}%
\bibitem [{\citenamefont {Mart{\'\i}nez-Pinedo}\ and\ \citenamefont
  {Vogel}(1998)}]{Martinez-Pinedo.Vogel:1998}%
  \BibitemOpen
  \bibfield  {author} {\bibinfo {author} {\bibfnamefont {G.}~\bibnamefont
  {Mart{\'\i}nez-Pinedo}}\ and\ \bibinfo {author} {\bibfnamefont
  {P.}~\bibnamefont {Vogel}},\ }\bibfield  {title} {\bibinfo {title} {{Shell
  Model Calculation of the $\beta^-$ and $\beta^+$ Partial Half-Lives of
  $^{54}$Mn and Other Unique Second Forbidden $\beta$ Decays}},\ }\href
  {https://doi.org/10.1103/PhysRevLett.81.281} {\bibfield  {journal} {\bibinfo
  {journal} {Phys. Rev. Lett.}\ }\textbf {\bibinfo {volume} {81}},\ \bibinfo
  {pages} {281} (\bibinfo {year} {1998})}\BibitemShut {NoStop}%
\bibitem [{\citenamefont {Suhonen}(2017)}]{suhonen2017}%
  \BibitemOpen
  \bibfield  {author} {\bibinfo {author} {\bibfnamefont {J.~T.}\ \bibnamefont
  {Suhonen}},\ }\bibfield  {title} {\bibinfo {title} {{Value of the
  Axial-Vector Coupling Strength in $\beta$ and $\beta\beta$ Decays: A
  Review}},\ }\href {https://doi.org/10.3389/fphy.2017.00055} {\bibfield
  {journal} {\bibinfo  {journal} {Front. Phys.}\ }\textbf {\bibinfo {volume}
  {5}},\ \bibinfo {pages} {55} (\bibinfo {year} {2017})}\BibitemShut {NoStop}%
\bibitem [{\citenamefont {Singh}\ \emph {et~al.}(1998)\citenamefont {Singh},
  \citenamefont {Rodriguez}, \citenamefont {Wong},\ and\ \citenamefont
  {Tuli}}]{singh1998}%
  \BibitemOpen
  \bibfield  {author} {\bibinfo {author} {\bibfnamefont {B.}~\bibnamefont
  {Singh}}, \bibinfo {author} {\bibfnamefont {J.}~\bibnamefont {Rodriguez}},
  \bibinfo {author} {\bibfnamefont {S.}~\bibnamefont {Wong}},\ and\ \bibinfo
  {author} {\bibfnamefont {J.}~\bibnamefont {Tuli}},\ }\bibfield  {title}
  {\bibinfo {title} {{Review Of Log$ft$ Values In $\beta$ Decay}},\ }\href
  {https://doi.org/10.1006/ndsh.1998.0015} {\bibfield  {journal} {\bibinfo
  {journal} {Nucl. Data Sheets}\ }\textbf {\bibinfo {volume} {84}},\ \bibinfo
  {pages} {487 } (\bibinfo {year} {1998})}\BibitemShut {NoStop}%
\bibitem [{\citenamefont {Firestone}(2007)}]{firestone2007nuclear}%
  \BibitemOpen
  \bibfield  {author} {\bibinfo {author} {\bibfnamefont {R.~B.}\ \bibnamefont
  {Firestone}},\ }\bibfield  {title} {\bibinfo {title} {{Nuclear data sheets
  for A=24}},\ }\href {https://doi.org/10.1016/j.nds.2007.10.001} {\bibfield
  {journal} {\bibinfo  {journal} {Nucl. Data Sheets}\ }\textbf {\bibinfo
  {volume} {108}},\ \bibinfo {pages} {2319} (\bibinfo {year}
  {2007})}\BibitemShut {NoStop}%
\bibitem [{\citenamefont {Turner}\ and\ \citenamefont
  {Cavanagh}(1951)}]{turner1951lxvi}%
  \BibitemOpen
  \bibfield  {author} {\bibinfo {author} {\bibfnamefont {J.~F.}\ \bibnamefont
  {Turner}}\ and\ \bibinfo {author} {\bibfnamefont {P.~E.}\ \bibnamefont
  {Cavanagh}},\ }\bibfield  {title} {\bibinfo {title} {{Highly forbidden
  transitions in the decay of {Na$^{24}$}}},\ }\href
  {https://doi.org/10.1080/14786445108561278} {\bibfield  {journal} {\bibinfo
  {journal} {Philos. Mag.}\ }\textbf {\bibinfo {volume} {42}},\ \bibinfo
  {pages} {636} (\bibinfo {year} {1951})}\BibitemShut {NoStop}%
\bibitem [{\citenamefont {Antony}\ \emph {et~al.}(1997)\citenamefont {Antony},
  \citenamefont {Pape},\ and\ \citenamefont {Britz}}]{Antony.Pape.Britz:1997}%
  \BibitemOpen
  \bibfield  {author} {\bibinfo {author} {\bibfnamefont {M.~S.}\ \bibnamefont
  {Antony}}, \bibinfo {author} {\bibfnamefont {A.}~\bibnamefont {Pape}},\ and\
  \bibinfo {author} {\bibfnamefont {J.}~\bibnamefont {Britz}},\ }\bibfield
  {title} {\bibinfo {title} {{Coulomb displacement energies between analog
  levels for $3\leq A \leq 239$}},\ }\href
  {https://doi.org/10.1006/adnd.1997.0740} {\bibfield  {journal} {\bibinfo
  {journal} {At. Data Nucl. Data Tables}\ }\textbf {\bibinfo {volume} {66}},\
  \bibinfo {pages} {1} (\bibinfo {year} {1997})}\BibitemShut {NoStop}%
\bibitem [{\citenamefont {Basunia}(2011)}]{basunia2011nuclear}%
  \BibitemOpen
  \bibfield  {author} {\bibinfo {author} {\bibfnamefont {M.~S.}\ \bibnamefont
  {Basunia}},\ }\bibfield  {title} {\bibinfo {title} {{Nuclear data sheets for
  A=27}},\ }\href {https://doi.org/10.1016/j.nds.2011.08.001} {\bibfield
  {journal} {\bibinfo  {journal} {Nucl. Data Sheets}\ }\textbf {\bibinfo
  {volume} {112}},\ \bibinfo {pages} {1875} (\bibinfo {year}
  {2011})}\BibitemShut {NoStop}%
\bibitem [{\citenamefont {Paxton}\ \emph {et~al.}(2010)\citenamefont {Paxton},
  \citenamefont {Bildsten}, \citenamefont {Dotter}, \citenamefont {Herwig},
  \citenamefont {Lesaffre},\ and\ \citenamefont {Timmes}}]{paxton2010modules}%
  \BibitemOpen
  \bibfield  {author} {\bibinfo {author} {\bibfnamefont {B.}~\bibnamefont
  {Paxton}}, \bibinfo {author} {\bibfnamefont {L.}~\bibnamefont {Bildsten}},
  \bibinfo {author} {\bibfnamefont {A.}~\bibnamefont {Dotter}}, \bibinfo
  {author} {\bibfnamefont {F.}~\bibnamefont {Herwig}}, \bibinfo {author}
  {\bibfnamefont {P.}~\bibnamefont {Lesaffre}},\ and\ \bibinfo {author}
  {\bibfnamefont {F.}~\bibnamefont {Timmes}},\ }\bibfield  {title} {\bibinfo
  {title} {{Modules for experiments in stellar astrophysics (MESA)}},\ }\href
  {https://doi.org/10.1088/0067-0049/192/1/3} {\bibfield  {journal} {\bibinfo
  {journal} {Astrophys. J. Suppl.}\ }\textbf {\bibinfo {volume} {192}},\
  \bibinfo {pages} {3} (\bibinfo {year} {2010})}\BibitemShut {NoStop}%
\bibitem [{\citenamefont {Wolf}\ \emph {et~al.}(2013)\citenamefont {Wolf},
  \citenamefont {Bildsten}, \citenamefont {Brooks},\ and\ \citenamefont
  {Paxton}}]{wolf2013hydrogen}%
  \BibitemOpen
  \bibfield  {author} {\bibinfo {author} {\bibfnamefont {W.~M.}\ \bibnamefont
  {Wolf}}, \bibinfo {author} {\bibfnamefont {L.}~\bibnamefont {Bildsten}},
  \bibinfo {author} {\bibfnamefont {J.}~\bibnamefont {Brooks}},\ and\ \bibinfo
  {author} {\bibfnamefont {B.}~\bibnamefont {Paxton}},\ }\bibfield  {title}
  {\bibinfo {title} {{Hydrogen burning on accreting white dwarfs: stability,
  recurrent novae, and the post-nova supersoft phase}},\ }\href
  {https://doi.org/10.1088/0004-637X/777/2/136} {\bibfield  {journal} {\bibinfo
   {journal} {Astrophys. J.}\ }\textbf {\bibinfo {volume} {777}},\ \bibinfo
  {pages} {136} (\bibinfo {year} {2013})}\BibitemShut {NoStop}%
\bibitem [{\citenamefont {Brooks}\ \emph {et~al.}(2016)\citenamefont {Brooks},
  \citenamefont {Bildsten}, \citenamefont {Schwab},\ and\ \citenamefont
  {Paxton}}]{brooks2016carbon}%
  \BibitemOpen
  \bibfield  {author} {\bibinfo {author} {\bibfnamefont {J.}~\bibnamefont
  {Brooks}}, \bibinfo {author} {\bibfnamefont {L.}~\bibnamefont {Bildsten}},
  \bibinfo {author} {\bibfnamefont {J.}~\bibnamefont {Schwab}},\ and\ \bibinfo
  {author} {\bibfnamefont {B.}~\bibnamefont {Paxton}},\ }\bibfield  {title}
  {\bibinfo {title} {{Carbon Shell or Core Ignitions in White Dwarfs Accreting
  from Helium Stars}},\ }\href {https://doi.org/10.3847/0004-637X/821/1/28}
  {\bibfield  {journal} {\bibinfo  {journal} {Astrophys. J.}\ }\textbf
  {\bibinfo {volume} {821}},\ \bibinfo {pages} {28} (\bibinfo {year}
  {2016})}\BibitemShut {NoStop}%
\bibitem [{\citenamefont {Schwab}\ and\ \citenamefont
  {Rocha}(2019)}]{schwab2019residual}%
  \BibitemOpen
  \bibfield  {author} {\bibinfo {author} {\bibfnamefont {J.}~\bibnamefont
  {Schwab}}\ and\ \bibinfo {author} {\bibfnamefont {K.~A.}\ \bibnamefont
  {Rocha}},\ }\bibfield  {title} {\bibinfo {title} {{Residual Carbon in
  Oxygen--Neon White Dwarfs and Its Implications for Accretion-induced
  Collapse}},\ }\href {https://doi.org/10.3847/1538-4357/aaffdc} {\bibfield
  {journal} {\bibinfo  {journal} {Astrophys. J.}\ }\textbf {\bibinfo {volume}
  {872}},\ \bibinfo {pages} {131} (\bibinfo {year} {2019})}\BibitemShut
  {NoStop}%
\bibitem [{\citenamefont {Weidenm{\"u}ller}(1961)}]{weidenmuller1961}%
  \BibitemOpen
  \bibfield  {author} {\bibinfo {author} {\bibfnamefont {H.~A.}\ \bibnamefont
  {Weidenm{\"u}ller}},\ }\bibfield  {title} {\bibinfo {title} {{First-Forbidden
  Beta Decay}},\ }\href {https://doi.org/10.1103/RevModPhys.33.574} {\bibfield
  {journal} {\bibinfo  {journal} {Rev. Mod. Phys.}\ }\textbf {\bibinfo {volume}
  {33}},\ \bibinfo {pages} {574} (\bibinfo {year} {1961})}\BibitemShut
  {NoStop}%
\bibitem [{\citenamefont {Biedenharn}\ and\ \citenamefont
  {Rose}(1953)}]{biedenharn1953theory}%
  \BibitemOpen
  \bibfield  {author} {\bibinfo {author} {\bibfnamefont {L.~C.}\ \bibnamefont
  {Biedenharn}}\ and\ \bibinfo {author} {\bibfnamefont {M.~E.}\ \bibnamefont
  {Rose}},\ }\bibfield  {title} {\bibinfo {title} {{Theory of Angular
  Correlation of Nuclear Radiations}},\ }\href
  {https://doi.org/10.1103/RevModPhys.25.729} {\bibfield  {journal} {\bibinfo
  {journal} {Rev. Mod. Phys.}\ }\textbf {\bibinfo {volume} {25}},\ \bibinfo
  {pages} {729} (\bibinfo {year} {1953})}\BibitemShut {NoStop}%
\bibitem [{\citenamefont {Condon}\ and\ \citenamefont
  {Shortley}(1951)}]{condon1951theory}%
  \BibitemOpen
  \bibfield  {author} {\bibinfo {author} {\bibfnamefont {E.~U.}\ \bibnamefont
  {Condon}}\ and\ \bibinfo {author} {\bibfnamefont {G.}~\bibnamefont
  {Shortley}},\ }\href@noop {} {\emph {\bibinfo {title} {{The Theory of Atomic
  Spectra}}}}\ (\bibinfo  {publisher} {Cambridge University Press},\ \bibinfo
  {address} {Cambridge, England},\ \bibinfo {year} {1951})\BibitemShut
  {NoStop}%
\bibitem [{\citenamefont {Ward}\ and\ \citenamefont
  {Fowler}(1980)}]{ward1980thermalization}%
  \BibitemOpen
  \bibfield  {author} {\bibinfo {author} {\bibfnamefont {R.~A.}\ \bibnamefont
  {Ward}}\ and\ \bibinfo {author} {\bibfnamefont {W.~A.}\ \bibnamefont
  {Fowler}},\ }\bibfield  {title} {\bibinfo {title} {{Thermalization of
  long-lived nuclear isomeric states under stellar conditions}},\ }\href
  {https://doi.org/10.1086/157983} {\bibfield  {journal} {\bibinfo  {journal}
  {Astrophys. J.}\ }\textbf {\bibinfo {volume} {238}},\ \bibinfo {pages} {266}
  (\bibinfo {year} {1980})}\BibitemShut {NoStop}%
\bibitem [{\citenamefont {Paxton}\ \emph {et~al.}(2015)\citenamefont {Paxton},
  \citenamefont {Marchant}, \citenamefont {Schwab}, \citenamefont {Bauer},
  \citenamefont {Bildsten}, \citenamefont {Cantiello}, \citenamefont {Dessart},
  \citenamefont {Farmer}, \citenamefont {Hu}, \citenamefont {Langer} \emph
  {et~al.}}]{paxton2015modules}%
  \BibitemOpen
  \bibfield  {author} {\bibinfo {author} {\bibfnamefont {B.}~\bibnamefont
  {Paxton}}, \bibinfo {author} {\bibfnamefont {P.}~\bibnamefont {Marchant}},
  \bibinfo {author} {\bibfnamefont {J.}~\bibnamefont {Schwab}}, \bibinfo
  {author} {\bibfnamefont {E.~B.}\ \bibnamefont {Bauer}}, \bibinfo {author}
  {\bibfnamefont {L.}~\bibnamefont {Bildsten}}, \bibinfo {author}
  {\bibfnamefont {M.}~\bibnamefont {Cantiello}}, \bibinfo {author}
  {\bibfnamefont {L.}~\bibnamefont {Dessart}}, \bibinfo {author} {\bibfnamefont
  {R.}~\bibnamefont {Farmer}}, \bibinfo {author} {\bibfnamefont
  {H.}~\bibnamefont {Hu}}, \bibinfo {author} {\bibfnamefont {N.}~\bibnamefont
  {Langer}}, \emph {et~al.},\ }\bibfield  {title} {\bibinfo {title} {{Modules
  for experiments in stellar astrophysics (MESA): binaries, pulsations, and
  explosions}},\ }\href {https://doi.org/10.1088/0067-0049/220/1/15} {\bibfield
   {journal} {\bibinfo  {journal} {Astrophys. J. Suppl.}\ }\textbf {\bibinfo
  {volume} {220}},\ \bibinfo {pages} {15} (\bibinfo {year} {2015})}\BibitemShut
  {NoStop}%
\bibitem [{\citenamefont {Paxton}\ \emph {et~al.}(2016)\citenamefont {Paxton},
  \citenamefont {Marchant}, \citenamefont {Schwab}, \citenamefont {Bauer},
  \citenamefont {Bildsten}, \citenamefont {Cantiello}, \citenamefont {Dessart},
  \citenamefont {Farmer}, \citenamefont {Hu}, \citenamefont {Langer} \emph
  {et~al.}}]{paxton2016erratum}%
  \BibitemOpen
  \bibfield  {author} {\bibinfo {author} {\bibfnamefont {B.}~\bibnamefont
  {Paxton}}, \bibinfo {author} {\bibfnamefont {P.}~\bibnamefont {Marchant}},
  \bibinfo {author} {\bibfnamefont {J.}~\bibnamefont {Schwab}}, \bibinfo
  {author} {\bibfnamefont {E.~B.}\ \bibnamefont {Bauer}}, \bibinfo {author}
  {\bibfnamefont {L.}~\bibnamefont {Bildsten}}, \bibinfo {author}
  {\bibfnamefont {M.}~\bibnamefont {Cantiello}}, \bibinfo {author}
  {\bibfnamefont {L.}~\bibnamefont {Dessart}}, \bibinfo {author} {\bibfnamefont
  {R.}~\bibnamefont {Farmer}}, \bibinfo {author} {\bibfnamefont
  {H.}~\bibnamefont {Hu}}, \bibinfo {author} {\bibfnamefont {N.}~\bibnamefont
  {Langer}}, \emph {et~al.},\ }\bibfield  {title} {\bibinfo {title} {{Erratum:
  ``Modules for experiments in stellar astrophysics (MESA): binaries,
  pulsations, and explosions''}},\ }\href
  {https://doi.org/10.3847/0067-0049/223/1/18} {\bibfield  {journal} {\bibinfo
  {journal} {Astrophys. J. Suppl.}\ }\textbf {\bibinfo {volume} {223}},\
  \bibinfo {pages} {18} (\bibinfo {year} {2016})}\BibitemShut {NoStop}%
\end{thebibliography}%

\end{document}